\documentclass[letter,11pt]{article}
\pdfoutput=1
\usepackage{heppub} 
\allowdisplaybreaks
\usepackage{hyperref}
\usepackage{graphicx}
\usepackage{amsmath}
\usepackage{amssymb} 
\usepackage[normalem]{ulem}
\usepackage{xspace}
\usepackage{slashed}
\usepackage[utf8]{inputenc}
\usepackage[T1]{fontenc}
\usepackage{subcaption}
\usepackage{scalerel,stackengine}
\usepackage{xfrac}
\usepackage{braket}
\usepackage[cal=cm,scr=euler]{mathalpha}
\usepackage{enumitem}
\usepackage[usenames,dvipsnames]{xcolor}
\usepackage[export]{adjustbox}

\newcommand{\T}{\mathcal{T}}

\usepackage{marginnote}
\usepackage[normalem]{ulem}

\preprint{}

\title{Stochastic Dynamics of Heavy Quarks in Strongly Coupled Plasma}

\author[1]{Krishna Rajagopal,}

\author[2]{Bruno Scheihing-Hitschfeld,}

\author[3]{and Urs Achim Wiedemann}

\affiliation[1]{
Center for Theoretical Physics --- A Leinweber Institute,\\ Massachusetts Institute of Technology, Cambridge, Massachusetts 02139, USA
}

\affiliation[2]{
Kavli Institute for Theoretical Physics,\\ University of California, Santa Barbara, California 93106, USA
}

\affiliation[3]{Theoretical Physics Department, CERN, CH-1211 Gen\`eve 23, Switzerland}

\emailAdd{krishna@mit.edu}
\emailAdd{bscheihi@kitp.ucsb.edu}
\emailAdd{Urs.Wiedemann@cern.ch}

\abstract{We study the stochastic dynamics of heavy quarks propagating through the strongly coupled plasma of 
$\mathcal{N}=4$ supersymmetric Yang-Mills (SYM) theory at nonzero temperature in terms of the corresponding Kolmogorov equation, which correctly describes  their kinetic equilibration and the non-Gaussian fluctuations in their momenta without having to restrict their velocity to the non-relativistic regime. Leveraging the heavy quark limit, we show that the evolution of the momentum space distribution function can be reformulated as a Hamilton-Jacobi problem, and therefore can be solved in terms of first-order ordinary differential equations. We solve these evolution equations in an infinite thermal plasma with a constant temperature for initial conditions specified by spherically symmetric heavy quark momentum distributions with a phenomenologically motivated shape that is steeply falling at large momentum. To highlight the distinctive features of the kinetic equilibration process, we compare their solutions with Fokker-Planck dynamics with the same drag coefficient and fluctuations chosen by hand to guarantee equilibration. We find qualitatively similar dynamics at small momentum, and very different dynamics at large momentum, where, much like in jet quenching phenomena, the steepness of the momentum distribution gives a larger relevance to unlikely events in which a heavy quark 
loses little momentum in a given time step. Such events are much less unlikely in the Kolmogorov evolution than in Fokker-Planck evolution with the same mean energy loss, meaning that equilibration at large momentum 
is significantly delayed.
Our results provide a systematic description of heavy quarks propagating through strongly coupled plasma from the ultra-relativistic to the non-relativistic regime and 
point the way towards implementation in phenomenological studies. 
}

\preprint{MIT-CTP/6040, CERN-TH-2026-127}

\begin{document}

\maketitle

\section{Introduction}

How does a relativistic distribution of heavy quarks evolve toward equilibrium when embedded in quark–gluon plasma (QGP)? This question, which is at the center of the present work, is of direct phenomenological relevance for the analysis of data from current and future ultra-relativistic heavy-ion collisions~\cite{Rapp:2009my,Andronic:2015wma,Aarts:2016hap,Prino:2016cni,Rapp:2018qla,Cao:2018ews,Dong:2019byy,Apolinario:2022vzg}. In such collisions, heavy quarks are produced initially unaffected by the quark–gluon plasma, with a perturbative power-law transverse momentum distribution that differs strongly from a thermal one. As heavy quarks propagate subsequently through the rapidly forming QGP, they participate in incomplete equilibration processes that are experimentally accessible via measurements of heavy-flavored hadron observables in the final state of the collision, including anisotropic flow and nuclear modification factors. Further theoretical progress in heavy-quark transport seeks to fully exploit the wealth of information provided by extensive measurements of heavy-flavor flow and nuclear modification factors at RHIC~\cite{PHENIX:2006iih,
PHENIX:2010xji,
STAR:2014wif,
STAR:2017kkh,
STAR:2018zdy,
Shi:2024eyk} and at the LHC
\cite{ALICE:2013olq,
ALICE:2014qvj,
ALICE:2017pbx,
CMS:2017qjw,
CMS:2017uuv,
ATLAS:2018hqe,
ALICE:2018gif,
ATLAS:2020yxw,
ALICE:2020iug,
CMS:2020bnz,
ATLAS:2021xtw,
ALICE:2021rxa,
ALICE:2021kfc,
CMS:2022sxl,
ALICE:2022tji,
CMS:2022vfn,
ALICE:2023gjj,
ALICE:2023hou,CMS:2024vip}.
It also lays the groundwork for a precision era of heavy-quark transport studies in LHC Run 5, enabled by new detector technologies~\cite{ALICE:2022wwr,ALICE3Scoping2025} offering order-of-magnitude improvements in pointing resolution and low-transverse-momentum rate capabilities.

Over a wide range of momenta, the evolution of heavy quarks toward equilibrium can be described in terms of kinetic transport~\cite{Svetitsky:1987gq,Moore:2004tg}. For quarks with typical thermal momenta $p \sim \sqrt{M\, T} \gg T$ and larger, this is so since a substantial change in their momentum occurs only over time scales much longer than the thermal timescale $1/T$. This separation of scales underpins the central assumption of kinetic theory --- that the spatial extent of the propagating wave packet is much smaller than the time scale over which its momentum distribution undergoes significant modification. Consequently, the heavy quark momentum distribution can be evolved through a Markovian sequence of temporally ordered, mutually incoherent, processes.

In the phenomenology of ultra-relativistic heavy-ion collisions, heavy quark flow phenomena and nuclear modification factors at 
(mildly) relativistic transverse momenta (say $p/M < 20$) have received great attention over the last two decades~\cite{vanHees:2004gq,Moore:2004tg,vanHees:2005wb,vanHees:2007me,Xu:2017obm,He:2011qa,Cao:2013ita,Das:2013kea,
Berrehrah:2014tva,
Xu:2018gux,Gossiaux:2019mjc,Andronic:2024oxz,Krishna:2025bll,Beraudo:2025nvq}, 
as they are sensitive to QGP transport properties that can be calculated from first principles in finite temperature quantum field theory. The current state of the art of such calculations includes perturbative leading-order~\cite{Moore:2004tg,Burnier:2010rp,Eller:2019spw,Scheihing-Hitschfeld:2022xqx,Scheihing-Hitschfeld:2023tuz,delaCruz:2024cix} and next-to-leading-order~\cite{Caron-Huot:2007rwy,Caron-Huot:2008dyw} results in QCD at nonzero temperature, non-perturbative lattice evaluations~\cite{Petreczky:2005nh,Caron-Huot:2009ncn,
Meyer:2011gj,
Banerjee:2011ra,
Francis:2015daa,
Brambilla:2020siz,
Brambilla:2022xbd,
Altenkort:2023eav,
HotQCD:2025fbd}, 
calculations in strongly coupled quantum field theories based on gauge-gravity duality~\cite{Casalderrey-Solana:2006fio,Gubser:2006nz,
Casalderrey-Solana:2007ahi,
Casalderrey-Solana:2011dxg,
Rajagopal:2025ukd}, as well as
the derivation of a universal condition for the in-medium equilibration of heavy quarks in any quantum field theory~\cite{Rajagopal:2025rxr}. 
In the phenomenological practice, modeling and data comparison to heavy quark transport theory is currently done either in terms of Langevin dynamics, or in terms of Boltzmann transport formulations. Both approaches have specific advantages and specific limitations that may be summarized as follows:

Langevin (or, equivalently, Fokker-Planck) dynamics is formulated~\cite{Akamatsu:2008ge,Beraudo:2009pe,He:2013zua,Das:2013kea} in terms of the heavy quark drag coefficient $\eta_D$ and the diffusion coefficients $\kappa_L$ and $\kappa_T$ that describe Gaussian longitudinal and transverse momentum fluctuations. In Langevin-type heavy quark transport theory, as in formulations of relativistic viscous fluid dynamics, one therefore deals  with an evolution equation that is defined completely in terms of fundamental properties of QGP that are calculable in quantum field theory without further model assumptions. In principle, this allows for a simple strategy of determining these fundamental QGP properties from the analysis of experimental data on heavy-flavored hadron spectra and azimuthal anisotropies measured in ultra-relativistic heavy-ion collisions. In practice, however, there is an important conceptual complication. Most experimental data are for quarks with relativistic kinematics that propagate with boost factors $\gamma > 1$ through the QGP. For such non-thermal quarks, 
the heavy-quark transport coefficients become boost-dependent, and the resulting Langevin dynamics evolves to thermal equilibrium if and only if the Einstein relation $\kappa_L(v) = 2 M T \eta_D(v) \gamma$ is satisfied. 
It is known, however, that transport coefficients calculated rigorously in quantum field theory do not satisfy this relation. In perturbative finite temperature QCD, deviations from the Einstein relation arise beyond leading logarithmic order~\cite{Moore:2004tg,DuPlessis:2026pyr}, and in the non-abelian plasma of strongly coupled ${\cal N}=4$ Super-Yang-Mills (SYM) theory, Einstein’s relation is strongly violated by a factor that scales like $\gamma^{3/2}$ at large $\gamma$~\cite{Gubser:2006nz}. 
This implies that Langevin-type formulations of heavy quark transport must be incomplete for any heavy quark momentum distribution that includes relativistic heavy quarks.

Formulations of heavy quark transport based on the Boltzmann equation
\cite{Gossiaux:2008jv,Uphoff:2014hza,Cao:2016gvr,Li:2019wri,Liu:2021dpm} overcome the latter problem because --- by construction --- the Boltzmann equation drives distributions to equilibrium irrespective of the specific form of the collision kernel. In principle, if the collision kernel is known, the fundamental QGP transport properties, $\eta_D$, $\kappa_L$ and $\kappa_T$ are determined by momentum moments of that kernel; thus, constraining the collision kernel amounts to constraining these fundamental QGP transport properties. In practice, however, a conceptual complication arises: at least for the 
strongly coupled 
small momentum transfers that dominate in experimentally relevant scenarios, little is known {\it a priori} about the functional form of the collision kernel. One therefore inevitably works with phenomenologically informed, but model-dependent, parametrizations. Although such choices can be optimized to reproduce data, the resulting constraints on transport coefficients depend on the assumed form of the Boltzmann kernel. Compared to the situation in Langevin-type transport formulations, the relation between the parameters extracted from model-data comparisons and the quantities calculable from first principles in quantum field theory is thus arguably less direct in Boltzmann transport formulations.

In light of the respective advantages and limitations of Langevin-type and Boltzmann-type transport, it would clearly be beneficial to have a formulation of heavy quark transport that:
\begin{enumerate} 
\item 
is specified entirely in terms of fundamental QGP  properties that are calculable from first principles directly from the underlying quantum field theory, without model assumptions; 
\item is dynamically complete in that it implements detailed balance and satisfies Einstein's relation for non-relativistic heavy quarks and 
satisfies the relativistic generalization of Einstein's relation valid in any quantum field theory~\cite{Rajagopal:2025rxr};
\item unlike the Boltzmann equation, does not require the specification of microscopic details of $n\to m $ collision kernels, since specifying phenomenologically relevant (non-perturbative) collision kernels may involve choices that are model-dependent and/or conceptually questionable in plasmas that show close-to-perfect fluidity.
\end{enumerate}
In recent works~\cite{Rajagopal:2025ukd,Rajagopal:2025rxr}, we have derived a Kolmogorov transport equation valid in the heavy quark limit that satisfies these three criteria. For heavy quarks with velocity ${\bf v}$,  Kolmogorov transport is governed by a momentum transfer probability distribution  $P({\bf k};{\bf v},t)$ whose first and second moments define the Langevin transport coefficients $\eta_D$, $\kappa_L$ and $\kappa_T$. However, the entire distribution $P({\bf k};{\bf v},t)$ is needed in order 
to make the description of heavy quark transport dynamically complete in the relativistic regime~\cite{Rajagopal:2025ukd,Rajagopal:2025rxr},
in the sense that it is the higher non-Gaussian moments 
of $P({\bf k};{\bf v},t)$ that 
ensure that the evolution implements detailed balance and that the thermal equilibrium momentum distribution is the stationary solution of the evolution to which the heavy quark momentum distribution evolves at late times.
The momentum transfer distribution $P({\bf k};{\bf v},t)$ --- and thus all of its moments --- is, at least in principle, calculable without additional assumptions in any 
quantum field theory at nonzero temperature. And indeed, $P({\bf k};{\bf v},t)$ has been calculated explicitly at nonzero temperature both in strongly coupled ${\cal N}=4$ SYM theory in the limit of a large number of colors and a large 't Hooft coupling~\cite{Rajagopal:2025ukd} and in weakly coupled perturbative QCD and ${\cal N}=4$ SYM theory~\cite{DuPlessis:2026pyr}.

So far, what we know about the Kolmogorov transport equation is on the level of a proof-of-principle. We know, from general field theoretic reasoning that this transport formulation satisfies the above-mentioned properties which makes it well-suited for a direct extraction of heavy quark transport properties from data-theory comparison. We know that the proposed formalism overcomes the fundamental limitation to extending Langevin-dynamics to relativistic particles and that it is arguably more directly related to fundamental QGP properties than the Boltzmann equation. The present paper aims to take the next step in rendering these considerations suitable for phenomenological applications. Here, building on the previous derivation of the Kolmogorov evolution equation and on the previous characterization of its generic properties, we analyze how this equation can be solved for phenomenologically relevant heavy quark distributions, and what is its kinematic range of validity.

As for any novel tool, it is prudent to deal with developments towards an eventual phenomenological application step by step. In particular, we restrict our discussion in the present paper to the transport of spherically symmetric distributions of heavy quarks in a spatially homogeneous, time-independent, and arbitrarily extended volume of quark-gluon plasma with temperature $T$. 
According to the fluid dynamic picture of ultra-relativistic heavy ion collisions, the expanding cooling droplet of QGP formed in such collisions is described in terms of a collective-flow velocity field that connects locally comoving regions of the fluid and their gradients. 
This motivates us to study how the heavy quark momentum distribution evolves in an equilibrated fluid at rest, leaving to future work the inclusion of many further phenomenologically relevant effects including the variation of $T$ with position and time in the finite-sized, rapidly expanding and cooling, droplets of QGP created in heavy ion collisions as well as hadronization.

Technically, the Kolmogorov transport equation is a relativistic evolution equation for the heavy quark momentum distribution $\mathscr{P}$ associated with a Markov process. 
It admits a Kramers-Moyal expansion, in which the dynamics remains linear in $\mathscr{P}$ but is expressed as an infinite series of derivative terms. In this work, we begin from the observation that -- in the heavy quark limit -- the Kolomogorov equation can be reformulated as a Hamilton-Jacobi problem involving only first order partial differential equations. After briefly reviewing the main results of our previous work~\cite{Rajagopal:2025ukd} in Section~\ref{sec2}, we make this Hamilton-Jacobi formulation of heavy quark transport explicit in Section~\ref{sec:K-as-HJ}. We also  discuss the truncation in which the Kolmogorov equation reduces to a Fokker-Planck equation upon dropping all higher non-Gaussian moments of $P(\bf{k},\bf{v},t)$ and artificially imposing the Einstein relation by hand. In addition, we introduce the computational techniques used to determine the kernel of the Kolmogorov equation  and examine the structure of the associated Hamiltonian flow and the construction of physical solutions, doing so explicitly for the specific case of the ${\cal N}=4$ SYM plasma.

Before turning to the discussion of explicit solutions of the Kolmogorov equation for the heavy quark momentum distribution $\mathscr{P}$, in Section~\ref{sec:interplay} we provide a broader overview of the aspects of heavy quark effective theory and gauge/string duality on which our formulation of the Kolmogorov equation is based. This sets the stage for discussing the 
validity of the approximations employed in deriving the Kolmogorov equation in depth and specifying the range of heavy quark velocities over which the Kolmogorov equation can be trusted. Notably, while calculations of the drag force in AdS/CFT are known to be restricted to velocities below a well-known ``speed limit'', we find that the validity of the Kolmogorov equation is instead subject to a refined bound that depends on the shape of 
$\mathscr{P}$ itself.
This bound is larger than the previously known speed limit by a factor given by the ratio between the most likely momentum transfer that a heavy quark with current Lorentz boost $\gamma$ will experience over the next time step and the most likely momentum transfer that this heavy quark experienced during the previous time step.
We discuss this subtle but quantitatively important distinction in detail in Section~\ref{sec:interplay} because it provides insight into the physical mechanism of heavy quark transport. We show that, precisely because of this distinction, the ``survivor bias'' effect noted by many authors is intrinsic to the Kolmogorov dynamics that we have formulated and is crucial for understanding how $\mathscr{P}$ evolves with time.

We present solutions of the Kolmogorov equation for heavy quark momentum distributions in Section~\ref{sec5}. As our analysis is restricted to spherically symmetric heavy quark momentum distributions evolving in a uniform and time-independent strongly coupled medium, the results are not yet suitable for phenomenological applications. Therefore, our discussion will focus on generic features of the solutions that may be expected to persist qualitatively in more realistic scenarios. In particular, our results illustrate how a heavy quark distribution that is initially far from equilibrium evolves dynamically toward equilibrium as a function of time or, equivalently, in-medium path length. We  
highlight central, and generic, features of our results for the evolution of the heavy quark momentum distribution 
at high momentum when 
(as for the $p_T$ distribution
of high-$p_T$ heavy quarks produced in a collision)
that
distribution is
a steeply falling function of momentum whose tail is nevertheless overoccupied compared to what it would be in equilibrium.
In such a regime,
the evolution of $\mathscr{P}$ is substantially slowed and its equilibration is greatly delayed
relative to what is 
predicted by the truncated
Fokker-Planck equation.
This effect originates directly from  the enhanced non-Gaussian fluctuations encoded in the full Kolmogorov equation and the resulting enhanced survivor bias.

As emphasized throughout this Introduction, we regard the present manuscript as only the first step toward rendering the Kolmogorov equation amenable to phenomenological studies. Our concluding discussion in Section~\ref{sec:ConclusionsOutlook} therefore provides not only a summary of the results obtained, but also an outlook toward several directions for future work that appear within reach.

\section{From Strongly Coupled $\mathcal{N}=4$ SYM to a Kolmogorov Equation}
\label{sec2}

The aim of this short introductory Section is two-fold. First, we recall key results and notation from Ref.~\cite{Rajagopal:2025ukd} that will be needed in the following. Second, we provide a novel derivation of the Kolmogorov equation obtained in Ref.~\cite{Rajagopal:2025ukd} that describes the evolution of the heavy quark momentum distribution in (strongly coupled) plasma.

\subsection{Recapitulation of recent results}

Throughout this work, we consider heavy quarks of mass $M$ 
and momentum ${\bf p}$ that propagate with velocity ${\bf v}({\bf p}) = \frac{\bf p}{\sqrt{{\bf p}^2 + M^2}}$ through plasma whose temperature is $T$. Our starting point is the probability $P({\bf k};{\bf v},t)$ that such a heavy quark transfers momentum ${\bf k}$ to the plasma during a time interval $t$, which is given by~\cite{Rajagopal:2025ukd,Rajagopal:2025rxr}
\begin{eqnarray}
    P({\bf k};{\bf v},t)
&=& \frac{1}{(2\pi)^3} \int d^3 {\bf L} \, e^{ - i {\bf k} \cdot {\bf L} } 
\underbrace{\exp \left( - \sqrt{\lambda} T t S_{\rm tot}({\bf L};{\bf v}) \right)}_{\equiv \langle W\left[{\cal C} \right]\rangle_T({\bf L})} \nonumber \\
&\propto& \exp \left[ - \sqrt{\lambda} T t \tilde{S}_{\rm tot}\left(  \frac{2 {\bf k} }{  \pi  \sqrt{\lambda}  T^2 t}; {\bf v} \right) \right] \, . \label{eqU1}
\end{eqnarray}
To leading order in $1/M$, this probability is defined in Heavy Quark Effective Theory (HQET) via the thermal expectation value $\langle W\left[{\cal C} \right]\rangle_T({\bf L})$  of a Wilson loop whose contour ${\cal C}$ has two long straight segments with spacetime slope $v$ and spatial separation ${\bf L}$~\cite{Rajagopal:2025ukd,Rajagopal:2025rxr}. The logarithm of $\langle W\left[{\cal C} \right]\rangle_T({\bf L})$ is proportional to the generating functional $S_{\rm tot}({\bf L};{\bf v})$ of connected moments of  $P({\bf k};{\bf v},t)$, the probability distribution for the momentum transfer ${\bf k}$. Alternatively, one can define $\tilde{S}_{\rm tot}$ via writing
\begin{equation}
    \langle W\left[{\cal C} \right]\rangle_T({\bf L})
    = \int d{\bf C}\, \exp\left(- \sqrt{\lambda} T t 
    \left[ \tilde{S}_{\rm tot}({\bf C};{\bf v}) - \frac{i  \pi T }{2} {\bf C} \cdot {\bf L} \right] \right) \, .
    \label{eqU2}
\end{equation}
Here, inserting \eqref{eqU2} into \eqref{eqU1} fixes the definition of ${\bf C}$, namely ${\bf C}\equiv\frac{2 {\bf k} }{ \pi \sqrt{\lambda} T^2 t}$\,.

In Ref.~\cite{Rajagopal:2025ukd}, we computed $\tilde{S}_{\rm tot}({\bf C};{\bf v})$ explicitly 
to leading order in $T/M$ at nonzero temperature in ${\cal N}=4$ SYM theory in the limit of large $N_c$ and large 't Hooft coupling $\lambda$.
The result takes a remarkably compact and explicit analytic form 
if expressed in terms of the Appell hypergeometric series $F_1(a,b_1,b_2;c;x,y)$:
\begin{align}
    \tilde{S}_{\rm tot}({\bf C}) &= \sqrt{1+{\bf C}^2} \frac{\pi}{32}  \frac{(z_+^4 - z_-^4 )^2 }{z_-^5 (1 - z_-^4) }  F_1 \! \left( \frac32 , \frac54 , 1 ; 3 ; -\frac{z_+^4 - z_-^4}{z_-^4} , \frac{z_+^4 - z_-^4}{1 - z_-^4} \right) + \frac{\pi}{2} |v C_3| \theta( - v C_3 )  
    \label{eqU3}
\end{align}
with
\begin{equation}
    z_{\pm}^4 \equiv \frac{1+C_\perp^2+ (1-v^2)(1+C_3^2) }{2(1+C_\perp^2+C_3^2)} \pm \frac{\sqrt{ C_\perp^4 + 2 C_\perp^2 (v^2 + C_3^2 (1 - v^2) ) + (v^2 - (1-v^2) C_3^2 )^2 } }{2(1+C_\perp^2+C_3^2)} \, . 
    \label{eqU4}
\end{equation}
Here, the $C_3$ component of the three-vector ${\bf C}= (C_\perp,C_3)$ points along ${\bf v}$.

To relate $S_{\rm tot}({\bf L};{\bf v})$ to $\tilde{S}_{\rm tot}({\bf C};{\bf v})$, we note that for large $\lambda$ Eq.~\eqref{eqU2} can be solved in the saddle point approximation. The saddle point $\bar{\bf C}({\bf L})$ is the solution to
\begin{equation}
\tfrac{\partial}{\partial {\bf C}} \left[ \tilde{S}_{\rm tot}({\bf C};{\bf v})  - \tfrac{i \pi T }{2} {\bf C} \cdot {\bf L} \right] = 0\ . 
\end{equation}
This yields the implicit relation 
\begin{equation}
{\bf L} = - \frac{2i}{ \pi T} \frac{\partial \tilde{S}_{\rm tot} }{\partial {\bf C} }
\Big\vert_{{\bf C} = {\bf \bar{C}}({\bf L})} 
\label{eqU5}
\end{equation} 
between ${\bf L}$ and ${\bf C}$ and yields
\begin{align}
    S_{\rm tot}({\bf L};{\bf v}) = \left[ \tilde{S}_{\rm tot}({\bf C};{\bf v}) - \frac{i \pi T}{2} {\bf C} \cdot {\bf L} \right]_{{\bf C} = {\bf \bar{C}}({\bf L}) } = \left[ \tilde{S}_{\rm tot}({\bf C};{\bf v}) - {\bf C} \cdot \frac{\partial \tilde{S}_{\rm tot} }{\partial {\bf C} } \right]_{{\bf C} = {\bf \bar{C}}({\bf L}) } \, . \label{eqU6}
\end{align}
In this way, we see that $S_{\rm tot}({\bf L}; {\bf v})$ and $\tilde{S}_{\rm tot}({\bf C};{\bf v})$ are Legendre transforms of each other. Throughout this work, we define the Legendre transform so as to include a prefactor $i\pi T/2$ in the conjugate variable, as doing so simplifies many subsequent expressions.  

\subsection{A novel derivation of the Kolmogorov equation} \label{sec:novel-derivation}

We want to describe the evolution of the momentum distribution $\mathscr{P}({\bf p}, t )$ of an ensemble of heavy quarks with time $t$. Anticipating that we shall work to leading order in $T/M$ and that the normalization of $\mathscr{P}({\bf p}, t )$ is subleading, we parametrize this distribution in terms of a function $f$ as
\begin{equation}
    \mathscr{P}({\bf p}, t ) = \exp\left( \frac{M}{T} f({\bf p},t) \right)\, .
    \label{eqU7}
\end{equation}
Convoluting $\mathscr{P}({\bf p}, t )$ with the probability \eqref{eqU1} that a heavy quark transfers momentum ${\bf k}$ within a small time step $\Delta t$, we can write the quark momentum distribution at the later time $t+\Delta t$ as
\begin{eqnarray}
    \mathscr{P}({\bf p}, t+\Delta t )  &=&
    \int d^3{\bf k}\, P({\bf k};{\bf v}({\bf p}+{\bf k}),\Delta t)\, 
    \mathscr{P}({\bf p}+{\bf k}, t )\, .
    \label{eqU8}
\end{eqnarray}
To analyze this expression, we introduce dimensionless variables, including the dimensionless momentum ${\bf u}$ of heavy quarks and the dimensionless momentum transfer ${\bf C}$ from those quarks to the plasma, and at the same time introduce
the natural dimensionless timescale $\tau $ over which the distribution $\mathscr{P}$ evolves:
\begin{eqnarray}
    \tau \equiv t \frac{\sqrt{\lambda}T^2}{M}\, ,\qquad {\bf u}\equiv \frac{\bf p}{M}
    \, ,\qquad {\bf C} \equiv \frac{2{\bf k}}{\pi \sqrt{\lambda} T^2 \Delta t} 
    = \frac{2{\bf k}}{\pi M \Delta \tau} \, .
    \label{eqU9}
\end{eqnarray}
Note that $u\equiv |{\bf u}| = \gamma v$. 
Up to prefactors of order ${\cal O}(1)$, Eq.~\eqref{eqU8} then takes the form
\begin{align}
    &\mathscr{P}({\bf u}, \tau+\Delta \tau ) 
    = \exp\left( \frac{M}{T} f({\bf u},\tau + \Delta \tau) \right)
    \nonumber \\
    &=
    \int d^3{\bf C} \exp \left[ \frac{M}{T} \left( f \! \left({\bf u} + \Delta \tau \frac{\pi}{2} {\bf C}, \tau \right) - \Delta \tau \, \tilde{S}_{\rm tot} \! \left( {\bf C} ; {\bf v} \left({\bf u} + \Delta \tau \frac{\pi}{2} {\bf C}  \right)  \right) \right) \right] \, . \nonumber\\
    &= \int d^3{\bf C} \exp \left[ \frac{M}{T} \left( f({\bf u},\tau) + \Delta \tau \left[ \frac{\pi}{2} {\bf C} \cdot \frac{\partial f}{\partial {\bf u}}({\bf u},\tau) - \tilde{S}_{\rm tot} \! \left( {\bf C} ; {\bf v} ({\bf u}) \right) \right] + \mathcal{O}(\Delta \tau^2) \right) \right] \nonumber \\
    &=  \exp \left[ \frac{M}{T} \left( f({\bf u},\tau) - \Delta \tau \left[  \tilde{S}_{\rm tot} \! \left( {\bf C} ; {\bf v} ({\bf u}) \right) -  {\bf C} \cdot \frac{\partial  \tilde{S}_{\rm tot}}{\partial {\bf C}}  \right]_{{\bf C} = \bar{\bf C}(\partial_{\bf u} f ) } + \mathcal{O}(\Delta \tau^2) \right) + \mathcal{O} \! \left( \left( \frac{T}{M}\right)^0 \right) \right] \nonumber \\
    &= \exp \left[ \frac{M}{T} \left( f({\bf u},\tau) - \Delta \tau S_{\rm tot}(-i T^{-1} \partial_{\bf u} f ;{\bf v} ({\bf u})) + \mathcal{O}(\Delta \tau^2) \right) + \mathcal{O} \! \left( \left(\frac{T}{M}\right)^0 \right) \right] \, .
    \label{eqU10}
\end{align}
Here, the second line  results from rewriting 
Eq.~\eqref{eqU8} in terms of the dimensionless variables \eqref{eqU9}, the third line follows from assuming that $\Delta\tau \tfrac{\pi}{2} {\bf C}$ is small enough that $f \! \left({\bf u} + \Delta \tau \frac{\pi}{2} {\bf C}, \tau \right)$ can be approximated to first order in a Taylor expansion, the fourth line uses the fact that $M/T$ is large so that the integral can be done via the saddle point approximation, and the last equality follows from Eq.~\eqref{eqU6}. Interestingly, the parametric requirement that must be satisfied
if the integral is to be carried out in the saddle point approximation is
$\Delta \tau M/T = \sqrt{\lambda} T \Delta t \gg 1$, 
which is independent of $M$ when expressed in terms of $\Delta t$. This observation serves to emphasize the importance of
the strong coupling limit. The saddle point $\bar{\bf C}(\partial_{\bf u} f )$ at which the integral is evaluated is found by requiring that the ${\bf C}$-derivative of the exponent in the integrand (see the third line of Eq.~\eqref{eqU10}) vanishes. This yields the implicit equation
\begin{equation}
    \frac{\partial f}{\partial {\bf u}} = \frac{2}{\pi} \frac{\partial  \tilde{S}_{\rm tot}}{\partial {\bf C}} \Big\vert_{{\bf C} = \bar{\bf C}(\partial_{\bf u} f )}\, ,
    \label{eqU11}
\end{equation}
which coincides with Eq.~\eqref{eqU5} for ${\bf L}=-i  T^{-1} \partial_{\bf u} f$. This justifies our use of Eq.~\eqref{eqU6} in the last equality in Eq.~\eqref{eqU10} to write the exponent in terms of $S_{\rm tot}(-i T^{-1} \partial_{\bf u} f;{\bf v}({\bf u}))$. 
Finally, comparison of the first and last line of \eqref{eqU10} yields the time evolution equation
\begin{equation}
    \partial_\tau f = - 
    S_{\rm tot}\left(-\frac{i}{T} \frac{\partial f}{\partial {\bf u}};{\bf v} ({\bf u})\right)\, .
    \label{eqU12}
\end{equation}
Eq.~\eqref{eqU12} is the Kolmogorov equation
\begin{equation}
    \partial_t \mathscr{P}({\bf p},t) = - \sqrt{\lambda} T 
    S_{\rm tot}\left(- i \frac{\partial}{\partial {\bf p}};{\bf v} ({\bf p})\right)  \mathscr{P}({\bf p},t) \, ,
    \label{eqU13}
\end{equation}
introduced in Ref.~\cite{Rajagopal:2025ukd}, and generalized in Ref.~\cite{Rajagopal:2025rxr}, if restricted to leading order in $T/M$. 

We make several remarks about the Kolmogorov equation \eqref{eqU12} and the new derivation that we have provided. 
\begin{enumerate}
\item Equilibrium is attained when $\tfrac{\partial f}{\partial {\bf u} } = - {\bf v}$, i.e., when $\mathscr{P}$ takes the form of a Boltzmann distribution.  Therefore, in terms of Eq.~\eqref{eqU12}, the equilibration condition is $S_{\rm tot}\left(\tfrac{i {\bf v} }{T} ;{\bf v} ({\bf u})\right) = 0$. This property is indeed satisfied by $S_{\rm tot}$ as specified by Eq.~\eqref{eqU6}. 
This is not a fortuitous consequence of this calculation, but rather the general equilibration condition heavy quarks satisfy in any quantum field theory~\cite{Rajagopal:2025rxr}.
\item Eq.~\eqref{eqU12} is a first-order, nonlinear, partial differential equation. It takes the form of a Hamilton-Jacobi equation. We shall exploit this fact in subsequent Sections.
\item In writing Eq.~\eqref{eqU8}, we have assumed a Markovian stochastic process for which the momentum transfer probabilities \eqref{eqU1} satisfy the addition property of independent random variables, 
$P({\bf k};{\bf v},t_1+t_2) = 
\int d{\bf q}\, P({\bf k}-{\bf q},{\bf v},t_1)\, P({\bf q},{\bf v},t_2)$. This ensures that the dynamics is independent of the choice of $\Delta t$. Short term memory effects that may or may not play a role in the evolution of heavy quarks in the ${\cal N}=4$ SYM plasma are neglected in this way,  and the Kolmogorov equation \eqref{eqU12} based on \eqref{eqU1} does not account for them.
\item The physical interpretation of ${\bf C}$ becomes apparent in the above derivation: it encodes the most likely rate of momentum transfer that a heavy quark with momentum ${\bf p}$ experienced in the previous (infinitesimal) time interval in order to reach its present momentum. While ${\bf C}$ will often be positive, indicating that the heavy quark underwent energy loss, as one would intuitively expect, it is also possible for ${\bf C}$ to be negative --- indicating that it is more likely that said heavy quark gained energy in the immediately prior time step. In fact, this is a feature of the equilibrium (Boltzmann) state, as we discuss further in Section~\ref{sec:H-Flow}.
\item The derivation of Eq.~\eqref{eqU10} relies on taking $\Delta\tau \tfrac{\pi}{2} {\bf C} = \tfrac{\bf k}{M}$ sufficiently small. Since  ${\bf k}$ is the momentum transferred from the heavy quark to the plasma during the time $\Delta t$, and since we obtain the Kolmogorov equation \eqref{eqU12} from \eqref{eqU10} in the limit $\Delta\tau \to 0$, this approximation is justified. We note, however, that the starting point \eqref{eqU1} is written in terms of a Wilson loop $\langle W\left[{\cal C} \right]\rangle_T({\bf L})$ whose contour extends over a long time $t$. We shall discuss this further in Section~\ref{sec:general-validity}. 
\item
Remarkably, while our derivation requires ${\bf C}\Delta\tau \tfrac{\pi}{2} = {\bf k}/M$ to be sufficiently small, none of the transformations made in Eq.~\eqref{eqU10} requires ${\bf u}= {\bf p}/M$ to be small. However, the most likely value ${\bf k}_{\rm max}$ of $P({\bf k};{\bf v}({\bf p}+{\bf k}),\Delta t)$ increases with ${\bf p}$. Depending on the functional shape of $\mathscr{P}({\bf p}, t )$, this may limit the range of validity of the Kolmogorov equation at large $\vert {\bf u}\vert$. 
Any discussion of the range of validity of the Kolmogorov equation in ${\bf p}$ is thus inevitably tied to assumptions about the functional shape of $\mathscr{P}({\bf p}, t )$. 
We shall discuss this further in Sections~\ref{sec:k-validity} and~\ref{sec:sym-validity}.
\end{enumerate}

\section{The Kolmogorov equation as a Hamilton-Jacobi equation --- and its solution} 
\label{sec:K-as-HJ}

In classical mechanics, a standard way to solve a system described by 
a distribution $f$ that is a function of generalized 
coordinates ${\bf u}$, with generalized momenta 
${\bf x} \equiv \tfrac{\partial f}{\partial {\bf u}}$,
whose evolution is governed by
a Hamiltonian 
$K({\bf x},{\bf u})$ 
is to find the canonical transformation that relates the coordinates $({\bf x},{\bf u})$ to the initial coordinates $({\bf x}_0,{\bf u}_0)$ at time $\tau_0=0$. Finding this canonical transformation 
\begin{eqnarray}
    {\bf u} &=& {\bf u}\left({\bf x}_0,{\bf u}_0,t\right)\, ,
    \label{eqHJ1} \\
    {\bf x} &=& {\bf x}\left({\bf x}_0,{\bf u}_0,t\right)
    \label{eqHJ2}
\end{eqnarray}
amounts to solving the system. The requirement that such canonical transformations preserve the Hamiltonian variational principle 
$\delta \int_{t_1}^{t_2} \left( {\bf x}\cdot{\bf u} - K({\bf x},{\bf u})\right) dt = 0$ implies the Hamilton-Jacobi equation 
\begin{equation}
    \partial_\tau f = - K\left( \frac{\partial f}{\partial {\bf u}} , {\bf u}\right) \, .
    \label{eqHJ3}
\end{equation}
The Kolmogorov equation \eqref{eqU12} is a Hamilton-Jacobi equation with Hamiltonian 
\begin{equation}
    K({\bf x}, {\bf u}) \equiv S_{\rm tot} \left(- \frac{i{\bf x}}{ T}; {\bf v}({\bf u}) \right)\, .
    \label{eqHJ4}
\end{equation}
It can thus be solved by solving the equivalent problem defined by the Hamiltonian evolution equations
\begin{eqnarray}
    \dot{\bf x} &=& - \frac{\partial K}{\partial {\bf u}}\, , \label{eqHJ5} \\
    \dot{\bf u} &=&  \frac{\partial K}{\partial {\bf x}}\, ,
    \label{eqHJ6}
\end{eqnarray}
which determine trajectories in the Hamiltonian phase space $\left({\bf x}(\tau),{\bf u}(\tau) \right)$. To avoid confusion, we stress that in our physics problem, ${\bf u}$ denotes the dimensionless momentum defined in Eq.~\eqref{eqU9}. However, when setting up a solution to the Hamilton-Jacobi equation~\eqref{eqHJ3}, ${\bf u}$ plays the role of a generalized coordinate, and it is ${\bf x} = \tfrac{\partial f}{\partial {\bf u}}$ that assumes the role of the generalized momentum.

For our problem, the initial condition for Eq.~\eqref{eqU7} is given by $f({\bf u}_0,\tau_0)$. We can therefore label the trajectories \eqref{eqHJ1}, \eqref{eqHJ2} by their initial generalized coordinates ${\bf u}_0$ only. Then, using ${\bf u}_0$ to identify each trajectory and finding the inverse map ${\bf u}_0 = {\bf u}_0({\bf u},\tau)$ at any fixed $\tau$, we can construct 
\begin{equation}
    \frac{\partial f}{\partial {\bf u}} ({\bf u},\tau) = {\bf x} \left({\bf u}_0({\bf u},\tau),\tau \right)\, ,
    \label{eqHJ7}
\end{equation}
from which we can determine $f$. This way of solving first order partial differential equations is sometimes called the method of characteristics. We shall employ it for the numerical solution of the Kolmogorov equation later in this Section.

\subsection{The Kolmogorov equation for spherically symmetric distributions}

In this subsection, we introduce the simpler but still 
nontrivial problem of studying the 
equilibration of a heavy quark momentum distribution function that is spherically symmetric at all times, $f({\bf u})=f(u)$, to which we shall restrict our numerical analysis in the present work. 
We then discuss two important technical points: (i) for the result \eqref{eqU3} explicitly derived in 
${\cal N}=4$ SYM theory, the domain of $K$ is smaller than what one may naively expect; and (ii) how to proceed for cases where, as in Eq.~\eqref{eqU3}, we have an explicit expression for $\tilde{S}_{\rm tot}$ but only an implicit expression for the Hamiltonian $S_{\rm tot}$ that enters the Kolmogorov equation.

Since $S_{\rm tot}\left({\bf L};{\bf v}({\bf u})\right)$ is the generating functional for connected momentum moments, the Kolmogorov kernel $K$ can be written as a moment expansion
\begin{align}
    K({\bf x},{\bf u}) &= -  \sum_{m,n=0}^\infty \frac{c_{mn}(u)}{(2m)!n!} \left({\bf x}^2 - \left(\frac{{\bf u} \cdot {\bf x} }{u}  \right)^2 \right)^m \!\! \left(\frac{{\bf u} \cdot {\bf x}}{u} \right)^n
    \label{eqHJ8} \\
    &= -  \sum_{m,n=0}^\infty \frac{c_{mn}(u)}{(2m)!n!} \left(\frac{ (\nabla_{\Omega} f)^2 }{u^2}  \right)^m \!\! \left(\frac{\partial f}{\partial u} \right)^n\, ,
    \label{eqHJ9}
\end{align}
where we have changed to spherical coordinates with $\nabla_{\Omega} f \equiv \hat{\theta} \partial_\theta f + \tfrac{\hat{\phi}}{\sin \theta} \partial_\phi f $ in the last expression. Here, $c_{mn}(u) = \langle k_1^{2m} k_3^n \rangle_c / ( \sqrt{\lambda} \, T^{2m+n+1} \Delta t) $ are directly determined by the cumulants $\langle k_1^{2m} k_3^n \rangle_c$ of $P({\bf k};{\bf v},\Delta t)$. 

If the initial distribution of heavy quarks is spherically symmetric, $f({\bf u}_0) = f(u_0)$, then evolution with the Kolmogorov equation 
preserves this spherical symmetry, and the problem of solving  \eqref{eqU12} reduces to a one-dimensional isotropic Kolmogorov equation, 
\begin{align}
    \partial_\tau f = - K \! \left( \frac{\partial f}{\partial u} , u \right)
    \, . \label{eqHJ10}
\end{align}
In the remainder of this work, we focus on solving this simpler problem. The solution of \eqref{eqHJ10} will inform us about the thermalization process of initially isotropic distributions of heavy quarks whose initial momentum distribution can be far from thermal equilibrium. 
We note as an aside that the problem of how a non-isotropic distribution with non-vanishing $\nabla_{\Omega} f $ isotropizes while thermalizing can be studied, in principle, with the same techniques that we use here. However, doing so would require solving the full equation \eqref{eqU12}, or the equivalent problem defined by the three-dimensional Hamiltonian evolution equations \eqref{eqHJ5} and \eqref{eqHJ6}.

\subsubsection{Fokker-Planck truncation of the Kolmogorov equation}

Our numerical solution of the isotropic Kolmogorov equation \eqref{eqHJ10}
will be based on knowledge of the full $K(x,u)$; it will not rely on truncating its moment expansion  \eqref{eqHJ9}. However, for some qualitative remarks and comparisons, the moment expansion will be useful in the following. For ${\cal N}=4$ SYM, explicit expressions for all moments 
$c_{mn}$ were given in Ref.~\cite{Rajagopal:2025ukd}. For the evolution \eqref{eqHJ10} of isotropic systems, only the coefficients $c_{0n}$  matter. And, the two lowest order isotropic coefficients serve to specify the drag coefficient $\eta_D$ and the longitudinal momentum broadening coefficient $\kappa_L$ according to
\begin{align}
    c_{01} &= \frac{M}{\sqrt{\lambda}T^2}    \eta_D \gamma v\, , \label{eqHJ11}\\
    c_{02} &= \frac{M}{\sqrt{\lambda}T^2}  \frac{\kappa_L}{M\, T}\, .
    \label{eqHJ12}
\end{align}
To the extent that higher moments can be neglected,
$\eta_D$ specifies the mean momentum that a heavy
quark with momentum ${\bf p}$ transfers to the plasma in a time $\Delta t$ via
$\langle {\bf k}\rangle = \eta_D {\bf p} \Delta t$
and $\kappa_L$ specifies the fluctuations in the longitudinal momentum transfer (Gaussian, if higher moments are neglected) via
$\langle k_3^2\rangle - \langle k_3 \rangle^2 = \kappa_L \Delta t$. 
Neglecting higher moments 
in the momentum transfer probability distribution $P({\bf k};{\bf v},\Delta t)$ is equivalent to
 truncating the Kolmogorov equation to second order in derivatives which leads to an evolution equation of the Fokker-Planck type. For the isotropic Kolmogorov equation~\eqref{eqHJ10}, this truncation simplifies the kernel on the right-hand side, yielding 
\begin{equation}
    K_{\rm FP}(x,u) = -  \frac{M}{\sqrt{\lambda}T^2} \left(
    \eta_D \gamma\, v\, x + \frac{1}{2} \frac{\kappa_L}{M\, T} x^2 \right)\, .
        \label{eqHJ13}
\end{equation}
Evolution governed by this truncated kernel will drive distributions to equilibrium if and only if drag and longitudinal momentum broadening satisfy the Einstein relation $\kappa_L = 2\, T\, M\, \eta_D\, \gamma$. However, it is known that this relation is typically violated for heavy quarks with any nonzero $v$ in gauge theory plasmas at weak or strong coupling. 
In particular, in the strong t'Hooft coupling limit of ${\cal N}=4$ SYM theory, the drag and longitudinal diffusion coefficients take the explicit forms $\eta_D = \tfrac{1}{2}\pi\sqrt{\lambda} T^2/M$ and $\kappa_L = \pi \sqrt{\lambda} \gamma^{5/2} T^3$, and the Einstein relation is 
badly violated, by a factor of $\gamma^{3/2}$.
Describing heavy quark thermalization then requires employing the full Kolmogorov kernel --- including all higher moments in \eqref{eqHJ9}~\cite{Rajagopal:2025ukd} --- and generalizing  the Einstein relation~\cite{Rajagopal:2025rxr}.   

Finally, we note that it is common phenomenological practice when describing the equilibration of heavy quarks in strongly coupled plasma to proceed as if the momentum transfer fluctuations were Gaussian, take the strongly coupled ${\cal N}=4$ SYM theory result $\eta_D=\frac{\pi}{2}\sqrt{\lambda}T^2/M$ for the drag coefficient, and enforce the Einstein relation by defining $\kappa_L \equiv 2\, T\, M\, \eta_D\, \gamma$ in terms of $\eta_D$ even though this disagrees with the known result for $\kappa_L$ calculated explicitly
in ${\cal N}=4$ SYM theory. With these assumptions, the Fokker-Planck kernel \eqref{eqHJ13} takes the form 
\begin{equation}
    K_{\rm FP-E}(x,u) = - \frac{\pi}{2} \gamma\, x\, \left(v + x\right)\, .
        \label{eqHJ14}
\end{equation}

\subsubsection{Domain of the isotropic Kolmogorov kernel in strongly coupled ${\cal N}=4$ SYM theory}

We focus henceforth on the full isotropic Kolmogorov kernel $K(x,u) = S_{\rm tot}(-ix,u)$ of strongly coupled ${\cal N}=4$ SYM theory, including all of its moments. Correspondingly, its
calculation is more involved than for the Fokker-Planck kernel. In this subsection, we present its evaluation and analyze the domain of $(x,u)$ in which it is defined.

For this quantum field theory,  we have an explicit expression \eqref{eqU3} for the Legendre transform $\tilde{S}_{\rm tot}$ of $K(x,u)$. 
It proves convenient to define
\begin{equation}
\tilde{K}(C_3,u) \equiv \tilde{S}_{\rm tot}(C_3) - C_3\frac{\partial \tilde{S}_{\rm tot} }{\partial C_3}\ ,
\label{eq:new-Ktilde}
\end{equation}
as with this definition the Legendre transform relationship 
\eqref{eqU6} takes the simplified form
\begin{equation}
K(x,u) =  \tilde{K}(C_3,u) \Big\vert_{C_3 = \bar{C}_3(x,u)}\, ,
\label{eq:new-Legendre}
\end{equation}
where $\bar{C}_3(x,u)$ is defined implicitly via the condition \eqref{eqU5} of the Legendre transform that, as in Eq.~\eqref{eqU11}, we can rewrite as
\begin{equation}
    x = \frac{2}{\pi} \frac{\partial \tilde{S}_{\rm tot} }{\partial C_3} \Bigg\vert_{C_3 = \bar{C}_3(x,u)}\, .\label{eqHJ16}
\end{equation}
(Recall that $u=\gamma v$ and that $L_3 = - i x / T$.)
The expression $\tilde{K}(C_3,u)$ defined via  \eqref{eq:new-Ktilde} takes a simpler form than $\tilde{S}_{\rm tot}$ in \eqref{eqU3}. First, the term proportional to the $\Theta$-function  in $\tilde{S}_{\rm tot}$ is linear in $C_3$ and therefore does not contribute to $\tilde{K}(C_3,u)$. Second, the $C_3$-symmetric term can be expressed in terms of the ordinary hypergeometric function ${}_2 F_1(a,b;c;x)$. We find
\begin{align}
    \tilde{K}(C_3,u)
    &= \frac{\pi}{4 z_- \sqrt{1 + C_3^2} } \left[ {}_2 F_1 \! \left( \frac14, \frac12, 1, -\frac{z_+^4-z_-^4}{z_-^4}  \right) - \frac{1}{(1+u^2)z_-^4} {}_2 F_1 \! \left( \frac54, \frac12, 1, -\frac{z_+^4-z_-^4}{z_-^4}  \right) \right]\, ,
    \label{eqHJ17}
\end{align}
where
\begin{equation}
    z_\pm^4 \equiv \left[ 1 + \frac{u^2+C_3^2 \mp \sqrt{(u^2-C_3^2)^2} }{2} \right]^{-1} \, .
        \label{eqHJ18}
\end{equation}
The Kolmogorov kernel is then defined by evaluating $\tilde{K}$ at the saddle point $\bar{C}_3(x,u)$ as in Eq.~\eqref{eq:new-Legendre}.
For the purpose of numerical calculations, rather than determining $\bar{C}_3(x,u)$ from condition \eqref{eqHJ16}, 
it is more efficient to rewrite this conditions in terms of $\tilde{K}$. We do so by noting that the definition of $\tilde{K}(C_3,u)$ \eqref{eq:new-Ktilde} implies that
\begin{equation}
    \frac{\partial \tilde{K}}{\partial C_3} = -C_3 \frac{\partial^2 \tilde{S}_{\rm tot} }{\partial C_3^2} \, ,
    \label{eqHJ20}
\end{equation}
and then the condition \eqref{eqHJ16} 
for $\bar{C}_3(x,u)$ --- in terms of the one for the inverse map $x(C_3,u)$ satisfying $x(\bar{C}_3(x,u),u) = x$ --- reads
\begin{equation}
    \frac{\partial x}{\partial {C}_3} = - \frac{2}{\pi {C}_3} \frac{\partial \tilde{K}}{\partial {C}_3} \, .
    \label{eqHJ21}
\end{equation}
The right hand side of this expression is an explicitly known function determined from \eqref{eqHJ17} and $\bar{C}_3(x,v)$ is then obtained via integrating \eqref{eqHJ21} and inverting the resulting $x(C_3,v)$.

\begin{figure}
    \centering
    \includegraphics[width=0.49\linewidth]{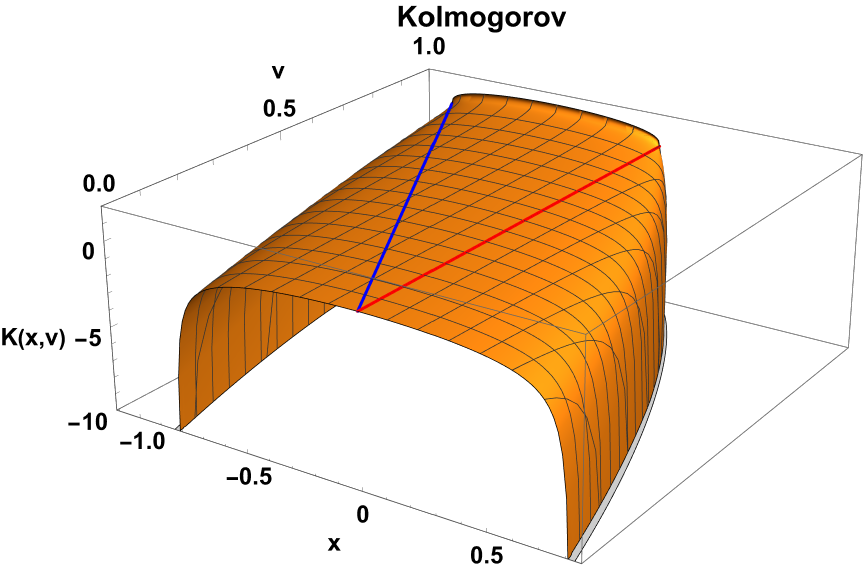}
    \includegraphics[width=0.49\linewidth]{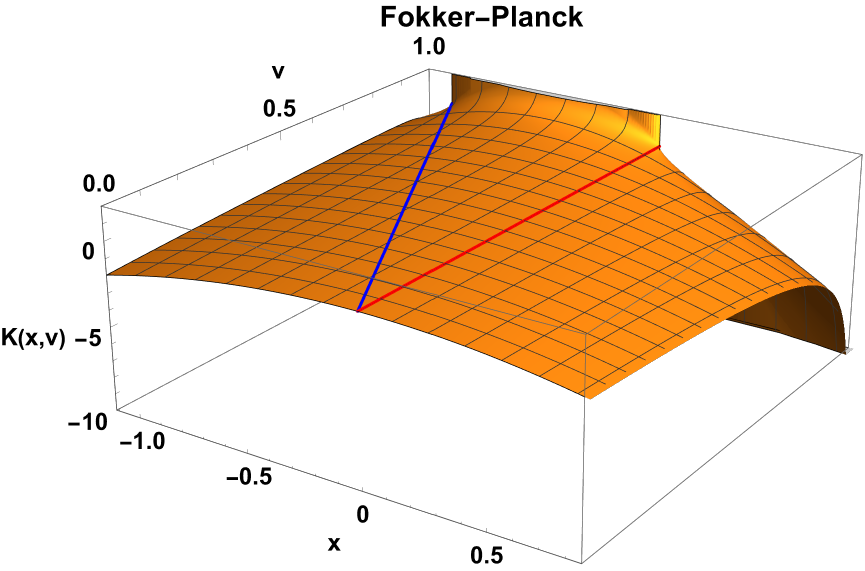}
    \caption{Left: The Kolmogorov evolution operator $K(x,u)$ 
    for strongly coupled ${\cal N}=4$ SYM theory, computed as described in the text, plotted as a function of $x$ and $v=u/\sqrt{1+u^2}$. $K(x,u)$ diverges along the curves $x=-\frac{v}{2}\pm x_b(v)$, with $x_b(v)$ given in Eq.~\eqref{eqHJ24}, which define the domain of the isotropic Kolmogorov kernel in strongly coupled ${\cal N}=4$~SYM theory. The red line corresponds to a homogeneous distribution
    and the blue line corresponds to the equilibrium Boltzmann distribution. 
    $K$ vanishes on these lines, and on these lines only. Right: Same plot for the Fokker-Planck truncation~\eqref{eqHJ14} of the Kolmogorov kernel in which the Einstein relation has been artificially enforced. At any fixed $v$, the surface in the right panel is described by a parabola centered at $x = -v/2$ with zeros at $x = 0$ and $x= -v$. By construction, the linearizations of this surface around the red and blue lines agree with the linearizations of the full Kolmogorov evolution operator in the left panel around those lines, meaning that the Fokker-Planck kernel~\eqref{eqHJ14} is guaranteed to be accurate to first order in gradients $x = f'$, either around a flat distribution or around a Boltzmann distribution. In general, however, it need not resemble the full Kolmogorov kernel in any other regime. 
    }
    \label{fig1}
\end{figure}

From Eq.~\eqref{eqHJ16}, one sees that $x=0$ corresponds to the minimum of $\tilde{S}_{\rm tot}$. At this minimum, the momentum transfer probability \eqref{eqU1} is maximal and the value of $C_3$ --- by virtue of the large $t$ and large $\sqrt{\lambda}$ limits --- coincides with the value that sets the drag coefficient, namely
$\langle C_3 \rangle = v \gamma = u$. (Recall that $\langle k_3 \rangle = \eta_D u M \Delta t = \Delta t \tfrac{\pi}{2} \sqrt{\lambda} T^2 u = \Delta t \tfrac{\pi}{2} \sqrt{\lambda} T^2 \langle C_3 \rangle$.)
One then obtains
\begin{equation}
    x(C_3,u) = -\frac{2}{\pi} \int_{\langle C_3 \rangle}^{C_3} \frac{dC_3'}{C_3'} \frac{\partial \tilde{K}}{\partial C_3'}(C_3',u) \, ,
        \label{eqHJ22}
\end{equation}
which fully specifies $x$. Either via this integral, or equivalently, via the Appell hypergeometric function, one can take the limit as $C_3 \to \pm \infty$ in order to determine the range of $x(C_3,u)$. In terms of $v = u/\sqrt{1+u^2}$, a direct calculation shows that
\begin{equation}
    \lim_{C_3 \to \pm \infty} x(C_3,u) = -\frac{v}{2} \pm x_b(v) \, ,
        \label{eqHJ23}
\end{equation}
where
\begin{equation}
    x_b(v) = \frac{v}2 + \frac{9 \sqrt{\pi} \left[ 4 (1-v^2)^{3/4} - v\, {\rm B} \left( 1 - v^2, \frac34, \frac12 \right) \right] }{64 \sqrt{2} \, \Gamma(7/4)^2 }\ ,
        \label{eqHJ24}
\end{equation}
with ${\rm B}$  the incomplete Beta function. One can also show that (again, with $v = u/\sqrt{1+u^2}$)
\begin{equation}
    \lim_{x \to -v/2 \pm x_b(v)} K(x,u) = \lim_{C_3 \to \pm \infty} \tilde{K}(C_3,u) = -\infty \, ,
        \label{eqHJ25}
\end{equation}
which means that $K$ diverges at the endpoints of the interval on which $x$ is defined.
We have evaluated the Kolmogorov kernel given in Eq.~\eqref{eq:new-Legendre} as described above. In the left panel of Fig.~\ref{fig1}, we present the result for $K(x,u)$, plotted as a function of $x$ and $v = u/\sqrt{1+u^2}$.
The curves $x=-\frac{v}{2}\pm x_b(v)$ where both $C_3$ and $K(x,u)$ diverge are apparent. In strongly coupled ${\cal N}=4$ SYM theory, the isotropic Kolmogorov kernel $K(x,u)$ is only defined in the domain of $(x,u)$ between these curves.

\subsubsection{Comparing the Kolmogorov kernel and its Fokker-Planck truncation} \label{sec:comparing-K-FP-conceptual}

In Fig.~\ref{fig1}, we compare the result for $K(x,u)$
in strongly coupled ${\cal N}=4$ SYM theory to the Fokker-Planck truncation $K_{\rm FP-E}(x,u)$ of equation~\eqref{eqHJ14}, where the Einstein relation has been imposed artificially.\footnote{Even though it is not the truncation of a series expansion
of the full Kolmogorov kernel, we will refer to this kernel as the Fokker-Planck ``truncation'' of the full Kolmogorov kernel throughout the rest of our exposition as a shorthand for ``truncation where the Einstein relation has been enforced by hand.''} Both the full Kolmogorov evolution equation and its Fokker-Planck truncation admit two stationary, time-independent solutions for which the evolution kernel $K$ vanishes. The first is the trivial solution, corresponding to a homogeneous distribution with $x = \partial f/\partial u = 0$ for all $u$, that samples the kernels around the red lines shown in Fig.~\ref{fig1}. The second is the thermal Boltzmann distribution with $f(u) = \sqrt{1+u^2}$ for which $x = \tfrac{\partial f}{\partial u} = \tfrac{u}{\sqrt{1+u^2}}  = v$. This samples the kernels along the blue lines in Fig.~\ref{fig1}.

As seen in Fig.~\ref{fig1}, the Fokker-Planck kernel resembles the full Kolmogorov kernel in the region between the red and blue lines, except for the asymptotic behavior as $v\to 1$. The resemblance is qualitative in nature for most of the region in between these two lines. Quantitatively, however, they are only guaranteed to agree in a small vicinity around the red line and a small vicinity around the blue line, because the Fokker-Planck kernel $K_{\rm FP-E}(x,v)$ is constructed
as a quadratic function of $x$, with a maximum at $x = -v/2$, whose first derivative matches that of the full Kolmogorov kernel at the line $x = 0$ and at the line $x = - v$. The KMS relation $K(x,v) = K(-v-x,v)$ established in Ref.~\cite{Rajagopal:2025rxr} guarantees that matching the first derivatives at either line 
automatically matches the first derivatives at the other line, and
gives the same result for $K_{\rm FP-E}(x,v)$. By construction, the Fokker-Planck kernel also satisfies the aforementioned KMS relation, and thus reaches the correct equilibrium distribution.

That being said, this resemblance is consistent with the expectation that the Fokker-Planck kernel is applicable at sufficiently small velocity.  As the region between the red and blue lines shrinks, the small momentum gradient approximation --- either small $\partial_u f$ (small gradients around a flat distribution), or small $v+\partial_u f$ (small gradients around equilibrium) --- becomes more reliable in this region, since its linear behavior near both the red and blue lines is reproduced by the truncated Fokker-Planck kernel.

One marked difference between the full Kolmogorov kernel and its Fokker-Planck truncation is that the latter remains finite for all values of $(x,u)$ whereas the former diverges when $x = \partial_u f \to -v/2 \pm x_b(v)$. This implies that for values of $x=f'$ that are larger (or smaller) than a certain critical value, the Kolmogorov equation~\eqref{eqHJ10} does not admit a solution. We can illustrate one possible reason for this
by analyzing the much simpler toy kernel
    \begin{equation}
        K_{\rm simple}(x,v) = \frac{1}{\sqrt{1 - v^2/4}} - \frac{1}{\sqrt{1 - (Tx + v/2 )^2}} \, .
        \label{eqHJ26}
    \end{equation}
This expression shares key qualitative features with the full Kolmogorov kernel plotted in the left panel of Fig.~\ref{fig1}. In particular, similar to the full $K$, it has a finite domain along $x$, namely  $-1-v/2 < T x < 1 -  v/2$. However, $K_{\rm simple}(x,v)$ can be viewed as the leading order term in the
expansion of a more complete expression in powers of $T/M$, 
with subleading $T/M$ power corrections having been resummed. One such more complete expression would be 
   \begin{equation}
        K_{\rm simple}^{\rm resum}(x,v) = \frac{\sqrt{2}}{\sqrt{1 - \frac{v^2}{4} + \sqrt{ (1-\frac{v^2}{4})^2 + \frac{T^2}{M^2} } }} - \frac{ \sqrt{2}  }{\sqrt{ 1 - (Tx + \frac{v}{2} )^2 + \sqrt{ (1 - (Tx + \frac{v}{2} )^2)^2 + \frac{T^2}{M^2} } }} \, .
        \label{eqHJ27}
    \end{equation}
This more complete expression is finite for any real value $x$ for any nonvanishing $T/M$, but in the limit $T/M\to 0$ it diverges everywhere except in the domain of the leading term of its $T/M$ expansion \eqref{eqHJ26}. Although artificial, this simple example points to the possibility that the finite domain of the Kolmogorov kernel $K(x,u)$ seen in the left panel of Fig.~\ref{fig1} is an artifact of the $T/M$ expansion, 
meaning that it is a consequence of
the fact that we are working in the heavy quark limit throughout. 

From a practitioner's point of view, the limited domain of the Kolmogorov kernel $K$ prevents us from propagating sharply peaked distributions, including delta-function–like distributions that may be viewed as proxies for individual heavy quarks injected into the plasma with some well-defined velocity $v$. 
In contrast, from a purely technical standpoint the Fokker-Planck truncation does admit solutions for such sharply peaked distributions --- but these are not physically meaningful, since truncating the kernel at second order in $x$ cannot be justified when the gradient $x=\partial_u f$ is large. That is to say, the fact that the domain of the Kolmogorov kernel is more restricted than that of its Fokker-Planck truncation should not be mistaken to mean that the kinematic regime in which the Kolmogorov equation describes the evolution of the heavy quark momentum distribution is more limited. In fact, the Fokker-Planck equation provides a good approximation to the full Kolmogorov dynamics only in a subset of the regime in which the Kolmogorov description is controlled.

An interesting avenue to pursue in future work is to use the Kolmogorov kernel $K$ as input to a \textit{non-Gaussian}, Langevin-like, stochastic description of an individual heavy quark, using the momentum change probability distribution in Eq.~\eqref{eqU1} to evolve the heavy quark momentum in discrete time. Since in such a formulation, one would be describing the evolution of the probability distribution ${\mathscr P}$ (that we describe here using the Kolmogorov equation) via the non-Gaussian stochastic behavior of individual heavy quarks in an ensemble, one at a time, it would seem that there should also be no obstruction to propagating sharply peaked probability distributions. Why is it, then, that the Kolmogorov kernel seems to prevent us from propagating sharply peaked distributions? The resolution to this apparent contradiction lies in the details: As a careful reader will have noticed, the ensuing Langevin-like dynamics differs from the Kolmogorov equation in its Hamilton-Jacobi form \eqref{eqHJ10} starting at the first subleading power in $T/M$, as Eq.~\eqref{eqU10} makes clear. It then follows that, as per our previous discussion in this Section, this is exactly a kind of modification that makes Kolmogorov dynamics well-defined for sharply peaked distributions --- and equivalent to Eq.~\eqref{eqHJ10} at leading order in $T/M$.

In the next three subsections, we turn to solving 
the Kolmogorov evolution equation \eqref{eqHJ10} for phenomenologically relevant heavy-quark momentum distributions with power-law tails, for which the equation admits well-defined solutions. Later, in Section~\ref{sec:interplay},
we shall return to our discussion of the range of validity of the formalism.

\subsection{Hamiltonian flow in phase space} \label{sec:H-Flow}

The solutions to Eqs.~\eqref{eqHJ5} and~\eqref{eqHJ6} define trajectories on the $(u,x)$ phase space, which completely characterize the evolution of the distribution function $f$ (to be precise, $f$ is the logarithm of the distribution function ${\mathscr P}$, see Eq.~\eqref{eqU7}) starting from any initial condition within the domain of $K$.
Concretely, the initial condition  $f(u_0,\tau_0)$ defines a curve in this phase space given by $(u,\partial_u f(u,\tau))$ with $u_0 = u_0(u,\tau)$. 
We shall evolve each point along this curve --- which we will label using a parameter $u_0$, coinciding with $u$ at the initial time ---  independently with Eqs.~\eqref{eqHJ5} and~\eqref{eqHJ6}. Doing so will define a curve $(u(\tau;u_0), x(\tau;u_0))$ parametrized by $u_0$ at each time $\tau$.
We shall describe how the 
full distribution function $\mathscr{P}$ may be reconstructed
from this curve
in Section~\ref{sec:P-reconstruction}.

In light of the mapping between $x$ and $C_3$ given in Eq.~\eqref{eqHJ22}, it is also possible to write these evolution equations in terms of the pair $(u,C_3)$, which defines an alternative parametrization of the phase space in which $f$ is evolved. This turns out to be especially convenient for the Kolmogorov equation in $\mathcal{N}=4$ SYM theory, where the Legendre transform of the evolution operator $K$ is known explicitly in terms of $(u,C_3)$. For the Fokker-Planck truncation there is no gain in simplicity by doing this, but it is nonetheless insightful to see the mapping take place. A straightforward calculation shows that the evolution equations in the $(u,C_3)$ phase space, equivalent to Eqs.~\eqref{eqHJ5} and~\eqref{eqHJ6} in the spherically symmetric case, are
\begin{align}
    \dot{u} &= - \frac{\pi C_3}{2} \, , \label{eqHJ28} \\ 
    \dot{C}_3 &= \frac{\pi C_3}{2} \frac{\frac{\partial \tilde{K}}{\partial u} }{ \frac{\partial \tilde{K}}{\partial C_3} } \, , \label{eqHJ29}
\end{align}
which make it apparent that $C_3$ encodes the rate of momentum transfer received by the element of the distribution labeled by $u$. While the expression for $\dot{C}_3$ looks more complicated than its analog for $\dot{x}$, it is actually advantageous because one may use the explicit expression in Eq.~\eqref{eqHJ17} to evaluate the r.h.s. and solve the equations in a way that is numerically more efficient.

Before solving these equations, it is helpful to have a qualitative look at their properties. To do so, we note that both pairs of equations, Eqs.~(\ref{eqHJ5},\,\ref{eqHJ6}) and Eqs.~(\ref{eqHJ28},\,\ref{eqHJ29}), can be viewed as vector fields $V^\alpha, W^\beta$ on the $(u,x)$ and $(u,C_3)$ planes, respectively, with components given by $(V^1,V^2) = (\dot{u},\dot{x})$ and $(W^1,W^2) = (\dot{u},\dot{C}_3)$. 
Because these vector fields are the tangent vectors to the trajectories induced by the Hamiltonian flow, visualizing them is equivalent to visualizing the trajectories that the solutions $(u(\tau;u_0), x(\tau;u_0))$ describe.

\begin{figure}
    \centering
    \includegraphics[width=0.50\linewidth]{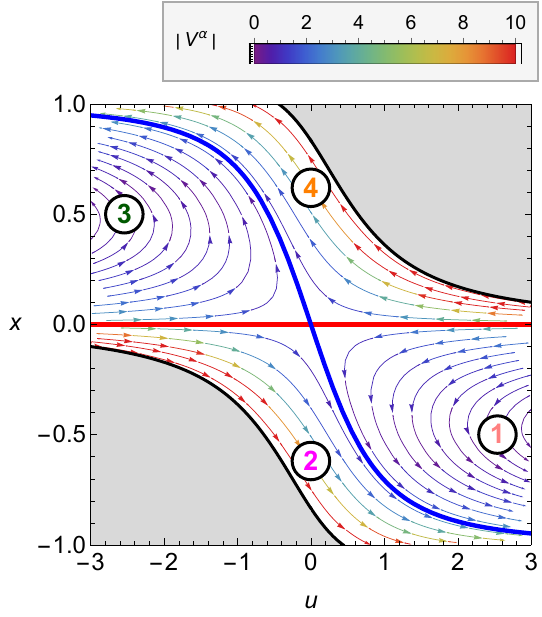}
    \includegraphics[width=0.48\linewidth]{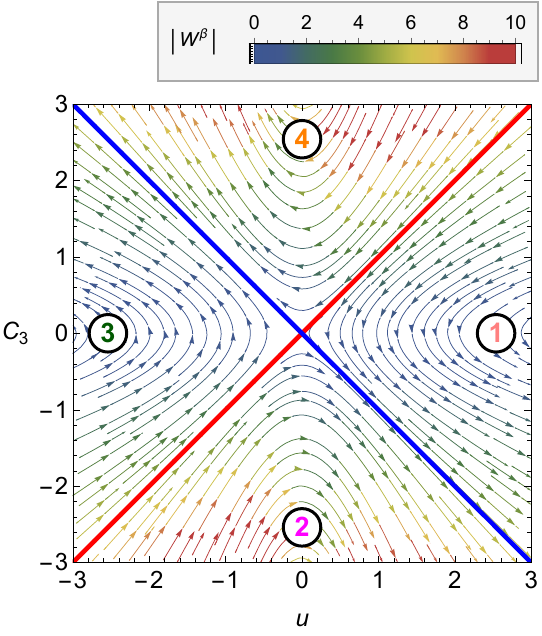}
    \caption{Flow diagrams in the $(u,x)$ and $(u,C_3)$ phase spaces for the Kolmogorov equation in $\mathcal{N}=4$ SYM theory. The arrows depict the trajectories followed by solutions $(u(\tau), x(\tau))$ to Hamilton's equations~\eqref{eqHJ5} and~\eqref{eqHJ6} in the spherically symmetric case or, equivalently, the trajectories followed by solutions $(u(\tau), C_3(\tau))$ to Eqs.~\eqref{eqHJ28} and~\eqref{eqHJ29} with $\tilde{K}$ given by Eq.~\eqref{eqHJ17}. There are four disjoint regions for these trajectories separated by the red and blue lines, which define stationary solutions. The blue line corresponds to the equilibrium Boltzmann distribution. The regions are labeled $1$ through $4$. Regions with the same label in the $(u,x)$ and $(u,C_3)$ phase spaces are equivalent descriptions of the same physical situation; each individual trajectory in one flow diagram corresponds to a unique trajectory in the other. The coloring along each trajectory indicates the magnitude of $(\dot u,\dot x)$ or $(\dot u,\dot C_3)$, which is to say the speed at which the trajectory is traversed as a function of $\tau$.}
    \label{fig:flow-K}
\end{figure}

\begin{figure}
    \centering
    \includegraphics[width=0.50\linewidth]{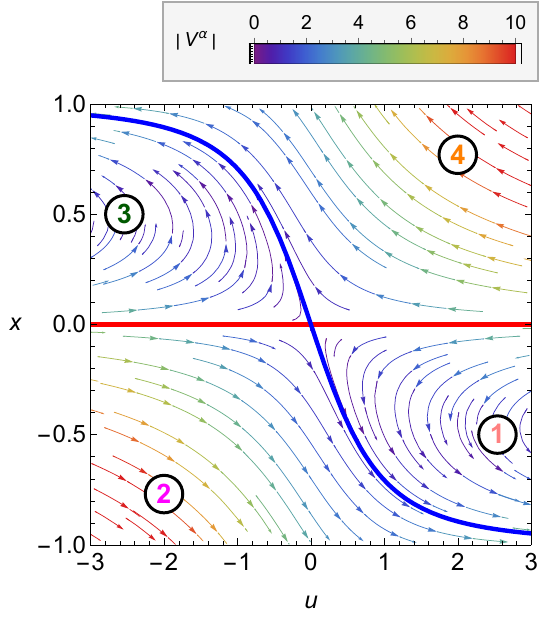}
    \includegraphics[width=0.48\linewidth]{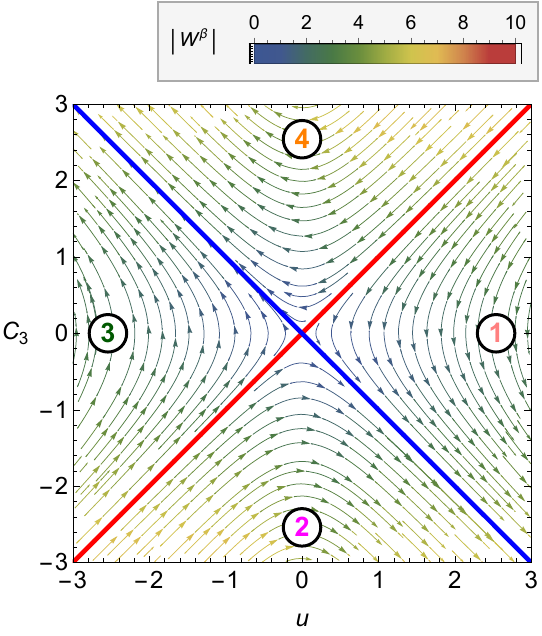}
    \caption{Flow diagrams in $(u,x)$ and $(u,C_3)$ phase spaces for the Fokker-Planck truncation, with the Einstein relation imposed artificially. The arrows depict the trajectories followed by solutions $(u(\tau), x(\tau))$ to Hamilton's equations~\eqref{eqHJ5} and~\eqref{eqHJ6} in the spherically symmetric case with $K$ given by Eq.~\eqref{eqHJ14} or, equivalently, the trajectories followed by solutions $(u(\tau), C_3(\tau))$ to Eqs.~\eqref{eqHJ28} and~\eqref{eqHJ29} with $\tilde{K}_{\rm FP-E}(C_3,u) = -\tfrac{\pi}{8} (u^2 - C_3^2)/\sqrt{1+u^2}$. As in Fig.~\ref{fig:flow-K}, there are four disjoint regions separated by the red and blue stationary solutions, with the blue line corresponding to the equilibrium Boltzmann distribution. Also as in Fig.~\ref{fig:flow-K}, regions with the same label $1$ through $4$ in the $(u,x)$ and $(u,C_3)$ phase spaces are equivalent descriptions of the same physical situation, with each trajectory in one flow diagram corresponding to a unique trajectory in the other. The coloring along each trajectory indicates the speed at which the trajectory is traversed as a function of $\tau$.}
    \label{fig:flow-FP}
\end{figure}

We show these vector fields $V^\alpha, W^\beta$ in Figures~\ref{fig:flow-K} and~\ref{fig:flow-FP} for the full $\mathcal{N}=4$ SYM kernel and for its Fokker-Planck truncation, respectively. Qualitatively, the flow patterns are fairly similar when one compares Kolmogorov and Fokker-Planck dynamics. This is due to the fact that they share the same fixed point and stationary configurations:
\begin{enumerate}
\item The point $(u,x) = (0,0)$ is a fixed point of the Hamiltonian flow --- in fact the only one. This is consistent with the fact that spherically symmetric distributions that are continuous at the origin satisfy $f'(u=0) = 0$ as a boundary condition, because if $\tfrac{\partial f}{\partial {\bf u} } \neq 0$ then a preferred direction would be selected, which would break the spherical symmetry. 
Thus, any solution to Hamilton's equations~\eqref{eqHJ28} and \eqref{eqHJ29}
that describes a spherically symmetric distribution that is continuous at the origin will be represented on a flow diagram as in Fig.~\ref{fig:flow-K} or \ref{fig:flow-FP} at any one time by a curve that begins at the origin, with the arrows in Figs.~\ref{fig:flow-K} or \ref{fig:flow-FP} then describing how this curve moves on the flow diagram at later times. 
For spherically symmetric distributions that are continuous at the origin, ${\bf p}\rightarrow -{\bf p}$ is a symmetry and 
therefore $(u,x)\rightarrow (-u,-x)$ and $(u,C_3)\rightarrow (-u,-C_3)$ are symmetries. Without loss of generality, therefore, we need only consider solutions that begin at the origin and extend to the right in the flow diagrams of Figs.~\ref{fig:flow-K} or \ref{fig:flow-FP}, where $u>0$.    We see from Figs.~\ref{fig:flow-K} and \ref{fig:flow-FP} that such solutions may lie in Regions 1, 2 or 4. However, phenomenologically relevant distributions decrease with increasing momentum $u$, meaning that $x=f'<0$, which means that their radial projection can only lie in Regions 1 or 2. We shall visualize such solutions to Hamilton's equations as curves on the flow diagrams
of Figs.~\ref{fig:flow-K} and \ref{fig:flow-FP} in Fig.~\ref{fig:K-FP-solution-sample} in the next subsection, where we will focus on a solution with
phenomenologically motivated initial conditions that lies in Region 1.

\item The red line denotes the uniform, momentum-independent distribution $x = \partial f/\partial u = 0$ (equivalently, $u = C_3$). This stationary distribution is unstable: an initial condition in its vicinity will eventually move away from it. 
\item The blue line $x = \partial f/\partial u  = -v = -u/\sqrt{1+u^2}$ (equivalently, $u = -C_3$) is the attractor that corresponds to thermal equilibrium, namely the Boltzmann distribution.  It is the only stable stationary configuration.  Initial conditions in its vicinity will evolve towards it. That is, they will thermalize.
\end{enumerate}

The stationary solutions separate the phase space into four distinct regions, as indicated in 
Figures~\ref{fig:flow-K} and~\ref{fig:flow-FP}. From the point of view of the Hamiltonian flow, these four regions are dynamically disjoint, as no trajectory crosses from one of the four regions into another.  
\begin{itemize}
\item {\bf Region 1}: $-v< x=f'<0$ and $u>0$\\
Since the production cross-section for heavy quarks is a decreasing function of heavy quark momentum,
phenomenologically relevant initial distributions of heavy quarks 
decrease with increasing momentum, meaning $x = f' < 0$.  Furthermore, 
this decrease is generically less 
steep than that of a thermal Boltzmann distribution because 
the heavy quark production cross-section does not fall off exponentially at large momentum, meaning that $x = f' > -v$.
For isotropic distributions, only the modulus $u = \vert{\bf u}\vert$ of the three-momentum matters, and $u\geq 0$. 
This means that the radial projection of any phenomenologically relevant initial condition is situated entirely within Region 1. Because the four regions are dynamically disjoint, the Hamiltonian
flow for a phenomenologically inspired spherically symmetric distribution evolves entirely within Region 1.
For this reason, in all the examples that we consider later in this work Region 1 is the only region that plays a role in the dynamics. Such initial conditions will evolve towards the blue line, which is to say they will thermalize.
\item {\bf Region 2}: $x=f'<-v$ and $u>0$\\
Even though doing so is not phenomenologically relevant, we can consider spherically symmetric ($u>0$) initial distributions that fall off more steeply than a thermal distribution in (some part of) momentum space ($f'< -v$). While the Hamiltonian flow in Regions 1 and 2 is disjoint, it is continuous across the equilibrium (blue) line. Put together with the fact that the flow arrows point toward increasing $u$, such an initial condition will not amount to any additional complication other than having to follow more flow lines. Such initial conditions will evolve towards the blue line, which is to say they will thermalize.
\item {\bf Region 2}: $x= f'< 0$ and $u<0$\\
All flow lines in this region eventually cross $u=0$. Since the original evolution operator $K$ in Eq.~\eqref{eqHJ3} is defined in three dimensions, this corresponds to a trajectory coming into the origin ${\bf u} = {\bf 0}$ from one direction on the unit sphere $S^2$ and emerging in the antipodal direction, which is to say a straight line through the origin. The solution would be smooth if we used Cartesian coordinates; however, spherical coordinates make this singular as one needs to send the polar and azimuthal angles to $\theta \to \pi - \theta$ and $\phi \to \pi + \phi$, respectively. For practical purposes, this means that the physical solution in this situation has to be constructed as a linear combination of the solution in the $u<0$ and $u>0$ regions as a function of $|u|$. Operationally, this construction would take place in the same way that we discuss  in Section~\ref{sec:P-reconstruction}, below. Because the initial conditions we will be interested in lie completely within Region 1, we will not discuss this case in more detail in this work; it may be of interest in future work where the full 3D dynamics is studied in detail. 
\item {\bf Region 3} \\
The phase space flow in Region 3 is obtained from that in Region 1 by the point symmetry $u\to - u$,  $x\to -x$ or $C_3 \to - C_3$, respectively. Region 3 therefore does not contain novel dynamical information. Any dynamics formulated in Region 1 could be equivalently formulated in Region 3.
\item {\bf Region 4} \\
Region 4 is related to Region 2 by the point symmetry $u\to - u$,  $x\to -x$ or $C_3 \to - C_3$. For spherically symmetric distributions initialized with $u > 0$ one could consider initial distributions that increase with increasing momentum in some momentum range  ($x= f' > 0$) and decrease with increasing momentum in some other momentum range ($x= f' < 0$). 
Such initial conditions would be initialized in Regions 1 and 4, and the trajectories in Region 4 would evolve towards negative $u$. As mentioned already, the physical solution in such a situation has to be constructed as a linear combination of the solution in the $u<0$ and $u>0$ regions as a function of $|u|$. We shall not elaborate on such cases in this work.
\end{itemize}

Next, we comment on the differences between the solutions for Kolmogorov and Fokker-Planck dynamics. A direct comparison between both formulations can only be done via the $(u,x)$ representations, as the mapping $x \longleftrightarrow C_3$ depends on the evolution operator $K$.  The greatest qualitative difference, visible in that $(u,x)$ flow diagrams, is that the Kolmogorov evolution equation has a bounded domain as a function of these coordinates, whereas the Fokker-Planck truncation's domain is unbounded. This is in line with our discussion in Section~\ref{sec:comparing-K-FP-conceptual}. 
The black curves that bound the domain of the Kolmogorov evolution equation in Regions 2 and 4 in the left panel of Fig.~\ref{fig:flow-K} correspond 
to $x=-\frac{v}{2}\pm x_b(v)$ with $x_b(v)$ given by Eq.~\eqref{eqHJ24}, with $u=\gamma v>0$.  We now see that these curves lie entirely outside the phenomenologically relevant Region 1.

To understand the important quantitative differences 
between Kolmogorov and Fokker-Planck dynamics, in particular in  Region 1, it is important to
ask how quickly solutions will evolve in time along the flow trajectories. This is determined by the norm of the vector field which is depicted via the colors along the trajectories and the color bars in Figs.~\ref{fig:flow-K} and~\ref{fig:flow-FP}. Detailed inspection reveals  that, in particular for sizeable $u$, the Fokker-Planck truncation evolves trajectories much more rapidly towards the equilibrium blue line than 
is the case for evolution according to the Kolmogorov dynamics.
This reflects the different behavior of $K(x,u)$ for large $u$, meaning $v\rightarrow 1$, that can be seen between the blue and red lines in Fig.~\ref{fig1}.
The direct display of sample solutions in Fig.~\ref{fig:K-FP-solution-sample} in the next subsection
will provide a different and more explicit visualization of this point.

\subsection{A solution to Hamilton's equations with phenomenologically motivated initial conditions }

Equipped with the flow diagrams in Figs.~\ref{fig:flow-K} and \ref{fig:flow-FP}, we now proceed to investigate the behavior of solutions to Hamilton's equations for a specific class of initial conditions. The  initial conditions we will use in this Section correspond to the FONLL-inspired $b$ quark production cross section discussed later in Section~\ref{sec:init-conds}. We will motivate the precise form of these initial conditions only there, as we will not need the exact expressions for our discussion in this Section. Here, we only aim to describe the features of these initial conditions and their subsequent evolution at a qualitative level, focusing entirely on features that are in fact common to all solutions that we discuss later in this work. 

\begin{figure}
    \centering
    \includegraphics[width=0.46\linewidth]{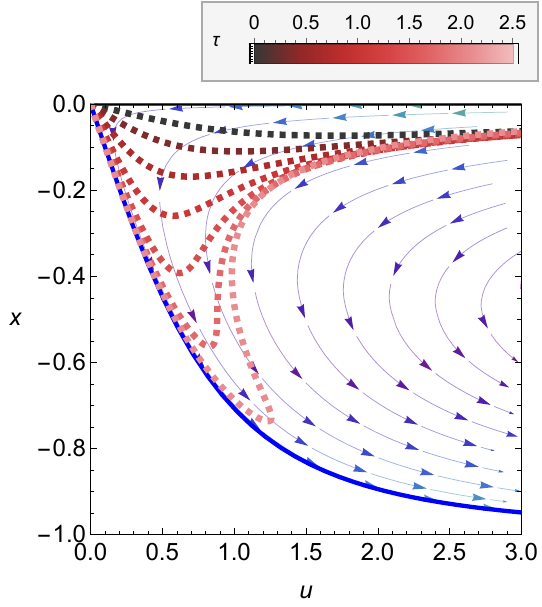}
    \includegraphics[width=0.4\linewidth]{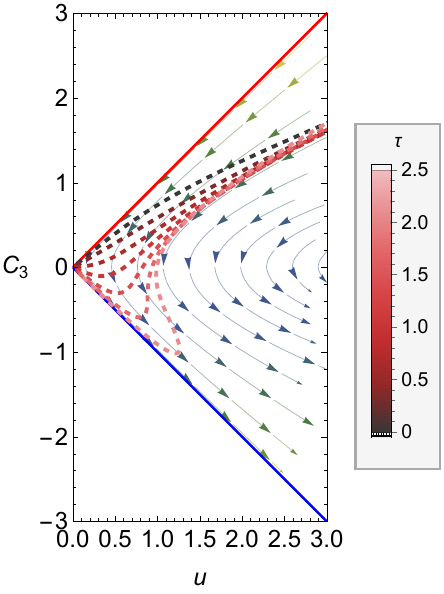}
        \centering
        {.}\vspace{0.5cm}
    \includegraphics[width=0.46\linewidth]{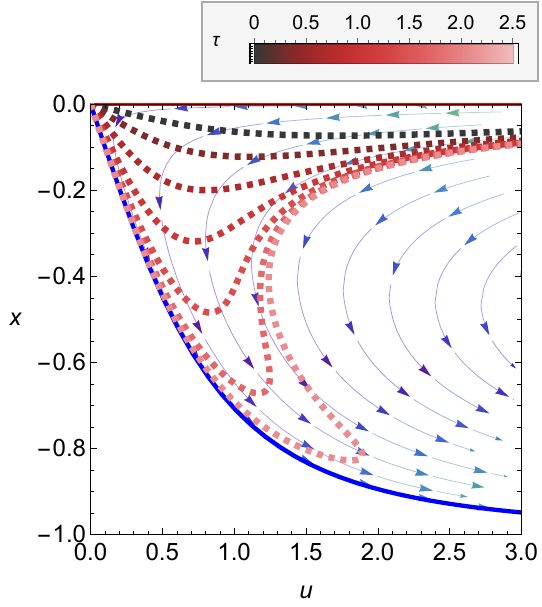}
    \includegraphics[width=0.4\linewidth]{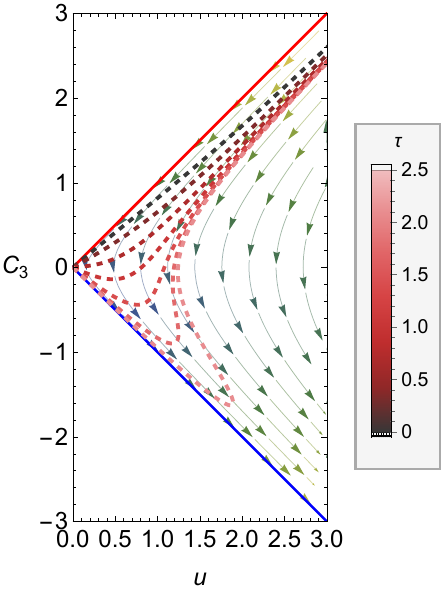}
    \caption{ Upper panels: A sample solution to the Kolmogorov equation in $\mathcal{N} = 4$ SYM theory via Hamilton's equations, in both the $(u,x)$ (upper left) and $(u,C_3)$ (upper right) representations of the phase space. The initial condition is represented via the dark grey dashed curve; subsequent times are represented by curves with lighter and lighter tones of red, with each curve $\Delta \tau =0.35$ later than the preceding one. Compared to Fig.~\ref{fig:flow-K}, we only display Region 1 since, as we have discussed, phenomenologically relevant initial conditions evolve entirely within this region.  Each point along a curve moves according to the Hamiltonian flow, represented by the same arrows as in Fig.~\ref{fig:flow-K} whose colors depict the magnitude of the vector field as in
    Fig.~\ref{fig:flow-K}. Lower panels: Same as upper panels, but for the 
    Fokker-Planck truncation of the Kolmogorov equation with the Einstein relation imposed artificially.}
    \label{fig:K-FP-solution-sample}
\end{figure}

Fig.~\ref{fig:K-FP-solution-sample} 
illustrates how a phenomenologically motivated 
spherically symmetric distribution function $\mathscr{P} = e^{M f(u)/T}$, fully characterized (up to its overall normalization) by $x = f'(u)$, evolves in time. Starting from the initial condition (dashed black line), the dynamics unfolds by evolving this line through the Hamiltonian $(u,x)$ phase space according to Eqs.~\eqref{eqHJ5} and~\eqref{eqHJ6} --- equivalently, in the $(u,C_3)$ phase space according to Eqs.~\eqref{eqHJ28} and~\eqref{eqHJ29} --- for which we show snapshots at later times. As 
mentioned already, the mapping $x \longleftrightarrow C_3$ depends on the evolution operator $K$. A direct comparison between the solutions for Kolmogorov and Fokker-Planck dynamics should therefore be done in the $(u,x)$ representations. However, if one is interested in the typical momentum transfer $C_3$ that heavy quarks experience (and that depends on $K$), inspecting the $(u,C_3)$ representation is more informative.

As we can see from Fig.~\ref{fig:K-FP-solution-sample},  
in both cases the Hamiltonian flow takes the initial condition (dashed black line) to a curve that is reasonably close to the equilibrium distribution (blue line) within $\tau = \frac{\sqrt{\lambda} T^2 t}{M} \sim 2 $, at least for $u = p/M \lesssim 1$. 
However, there are also significant differences.
For example, equilibration at higher momenta is clearly faster in the Fokker-Planck truncation than in the full Kolmogorov equation, as can be observed by following the dashed lines in lighter tones of red in each Figure. Correspondingly, the $(u,C_3)$ representation shows that at high heavy quark momentum the typical momentum transfer $C_3$ that heavy quarks experience is larger (almost by a factor of $2$) in the Fokker-Planck dynamics than in the Kolmogorov dynamics, which is also indicative of a slower equilibration process for large $u$. 
We will make this more visible later, for example in Fig.~\ref{fig:solutions-P-example} in the next subsection
where the $u \gg 1$ regime will be more prominent.

It is also clear from Fig.~\ref{fig:K-FP-solution-sample} that the shape of the initial condition at high momentum is left almost unchanged by the dynamics, as evidenced by the fact that all dashed curves in every panel are close to each other in this regime. Therefore, it is here where information about the initial condition is preserved for  longer times --- a fact that will be crucial later in our discussion.

\subsection{From Hamiltonian flow to solutions of the Kolmogorov equation} \label{sec:P-reconstruction}

In the $(u,x)$ representations of evolved initial conditions as in Fig.~\ref{fig:K-FP-solution-sample}, in general three distinct regions become visible at late times. There is a regime where $u$ is initially sufficiently small where the initial condition evolves rapidly towards the thermal distribution (blue line). 
There is a regime where $u$ is initially sufficiently large, $u \gtrsim 1$, where the solution evolves slowly and retains memory of the initial condition over a prolonged time. 
And, there is an intermediate regime that smoothly connects the low-$u$ and high-$u$ branches in such a way that the curves $(u(\tau;u_0), x(\tau;u_0))$ parametrized by $u_0$ are not single-valued functions $x(u)$ at all times.\footnote{It is interesting to note that parametrizing the dynamics in terms of $u_0$ is akin to a ``Lagrangian'' description of the distribution rather than an ``Eulerian'' description in the language of continuum mechanics, where functions may become similarly multi-valued.} 
In this subsection, we discuss how to reconstruct the time-evolved physical distribution $\mathscr{P}$ from the trajectories $(u(\tau;u_0), x(\tau;u_0))$ in such a generic situation.

As we discussed earlier around Eq.~\eqref{eqHJ7} (and upon specializing to the spherically symmetric case) once we have solved for all trajectories $u(\tau;u_0)$ and $x(\tau;u_0)$, we may reconstruct $\partial f / \partial u$ by making use of
\begin{equation}
    \frac{\partial f}{\partial u}(u,\tau) = x\left(\tau; u_0(  u,\tau ) \right) \, , \label{eq:inverse-map-H-dfdu}
\end{equation}
and determining $f$ via direct integration.
At any fixed time $\tau$, this is straightforward if $u(\tau;u_0)$ is a strictly increasing function of $u_0$. However,  
as illustrated in Figure \ref{fig:K-FP-solution-sample}, generic solutions of our Hamilton-Jacobi flow equation include cases with (at least) two trajectory labels $u_{0,1}$, $u_{0,2}$ such that at some time $\tau$
\begin{equation}
    u(\tau; u_{0,1} ) = u(\tau; u_{0,2} ) \, .
\end{equation}
Furthermore, it will generically be the case that for these trajectories
\begin{equation}
    x(\tau; u_{0,1} ) \neq x(\tau; u_{0,2} ) \, .
\end{equation}
This presents a problem, because the inversion problem $u_0 = u_0(u,\tau)$ that one needs to solve in order to determine $\partial f/\partial u$ from Eq.~\eqref{eq:inverse-map-H-dfdu} no longer has a unique solution. The approach to finding solutions, this problem notwithstanding, is clarified by two considerations:
\begin{enumerate}
    \item The evolution of $\mathscr{P}$ should be such that, as it starts from a continuous and differentiable function, $\mathscr{P}$ should remain continuous and differentiable. An immediate consequence of this is that all  trajectories with a given $u(\tau)$, each labeled by a different $u_0$, should be kept and should contribute to the final result.

    \item While Eq.~\eqref{eqHJ3} (equivalently, Eq.~\eqref{eqU12}) is highly nonlinear, Eq.~\eqref{eqU13} \textit{is} linear. Because the price to pay in going from the latter to the former is to lose accuracy beyond the leading order in $T/M$, it follows that if $f(u,\tau)$ and $\tilde{f}(u,\tau)$ are two solutions to Eq.~\eqref{eqHJ3}, then any linear combination
    \begin{equation}
        \mathscr{P}(u,\tau) = A e^{M f(u,\tau)/T} + B e^{M \tilde{f}(u,\tau)/T} \, ,
    \end{equation}
    is a solution to  Eq.~\eqref{eqU13}, up to power corrections that are subleading in $T/M$.
\end{enumerate}
It is then natural to \textit{construct} a solution, fully consistent with the trajectories determined via Hamilton's equations, via a linear combination of all the trajectories that at time $\tau$ pass through ``position'' $u$.

We now work out this construction explicitly for the spherically symmetric case upon making some mild assumptions. 
The initial domain of the function $f(u_0,\tau = 0)$ is $u_0 \in (0,\infty)$. We shall assume that this initial condition is a strictly decreasing function of $u_0$, and that it has a global maximum at $u_0 = 0$. These conditions are all satisfied by any phenomenologically reasonable initial conditions. They are also satisfied by the simplified initial conditions inspired by the FONLL calculation of  $b$ quark production that we shall employ in later Sections.

Inspecting the flow diagrams in Figs.~\ref{fig:flow-K} and~\ref{fig:flow-FP}, one can see that for such an initial condition the trajectories will never cross the line $C_3 = u$, which corresponds to $x = 0$, and that there will therefore always be a maximum at $u = 0$. Furthermore, the ``particle'' associated with the point of maximum $(u, C_3) = (0,0)$ is a time-independent solution to Hamilton's equations, and may therefore be used as a reference point from which to integrate $f'$ to get $f$.

With these considerations in mind, at any point in time $\tau$, we define a function $g(u_0, \tau)$ via
\begin{equation}
    g(\tau;u_0) \equiv \int_0^{u_0} du_{0,1} \frac{\partial u(\tau;u_{0,1} )}{ \partial u_{0,1}} x(\tau;u_{0,1}) \, . \label{eq:g-def}
\end{equation}
Loosely speaking, this equation states that $g = \int df$ because $x = \partial f/\partial u$. However, this assumes that $f$ is a single-valued function. In particular, if $u(\tau;u_0)$ is an invertible function at fixed $\tau$, i.e., if $u_0(\tau;u)$ is uniquely defined (and therefore $f$), then the integration is straightforward and one arrives at
\begin{equation}
    f(u,\tau) = g(\tau; u_0(\tau,u) ) \, .
    \label{eq:f-equals-g}
\end{equation}
However, if $u(\tau;u_0)$ is \textit{not} an invertible function, then Eq.~\eqref{eq:g-def} does not directly define a function $f(u,\tau)$. In general, constructing $f$ can be as complicated as the trajectories that solve the dynamics. For our present purposes, inspection of Fig.~\ref{fig:K-FP-solution-sample} tells us that it will be sufficient to consider the case where $u(\tau;u_0)$ is a strictly increasing function of $u_0$ from $\tau = 0$ until some time $\tau_1$, and starting from that time onwards there exist three ($\tau$-dependent) intervals where $u$ is a monotonic function of $u_0$. As $u_0$ is increased from $0$, $u$ first grows, then decreases, and finally grows again as $u_0 \to \infty$. In detail:
\begin{enumerate}
    \item For $0 < \tau < \tau_1$, $\partial u(\tau;u_0) / \partial u_0 > 0$ everywhere, and therefore $u_0(\tau;u)$ is uniquely defined. We simply have $f(u,\tau) = g(\tau; u_0(\tau,u) )$ as in Eq.~\eqref{eq:f-equals-g}.
    \item For $\tau > \tau_1$, there exist two special values of the label coordinate $u_0$, which we shall denote by $u_0^-(\tau)$ and $u_0^+(\tau)$, which characterize the intervals where $\partial u(\tau;u_0) / \partial u_0$ takes a definite sign. That is to say,
    \begin{align}
        &\frac{\partial u(\tau;u_0)}{\partial u_0} > 0 \, , \quad {\rm if} \quad u_0 < u_0^-(\tau) \, , \\
        &\frac{\partial u(\tau;u_0)}{\partial u_0} < 0 \, , \quad {\rm if} \quad u_0^-(\tau) < u_0 < u_0^+(\tau) \, , \\
        &\frac{\partial u(\tau;u_0)}{\partial u_0} > 0 \, , \quad {\rm if} \quad u_0^+(\tau) < u_0 \, .
    \end{align}
\end{enumerate}
We see in the upper panels of Fig.~\ref{fig:K-FP-solution-sample} that for the initial condition employed there if we evolve the distribution with the full Kolmogorov dynamics then $\tau_1$ is between the second-to-latest and latest curves shown, whereas we see in the lower panels that with the truncated Fokker-Planck dynamics $\tau_1$ is earlier.  For $\tau>\tau_1$, the intervals described above are apparent in Fig.~\ref{fig:K-FP-solution-sample}.

It is then natural to define three coordinate functions $u_0^{(1)}(\tau;u)$, $u_0^{(2)}(\tau;u)$, and $u_0^{(3)}(\tau;u)$, via
\begin{align}
    & u_0^{(1)}(\tau;u(\tau;u_0) ) = u_0 \, , \quad {\rm for} \quad u_0 < u_0^-(\tau) \, , \\
    & u_0^{(2)}(\tau;u(\tau;u_0) ) = u_0 \, , \quad {\rm for} \quad u_0^-(\tau) < u_0 < u_0^+(\tau) \, , \\
    & u_0^{(3)}(\tau;u(\tau;u_0) ) = u_0 \, , \quad {\rm for} \quad u_0^+(\tau) < u_0 \, ,
\end{align}
which are the local inverses of $u(\tau;u_0)$ and which in turn define three functions
\begin{align}
    & f_1(u,\tau) = g(\tau,u_0^{(1)}(\tau;u)) \, , \quad {\rm for} \quad u < u(\tau; u_0^-(\tau)) \, , \\
    & f_2(u,\tau) = g(\tau,u_0^{(2)}(\tau;u)) \, , \quad {\rm for} \quad u(\tau;u_0^+(\tau)) < u < u(\tau;u_0^-(\tau)) \, , \\
    & f_3(u,\tau) = g(\tau,u_0^{(3)}(\tau;u)) \, , \quad {\rm for} \quad u(\tau; u_0^+(\tau)) < u \, ,
\end{align}
all of which satisfy the Kolmogorov equation at leading order in $T/M$. That is to say, they all satisfy Eq.~\eqref{eqHJ3}. We then obtain the solution as
\begin{equation}
    \mathscr{P}(u,\tau) = \begin{cases}
        \exp \left( \frac{M}{T} f(u,\tau) \right) & {\rm if} \quad \tau < \tau_1 \\
        \exp \left( \frac{M}{T} f_1(u,\tau) \right) & {\rm if} \quad \tau > \tau_1 \quad \& \quad u < u(\tau; u_0^+(\tau)) \\
        \substack{ \textstyle \exp \left( \frac{M}{T} f_1(u,\tau) \right) \\  \textstyle - \exp \left( \frac{M}{T} f_2(u,\tau) \right) \\  \textstyle + \exp \left( \frac{M}{T} f_3(u,\tau) \right)} & {\rm if} \quad \tau > \tau_1 \quad \& \quad u(\tau;u_0^+(\tau)) < u < u(\tau;u_0^-(\tau)) \\
        \exp \left( \frac{M}{T} f_3(u,\tau) \right) & {\rm if} \quad \tau > \tau_1 \quad \& \quad u(\tau;u_0^-(\tau)) < u
    \end{cases} \, . \label{eq:P-solution-explicit-reco}
\end{equation}
Note that because $\frac{\partial u(\tau;u_0)}{\partial u_0} < 0$ if $ u_0^-(\tau) < u_0 < u_0^+(\tau)$, the ordering in $u$ in the third case $u(\tau;u_0^+(\tau)) < u(\tau;u_0^-(\tau))$ is opposite to the corresponding $u_0$ values.

In the process of constructing the solution \eqref{eq:P-solution-explicit-reco}, we have used the fact that since the equations of motion we solve do not specify the $\mathcal{O}(1)$ factors in front of $e^{Mf_i/T}$, some of these prefactors can  %principle 
be negative as long as the full linear combination that forms $\mathscr{P}$ is positive. And, while our analysis does not fix their magnitude, their sign \textit{does} follow from physical considerations. In all cases except for the third case in Eq.~\eqref{eq:P-solution-explicit-reco}, only one solution $e^{Mf_i/T}$ (or $e^{Mf/T}$) term is available, thus automatically fixing the sign of the prefactor to be positive.
In order to fix the sign in front of each term in the $ \tau > \tau_1$ \&  $u(\tau;u_0^+(\tau)) < u < u(\tau;u_0^-(\tau))$ case, we have required that the solution be smooth as a function of $u$ and continuous in time,  yielding $e^{Mf_1/T} - e^{Mf_2/T} + e^{Mf_3/T}$ as the answer. This is the unique linear combination of the $e^{Mf_i/T}$ solutions that matches to the $u < u(\tau; u_0^+(\tau))$ and $u(\tau;u_0^-(\tau)) < u$ regimes in a continuously differentiable manner. Furthermore, consistent with physical requirements, the distribution is always positive because $f_2$ is always bounded between $f_1$ and $f_3$. In this way, each of the $f_i$'s is a solution to the Hamilton-Jacobi equation~\eqref{eqHJ3}, and therefore $\mathscr{P}$ is the solution to the full Kolmogorov equation~\eqref{eqU13} (up to $T/M$ power corrections). 

Finally, to enforce number conservation in consistency with the original equation~\eqref{eqU13}, we multiply the solution that we have obtained by a $u$-independent, $\tau$-dependent, factor so as to enforce the condition that $\int_0^\infty dp \, p^2 \, \mathscr{P} = N$ stays constant.

\begin{figure}
    \centering
    \includegraphics[width=0.49\linewidth]{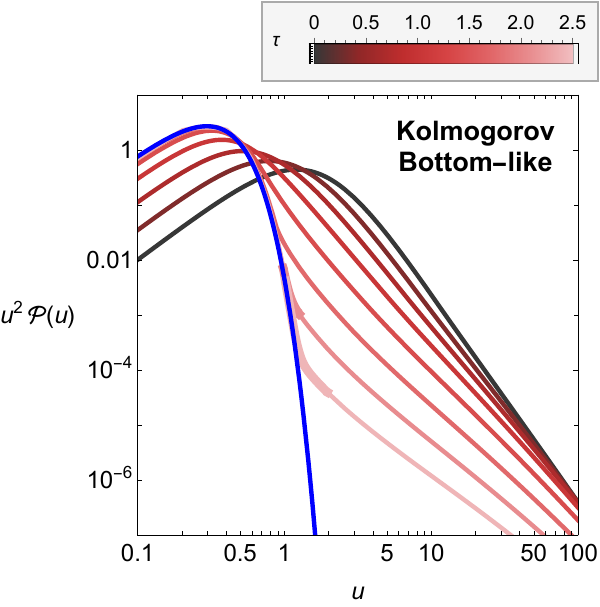}
    \includegraphics[width=0.49\linewidth]{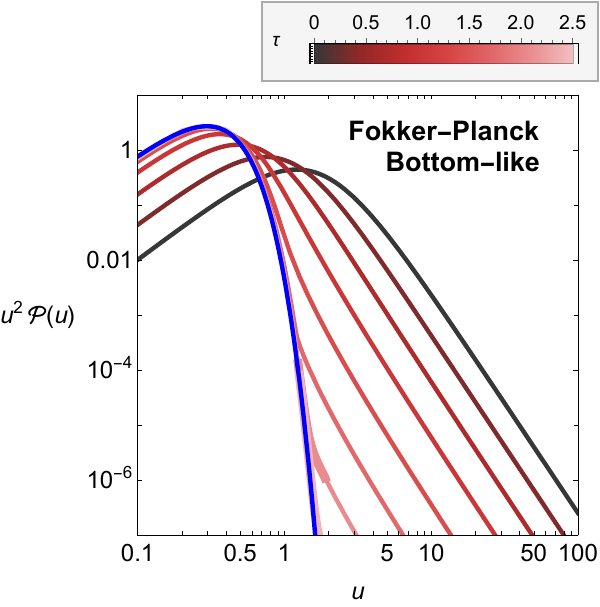}
    \caption{Physical momentum space distribution functions ${\mathscr P}(u)$ resulting from the time evolution 
    of the same initial distribution (dark grey curve) using the Kolmogorov equation for
    for $\mathcal{N}=4$ SYM theory (left panel) and its Fokker-Planck truncation (right panel), obtained as discussed in the main text, using $M/T = 4.75/0.2$. The equilibrium configuration is displayed in blue. Snapshots of the time evolution of the distribution function are shown from darker to lighter tones of red (same color scheme as for the dashed lines in Fig.~\ref{fig:K-FP-solution-sample}), starting from the same initial condition at $\tau = 0$, and spaced in time by $\Delta \tau = 0.35$. For the distributions at late times, we have used thicker lines to highlight the
    position of the intermediate interval in Eq.~\eqref{eq:P-solution-explicit-reco} where $\mathscr{P}$ is determined by the sum of three terms.
    The equilibration dynamics at small momenta $u < 1$ is almost indistinguishable for Kolmogorov and Fokker-Planck evolution. On the other hand, at large momenta $u \gg 1$ both the timescale of equilibration and the shape of the curves are quite different. Equilibration at $u\gg 1$ is much slower with Kolmogorov dynamics and the slope of the curves changes in time. This will have important consequences for our discussion in Section~\ref{sec:dynamics-solutions}.}
    \label{fig:solutions-P-example}
\end{figure}

To illustrate the result of carrying out these steps, in Fig.~\ref{fig:solutions-P-example} we show the resulting physical distribution functions $\mathscr{P}$ obtained from the solutions to Hamilton's equations shown in Fig.~\ref{fig:K-FP-solution-sample} via the procedures we have just described.
In Fig.~\ref{fig:solutions-P-example}, we have chosen
$M/T = 4.75/0.2$. 
The purpose of this figure is two-fold: (i) It illustrates how the curves in Fig.~\ref{fig:K-FP-solution-sample} translate
into physical distribution functions $\mathscr{P}$.
We have highlighted the parts of the solutions satisfying $ \tau > \tau_1 \quad \& \quad u(\tau;u_0^+(\tau)) < u < u(\tau;u_0^-(\tau))$ with thicker solid lines 
to highlight
that the matching between the low-momentum regime and the high-momentum regime, in addition to being mathematically sound, is physically sensible.
And, (ii) it sets the stage to begin a discussion of the physical features of the dynamics, including a discussion of the contrast between Kolmogorov evolution and evolution with the truncated Fokker-Planck kernel with the Einstein relation imposed artificially.
While we will only undertake this discussion in full once we get to Section~\ref{sec:dynamics-solutions}, we can remark at this point that the biggest difference between Kolmogorov and Fokker-Planck dynamics is how fast --- and how --- equilibration takes place at high momentum $u \gg 1$,
whereas the dynamics at small momentum $u \lesssim 1$ is quite similar. 
Specifically, the change in the slope of the high momentum tail of the distribution for Kolmogorov dynamics is most striking, and has consequences that we will discuss at length in Section~\ref{sec:dynamics-solutions}.

\section{The interplay between HQET, the Kolmogorov equation, and AdS/CFT}
\label{sec:interplay}

Before continuing onward in Section~\ref{sec:dynamics-solutions} with interpreting, and understanding the implications of our central results for the evolution of heavy quark momentum space distribution functions ${\mathscr P}(u)$,
as illustrated in Fig.~\ref{fig:solutions-P-example}, in this Section we look more broadly at the interplay between the effective theory (HQET) and gauge/gravity duality approaches from which our results draw input. In so doing we shall also understand up to what heavy quark velocity our results can be trusted, and how and why the answer to this question depends on the 
heavy quark momentum distribution itself. The goal of this exposition is to provide a conceptual discussion that shows how the various assumptions and/or approximations that enter into our framework are tied to physical considerations regarding the dynamics of heavy quarks, first in general theories and then specifically in $\mathcal{N}=4$ SYM, so that systematic improvements of the description of heavy quark transport from first-principles field theory may be carried out in the future.

Given that the possibility to formulate a Kolmogorov equation for heavy quarks does not rely on the specific theory to which heavy quarks are coupled, we shall first discuss the theory-independent considerations regarding the momentum transfer probability in Section~\ref{sec:general-validity}. While this discussion may seem unnecessarily technical on a first reading, it is necessary in order to give proper context to our results in subsequent sections. In Section~\ref{sec:k-validity} we do the same for the Kolmogorov equation. In particular, we show how the Kolmogorov equation is directly sensitive only to the most likely momentum transfer that a heavy quark experienced in order to attain a momentum ${\bf p}$ after a given time step --- which depends on the heavy quark momentum distribution. Note that this differs from the most likely momentum transfer that a heavy quark will experience in the next time step --- which defines the drag force. It follows that survivor bias effects are naturally encoded in our framework, as we illustrate explicitly in Section~\ref{sec:Survivor-Bias}.

Next, in Section~\ref{sec:sym-validity}, we relate these general considerations to the explicit results established previously in $\mathcal{N}=4$ SYM theory. Specifically, we discuss the ``speed limit'' for the validity of the holographic description of heavy quarks with a finite mass $M$ moving through the strongly coupled plasma of $\mathcal{N}=4$ SYM theory~\cite{Liu:2006he,Casalderrey-Solana:2007ahi,BitaghsirFadafan:2008adl,Chernicoff:2008sa}, given by $\gamma < {\cal O}(M^2/\lambda T^2)$, as well as our more recent calculation~\cite{Rajagopal:2025ukd} that reached the same conclusion. Our analysis in this Section reveals that the speed limit is increased by a factor given by the ratio $|\langle C_3 \rangle / \bar{C}_3|$  between the most likely momentum transfer a heavy quark whose current Lorentz boost is $\gamma$ will experience during the next time step ($\langle C_3 \rangle$, determined by the drag force) and the most likely momentum transfer that a heavy quark with a given momentum experienced in the previous time step in order to attain its present momentum ($\bar{C}_3$, determined from the present momentum distribution by the Kolmogorov equation). We shall derive this refined speed limit in Eq.~\eqref{eq:limit-C3}, but to be concrete we state it already here:
\begin{equation}
    \gamma < \left| \frac{\langle C_3 \rangle}{\bar{C}_3} \right| \gamma_{\rm drag}^{\rm limit} \, .
    \label{eq:new-speed-limit}
\end{equation}
In the course of deriving this equation, we shall introduce these two rates of momentum transfer $\langle C_3 \rangle$ and $\bar{C}_3$, and the distinction between them, conceptually in Sections~\ref{sec:k-validity} and~\ref{sec:Survivor-Bias} and quantitatively in Section~\ref{sec:sym-validity}. In Eq.~\eqref{eq:new-speed-limit}, $\gamma_{\rm drag}^{\rm limit}= \mathcal{O}(M^2/\lambda T^2)$ is the limiting speed beyond which the holographic (trailing string) description~\cite{Herzog:2006gh,Gubser:2006bz} of heavy quark propagation breaks down.

\subsection{General considerations regarding the momentum transfer probability} \label{sec:general-validity}

As we did in our analysis of the universal equilibration condition for heavy quarks~\cite{Rajagopal:2025rxr}, we begin from the HQET Lagrangian~\cite{Georgi:1990um} 
\begin{equation}
    \mathcal{L} = \bar{Q}_v i v \cdot D Q_v - \frac{1}{2M}  \bar{Q}_v \left[D_\perp^2 + \tfrac12 g \sigma^{\mu \nu} F_{\mu \nu} \right] Q_v + \mathcal{O} \left( \frac{1}{M^2} \right) \, .
    \label{eq5.1}
\end{equation}
To leading order in $1/M$, the propagator of a charged heavy particle is the Wilson line characterized by the velocity of the particle
\begin{equation}
    W_{\left[x_f,x_i \right]}^{ij} = {\cal P} \exp\left[ i g \int_{t_i}^{t_f} dt\, A_\mu \dot{x}^\mu\right]\, ,
    \label{eq5.2}
\end{equation}
where the integration path is a straight line $x^\mu(t) = x_i^{\mu} + \left(t-t_i,{\bf v} (t-t_i) \right)$, and $x_f = x^{\mu}(t_f)$. In this way, momentum transfer to the heavy quark is related to the position dependence of the gauge field $A_\mu$. Corrections to the leading order dynamics are controlled by the magnitude of $D_\perp^\mu = D^\mu - v^\mu (v \cdot D) \sim k^\mu - v^\mu (v \cdot k)$ relative to the mass of the heavy quark $M$, where $v^\mu =  (\gamma, \gamma{\bf v}) $.\footnote{Note that $v^\mu$ is the $4$-velocity, and here we shall refer to $v^\mu v_\mu$ as $v^2$.
Elsewhere in the text, we use a different notation, with $v$ representing the magnitude of the 3-velocity: $v = |{\bf v}|$.} 
The spin-dependent terms arising from the $\sigma_{\mu \nu} F^{\mu \nu}$ term in Eq.~\eqref{eq5.1} do not contribute at this order (unless we look at polarization-dependent observables, which we will not do).
The probability amplitude for the momentum ${\bf p}$ of the heavy quark to change by an amount $-{\bf k}$ is then
\begin{equation}
\langle {\bf p}-{\bf k}, i\vert_{\rm out} \vert {\bf p}, j\rangle_{\rm in}
    = \int d^3{\bf x}_f\, e^{i{\bf k} \cdot {\bf x}_f} W_{\left[x_f,x_i \right]}^{ij} \left( 1 + \mathcal{O}\left( t \frac{D_\perp^2 }{M} \right) \right) \, .
    \label{eq5.3}
\end{equation}
The two open indices $i,j$  indicate the color state of the ingoing and outgoing quark. Our sign convention for ${\bf k}$ is that ${\bf k}$ is the momentum transferred from the heavy quark to the medium, meaning that ${\bf k} \cdot {\bf p} > 0$ reflects loss of longitudinal momentum. Note that this amplitude automatically projects the dynamics onto states with one (and only one) heavy quark.

In a thermal equilibrium state, the probability for such a change of momentum to take place is essentially given by the absolute value squared of the amplitude \eqref{eq5.3}, averaged over the thermal density matrix of the gauge theory $e^{-\beta H}$, where $H$ is the gauge theory Hamiltonian and $\beta = 1/T$ is the inverse temperature. However, care is needed to render the physical probability gauge-invariant. In practice, to sum over color indices that are attached to Wilson lines ending at different spatial positions, we need to parallel transport this open index from one point in space to the other, which may be done with a Wilson line. For the sake of generality, let us denote the matrices that implement such parallel transport by $\rho_{ij}$ at the initial time and by $\tilde{\rho}_{ij}$ at the final time. It then follows that 
\begin{equation}
    P({\bf k};{\bf v},t) = \frac{1}{(2\pi)^3} \int d^3 {\bf L} \, e^{-i {\bf k} \cdot {\bf L} } \, \langle W_{\bf v} \rangle_T({\bf L}) \left[ 1 + \mathcal{O}\left( t \frac{ k^2 - (k \cdot v)^2 }{M} , t \frac{ T^2 }{M} \right) \right] \, , \label{eq5.4}
\end{equation}
where the combination $k^2 - (k \cdot v)^2$ appears because corrections due to a nonzero momentum transfer $k^\mu$ are controlled by its components orthogonal to the 4-vector $v^\mu$, and $\langle W_{\bf v} \rangle_T({\bf L})$ is the Wilson loop
\begin{equation}
    \langle W_{\bf v} \rangle_T({\bf L}) = \frac{1}{Z} {\rm Tr}_{\cal H} \left[W^{lk}_{[(0,{\bf L}),(t,{\bf v}t + {\bf L})]} \,
    \tilde{\rho}_{ki} \, W^{ij}_{[(t,{\bf v}t),(0,{\bf 0)}]} \, \rho_{jl}  \right]\, . \label{eq5.5}
\end{equation}
We note that the matrices $\tilde{\rho}$, ${\rho}$ depend on {\bf L}, but not on $t$. The normalization factor $Z$ is independent of ${\bf L}$ but it may depend on $t$ and ${\bf v}$; it is chosen such that $\langle W_{\bf v} \rangle_T({\bf 0}) = 1$. Eq.~\eqref{eq5.4} makes the two kinds of corrections explicit that arise from neglecting subleading orders $1/M$ in the HQET Lagrangian \eqref{eq5.1}. Corrections arise from:
\begin{itemize}
 \item   
contributions controlled by the magnitude of the momentum transfer relative to the mass, and 
\item contributions controlled by $T/M$. 
\end{itemize}

Eq.~\eqref{eq5.4} denotes the momentum transfer probability, that is the probability that a heavy quark with velocity ${\bf v}$ (relative to the rest frame of the thermal medium) will change its momentum by $-{\bf k}$ after a time $t$. This is the field theory input into the stochastic equation~\eqref{eqU8},
\begin{equation}
    \mathscr{P}({\bf p}, t+\Delta t )  =
    \int d^3{\bf k}\, P({\bf k};{\bf v}({\bf p}+{\bf k}),\Delta t)\, 
    \mathscr{P}({\bf p}+{\bf k}, t )\, , \label{eq5.6}
\end{equation}
which is the starting point of our formalism. It is the most general equation one can write for the evolution of the momentum distribution of an ensemble of individual heavy quarks.

Up to this point in our discussion, it may seem that it is sufficient to choose sufficiently small time steps $\Delta t$ in Eq.~\eqref{eq5.6} such that corrections from the discretization of \eqref{eq5.5} remain small. However, while Eq.~\eqref{eq5.6} is exact, one needs a concrete rather than a formal expression for the Wilson loop in \eqref{eq5.4}. This is what our calculation in~\cite{Rajagopal:2025ukd} provides for $\mathcal{N}=4$ SYM in the 't Hooft limit ($N_c \to \infty$, $\lambda = g^2 N_c \to \infty$). To arrive at this expression, we expand the Wilson loop \eqref{eq5.5} in the limit $t \gg 1/T$, $\vert {\bf L}\vert$, obtaining
\begin{equation}
    \langle W_{\bf v} \rangle_T({\bf L}) = \exp\left[ - \sqrt{\lambda}\, t\, T\, S_{\rm tot} ({\bf L};{\bf v}) + \dots\right]\, .\label{eq5.7}
\end{equation}
Here, ``\dots'' represents terms that are not extensive in $t$: they are of order ${\cal O}((tT)^{-n})$ with $n\geq 0$. It is only after neglecting these terms that we arrive at the starting point \eqref{eqU1} suitable for explicit calculation. 
Therefore
\begin{itemize}
    \item The expression \eqref{eq5.7} for the Wilson loop is based on requiring $T t \gg 1$. This condition also appears in the analysis for light quark energy loss~\cite{Chesler:2014jva,Chesler:2015nqz}. 
\end{itemize}
We emphasize that this is not only a technical but also a physical requirement for our formalism: keeping only the dominant,  extensive, contribution in Eq.~\eqref{eq5.7} guarantees that the momentum transfer probability satisfies the addition property of independent random variables, thus making the dynamics independent of the choice of $\Delta t$.

The approximations emphasized in the three bullets $\bullet$ above and made explicit in Eqs.~\eqref{eq5.4} and \eqref{eq5.7} refer to the parametric dependence of corrections to the $1/M$ expansion and to the identification of the extensive piece of $\langle W_{\bf v} \rangle_T({\bf L})$, respectively. In addition, $\langle W_{\bf v} \rangle_T({\bf L})$ has a (parametrically $\mathcal{O}(1)$) prefactor which can be large in value and may be velocity-dependent and may diverge in the limit $\gamma\to \infty$. Assessing the range of applicability of $P({\bf k};{\bf v})$ as a function of ${\bf k}$ thus requires a calculation in the theory of interest. For example, if after calculating $\langle W_{\bf v} \rangle_T({\bf L})$ at a given $v$ one finds that $P({\bf k};{\bf v})$ has a significant amount of probability outside the range of ${\bf k}$ in which the calculation is parametrically under control, then most expectation values (integral moments of $P({\bf k};{\bf v})$) will not be trustworthy. We shall examine all of the above considerations closely in Section~\ref{sec:sym-validity} for the concrete case of $\mathcal{N}=4$ SYM theory.

As a final remark in this qualitative discussion, we note that, provided the $T/M$ corrections are under control, there will always exist a regime around zero momentum transfer $k^\mu = 0$ in which the Wilson loop $ \langle W_{\bf v} \rangle_T({\bf L})$ will characterize the momentum transfer probability. One may ask how it is possible to characterize a probability distribution in one region of its domain without knowing its behavior in the rest of its domain. The answer is that in any quantum theory one can write
\begin{equation}
    P({\bf k};{\bf v},t) = \frac{ {\rm Tr}_{\mathcal{H}} \left[ \ket{{\bf p} - {\bf k} }_j  \bra{{\bf p} - {\bf k}}_j U(t) \ket{\bf p}_i e^{-\beta H_{\rm QGP}} \bra{\bf p}_i U^\dagger(t) \right] }{{\rm Tr}_{\mathcal{H}} \left[ U(t) \ket{\bf p}_i e^{-\beta H_{\rm QGP} } \bra{\bf p}_i U^\dagger(t) \right]} \, , \label{eq:P-formal}
\end{equation}
and this provides an explicit expression for $P({\bf k};{\bf v})$ in terms of matrix elements of the theory, where $e^{-\beta H_{\rm QGP}}$ describes a thermal state unperturbed by the heavy quark, and $U(t) = \exp (- i (H_{\rm QGP}  + H_{\rm int} + H_{\rm HQ} )t)$ contains the dynamics of the QGP and the HQ as well as their interactions. The trace ${\rm Tr}_{\mathcal{H}}$ goes over all states in the theory. 
Because the projection operator $\ket{{\bf p} - {\bf k} }_j  \bra{{\bf p} - {\bf k}}_j$ singles out specific matrix elements of the theory, the numerator may be evaluated independently at each value of ${\bf k}$, and HQET methods may be used if ${\bf k}$ is small. On the other hand, the normalization in the denominator involves a sum over all states, and so it contains information on all processes that can happen. However, because of unitarity this denominator ${\rm Tr}_{\mathcal{H}} \left[ U(t) \ket{\bf p}_i e^{-\beta H_{\rm QGP} } \bra{\bf p}_i U^\dagger(t) \right] = {\rm Tr}_{\mathcal{H}} \left[ \ket{\bf p}_i e^{-\beta H_{\rm QGP} } \bra{\bf p}_i  \right]$ is actually time-independent. This means that the ratios $P({\bf k}_1;{\bf v},t)/P({\bf k}_2;{\bf v},t)$, and
thus the prefactor $\tilde{S}_{\rm tot}$ of the exponential time dependence of $P({\bf k};{\bf v},t)$ 
may be reliably calculated using the $1/M$ expansion at small ${\bf k}$. Fittingly, this is all we need in the derivation of the Kolmogorov equation~\eqref{eqU10}.

\subsection{General considerations for a Heavy Quark Kolmogorov Equation} 
\label{sec:k-validity}

In addition to the assumptions highlighted by bullets $\bullet$ in the above subsection, the derivation of the Kolmogorov Equation set out in Eq.~\eqref{eqU10} introduces two important additional assumptions:
\begin{enumerate}
    \item $f$, the logarithm of the distribution function $\mathscr{P}$, is a sufficiently slowly-varying function  that the approximation
    \begin{equation}
        f \!  \left( {\bf u} + \Delta \tau \frac{\pi}{2} {\bf C} , \tau \right) \approx f \! \left( {\bf u} , \tau \right) + \Delta \tau \frac{\pi}{2} {\bf C} \cdot \frac{\partial f}{\partial {\bf u} } \left( {\bf u} , \tau \right) 
        \label{eq5.8}
    \end{equation}
    is consistent with the momentum transfer selected by the Kolmogorov dynamics which, we shall see, depends on the shape of $f$. 
    Recall here that ${\bf C}=2{\bf k}/(\pi M \Delta \tau)$.
    \item The velocity dependence of $\tilde{S}_{\rm tot}$, defined in Eq.~\eqref{eqU1}, at fixed ${\bf C}$ is sufficiently smooth such that
    \begin{equation}
        \tilde{S}_{\rm tot} \! \left( {\bf C} ; {\bf v} \left({\bf u} + \Delta \tau \frac{\pi}{2} {\bf C}  \right)  \right) \approx  \tilde{S}_{\rm tot} \! \left( {\bf C} ; {\bf v} \left({\bf u}  \right)  \right)  
        \label{eq5.9}
    \end{equation}
    is a good approximation, in the same sense as for the previous assumption.
\end{enumerate}
 We note  that   $M \Delta \tau / T = \sqrt{\lambda} T \Delta t \gg 1$ is necessary in order for the saddle point approximation 
 that we used to go from the third to 
 the fourth line of Eq.~\eqref{eqU10} to be justified, but at strong coupling $\lambda\gg 1$ this is already implied by the requirement that $T \Delta t \gg 1$, a requirement that we have already imposed in the previous subsection as it is necessary for the evaluation of $ \langle W_{\bf v} \rangle_T({\bf L})$ 
 in Eq.~\eqref{eq5.7}.
 
The approximation \eqref{eq5.9} is justified whenever the gradients $\partial {\bf v}_j / \partial {\bf u}_i $ are sufficiently small compared to the momentum transfer ${\bf C}$ 
that ${\bf v}({\bf u}) \gg  \tfrac{\partial {\bf v}_j}{\partial {\bf u}_i }\, \Delta \tau \frac{\pi}{2} {\bf C}_i $.
In the isotropic case on which we focus throughout, $\partial v/\partial u = \gamma^{-3}$, and this condition is easily satisfied in the relativistic limit. 
For the case of $\mathcal{N}=4$ SYM theory, we have validated these assumptions with numerical studies described in Appendix~\ref{app:Kolmogorov-checks}.

We shall see below that the value of the momentum transfer rate ${\bf C}=\bar{\bf C}(\partial_{\bf u}f)$ selected via the saddle point approximation, used to obtain the fourth line in the derivation~\eqref{eqU10}, plays a decisive role in determining 
the regime of applicability of our derivation of the Kolmogorov Equation.
This saddle point approximation relies on the  requirements \eqref{eq5.8} and \eqref{eq5.9}, which were used to obtain the third line of the derivation \eqref{eqU10}. These requirements are, essentially, that the effected momentum transfer $ \propto \Delta \tau C_3$ relevant for the dynamics is sufficiently small. For concreteness, we reproduce this step here for the isotropic case:
\begin{align}
    &f \! \left(u + \Delta \tau \frac{\pi}{2} C_3, \tau \right) - \Delta \tau \, \tilde{S}_{\rm tot} \! \left( C_3 ; v \left(u + \Delta \tau \frac{\pi}{2} C_3  \right)  \right) \nonumber \\ &\approx f(u,\tau) + \Delta \tau \left[ \frac{\pi}{2} C_3 \cdot \frac{\partial f}{\partial u}(u,\tau) - \tilde{S}_{\rm tot} \! \left( C_3 ; v(u) \right) \right] \, , \label{eq5.10}
\end{align}
where the global maximum of the right-hand side of this expression returns the saddle point solution $C_3 = \bar{C}_3(\partial_u f)$ as in \eqref{eqU11} and \eqref{eqHJ16}. We stress that, as our notation emphasizes, the momentum transfer rate that we denote by $\bar C_3$ depends
on the log of the momentum distribution, $f$, specifically on its slope $\partial_{\bf u} f$,
and not only on the momentum $\bf u$ of the heavy quark. In this subsection, we shall  see why this is so, and why this is important.

One fact that has been inconspicuous so far in our discussion is the \textit{time} at which $\bar{\bf C}$ appears in Eq.~\eqref{eqU10}. Note that the  left-hand side
of Eq.~\eqref{eqU10} is evaluated at a fixed momentum ${\bf u}$ and at a time $\tau + \Delta \tau$, whereas the right-hand side is defined in terms of a convolution at the time $\tau$. That is to say, from the point of view of the distribution $\mathscr{P}$ at time $\tau + \Delta \tau$, $\bar{\bf C}$ is the momentum transfer rate that connects the previous point in time, $\tau$, with the present --- sampled from the momentum transfer probability at the previous time $\tau$ conditional with the restriction that a heavy quark that 
picks up momentum 
$\bar{\bf  C}\Delta\tau$ between the previous time $\tau$ and the present time $\tau+\Delta\tau$  
lands in the momentum ``bin'' ${\bf u}$ of the momentum distribution $\mathscr{P}$. 
It is because of this restriction that $\bar{\bf C}$ is intrinsically tied to the steepness of the distribution in the previous time step, encoded in the Kolmogorov equation through the value of $\partial_{\bf u} f$.

If $f$ were a flat (unphysical) distribution, i.e., independent of $u$, then the selected value of $\bar{C}_3$ at the saddle point 
would be the minimum of $\tilde{S}_{\rm tot} \! \left( C_3 ; v(u) \right)$. This minimum 
is given by $\langle C_3 \rangle$, which determines the average energy loss $\langle C_3\rangle \Delta\tau$ that a single heavy quark with momentum $M u$ will undergo during the
next time step. However, for a physically sensible, normalizable, $\mathscr{P}$, $f$ must must not be $u$-independent and must  
be a decreasing function of $u$ at large $u$.
In fact, the phenomenologically inspired distributions that we shall consider in this work are all strictly decreasing functions of $u$. 
And, as discussed in Section~\ref{sec:H-Flow}, $f'<0$ is guaranteed at all times if it is true at the initial time. We are thus left with the generic case $\partial f/\partial u < 0$, where the value of the momentum transfer rate
$\bar{C_3}$ at the global maximum of the right-hand side of Eq.~\eqref{eq5.10} 
is generically smaller than the average momentum transfer rate $\langle C_3\rangle$ 
governed by the momentum transfer distribution
$P({\bf k};{\bf v})$:
\begin{equation}\label{eq:Cbar-lessthan-meanC}
    \bar{C_3} < \langle C_3\rangle\, .
\end{equation}
Here, the $\bar C_3$ that maximizes \eqref{eq5.10} is the longitudinal momentum transfer at the saddle point 
that dominates the integral in Eq.~\eqref{eqU10}, which means that it is the most likely momentum transfer rate experienced in the previous time step by a heavy quark that has just landed in the momentum bin ${\bf u}$.
In contrast, the average momentum transfer rate $\langle C_3\rangle$ is the minimum of $\tilde{S}_{\rm tot}$ and is the most likely momentum transfer (according to $P({\bf k};{\bf v})$) which a heavy quark in the momentum bin ${\bf u}$ will experience next.

In the case of the thermal equilibrium
distribution to which the Kolmogorov dynamics will drive any distribution $\mathscr{P}$,
the inequality~\eqref{eq:Cbar-lessthan-meanC}
becomes both striking and quantitative.
We have seen that in thermal equilibrium,
$f' = - v$. In this case, the $\bar C_3$
that maximizes the right-hand side of Eq.~\eqref{eq5.10} is:
\begin{equation}\label{eq:Cbar-equals-minus-meanC3}
    \bar{C}_3 = - \langle C_3 \rangle\, \qquad \hbox{in thermal equilibrium.}
\end{equation}
Below we shall explain in qualitative terms why this must be so, but first we note 
that Eq.~\eqref{eq:Cbar-equals-minus-meanC3} 
is a direct consequence of a result from
Ref.~\cite{Rajagopal:2025rxr}, shown there to be universal in the sense that it is valid in any quantum field theory:
\begin{equation}
    \tilde{S}_{\rm tot}(C_3,v) = \tilde{S}_e(C_3,v) - \pi v C_3/4 \, , \label{eq:Se}
\end{equation}
which we have suitably rewritten due to the different prefactor we use in our definition of ${\bf C}$ in this work.
In Eq.~\eqref{eq:Se},
$\tilde{S}_e$ is an even function of $C_3$, which has the immediate consequence that $\tilde{K}(C_3,v)$ defined in Eq.~\eqref{eq:new-Ktilde} takes the form
\begin{equation}
    \tilde{K}(C_3,v) = 
    \tilde{S}_{\rm tot}(C_3,v) - C_3 \frac{\partial \tilde{S}_{\rm tot}}{\partial C_3} = \tilde{S}_e(C_3,v) - C_3 \frac{\partial \tilde{S}_e}{\partial C_3} \, 
\end{equation}
and is guaranteed to be even in $C_3$. Therefore, in addition to the minimum of $\tilde{S}_{\rm tot}$, where $\tilde{K} = 0$ (corresponding to $x=f'=0$), the evolution operator also vanishes on the opposite momentum transfer rate, which gets mapped to $f' = -v$ in light of Eqs.~\eqref{eqHJ16} and~\eqref{eq:Se}.
That is to say, when Eq.~\eqref{eqHJ16} is $f' = \frac{2}{\pi} \frac{\partial \tilde{S}_{\rm tot}}{\partial C_3} = - v$, it follows by taking a derivative of Eq.~\eqref{eq:Se} that
\begin{equation}
    -v = - \frac{v}{2} + \left. \frac{2}{\pi} \frac{\partial \tilde{S}_{e}}{\partial C_3} \right|_{C_3 = \bar{C}_3(\partial_u f = -v) }  \implies  -v/2 = \left. \frac{2}{\pi} \frac{\partial \tilde{S}_{e}}{\partial C_3} \right|_{C_3 = \bar{C}_3(\partial_u f = -v) }  \, . \label{eq:solving-barC3}
\end{equation}
We know that $C_3 = \langle C_3 \rangle$ solves the same equation with the opposite sign of $v/2$ on the left-hand side because, by comparison with Eq.~\eqref{eq:Se}, this corresponds to $\partial \tilde{S}_{\rm tot}/\partial C_3 = 0 $ and this \textit{defines} $\langle C_3 \rangle$. Because $\tilde{S}_e$ is an even function of $C_3$, it follows that $\bar{C}_3 = -\langle C_3 \rangle$ is the solution to Eq.~\eqref{eqHJ16} for a distribution in equilibrium.

We turn now to understanding in qualitative terms how the inequality \eqref{eq:Cbar-lessthan-meanC} arises as a consequence of the fact that
$\partial f/\partial u<0$, meaning that the heavy quark momentum distribution $\mathscr{P}$ is a decreasing function of $u$.  
Consider the heavy quarks with momentum $u$
at time $\tau+\Delta\tau$.
Since $\langle C_3\rangle>0$, 
which corresponds to momentum {\it loss}, in the next time step more of those heavy quarks (almost all of them at large $u$) will lose momentum, although the momentum transfer probability distribution $P({\bf k};v)$ has enough of a tail that a minority (a small minority at large $u$) will gain momentum.  
However, if we look at the heavy quarks with momentum $u$ at time
$\tau+\Delta \tau$ and instead ask what momentum they had at the previous time $\tau$,
then the slope $\partial f/\partial u<0$ in the momentum distribution skews the answer to this question in a way that favors heavy quarks that lost less momentum than the mean.
The inequality \eqref{eq:Cbar-lessthan-meanC} follows! It is therefore a consequence of  selection bias, sometimes referred to in this case as ``survivor bias''.  As we shall discuss in Section~\ref{sec:Survivor-Bias} and illustrate
explicitly in Fig.~\ref{fig:comparison-zero-change-probability}, this effect is much larger for the complete non-Gaussian dynamics than for the Fokker-Planck truncation.

The equality \eqref{eq:Cbar-equals-minus-meanC3} valid in thermal equilibrium
highlights a crucial aspect of the  momentum transfer rate $\bar C_3$ 
that dominates the integral in Eq.~\eqref{eqU10}: it doesn't have to be close to the most likely value specified by the momentum transfer distribution  $P({\bf k};{\bf v})$ --- it can in fact take the exact same value with the opposite sign. 
This is a manifestation of the difference between $\bar{C}_3 \Delta\tau$, the most likely momentum transfer that a heavy quark with (dimensionless) momentum $u$  experienced in the previous time step, and $\langle C_3 \rangle\Delta\tau$, the most likely momentum transfer that this heavy quark will experience next.  
The latter depends on the slope of the distribution function at $u$ which
generates a survivor bias that is stronger for more steeply falling momentum
distributions.
Because there are fewer heavy quarks with 
momenta above $u$ than with momenta below $u$, the product of the number of heavy quarks with momentum ${\bf p} + {\bf k}$ times their momentum change probability has a maximum at a value of ${\bf k}$ that is \textit{smaller} than the most likely value of $P({\bf k};{\bf v})$, thus selecting a most likely momentum transfer rate $\bar{C}_3$ to reach momentum $u$ to be \textit{always} smaller than $\langle C_3 \rangle$. In essence, this is the same argument and conclusion that was first given in Ref.~\cite{Baier:2001yt} in the context of understanding 
why a description of the quenching of a QCD jet in any QCD medium in terms of the mean energy loss of hard partons is inadequate, a conclusion that is central to all Monte Carlo descriptions of jet quenching. 

The thermal limiting case $\bar{C}_3 = - \langle C_3 \rangle$ in Eq.~\eqref{eq:Cbar-equals-minus-meanC3} encodes a very simple physical statement: in thermal equilibrium, there must be a balance between heavy quarks moving out of a given momentum ``bin'' and into that momentum bin. In particular, a quark with dimensionless momentum ${\bf u}$ at time $\tau+\Delta\tau$ will most likely have had \textit{less} momentum in the previous time step rather than more, because in equilibrium momentum fluctuations of other heavy quarks have to compensate for the fact that the heavy quarks that had momentum ${\bf u}$ at time $\tau$ will (on average) have lost momentum. The requirement of having a stationary distribution then implies that the typical heavy quark that moved into the ${\bf u}$ momentum bin actually gained energy in the previous time step --- with exactly the opposite momentum transfer rate $\bar{C}_3 = - \langle C_3 \rangle$.
That is, if we ask where the 
heavy quarks in the ${\bf u}$ momentum bin
came from, it is likely that they 
were among the small minority of heavy quarks in a lower momentum bin at the previous time step that gained momentum, since there are more heavy quarks with lower momentum. This is a particularly striking example of the consequences of selection bias, one that goes beyond survivor bias. Finally, if we ask where the heavy quarks in the ${\bf u}$ momentum bin will be at the next time step, most of them will lose momentum --- but when they arrive at their lower momentum bin they will be out-numbered in that bin by heavy quarks
that arrived there from below.

\subsection{Non-Gaussian vs.~Gaussian survivor bias} 
\label{sec:Survivor-Bias}

\begin{figure}
    \centering
    \includegraphics[width=\linewidth]{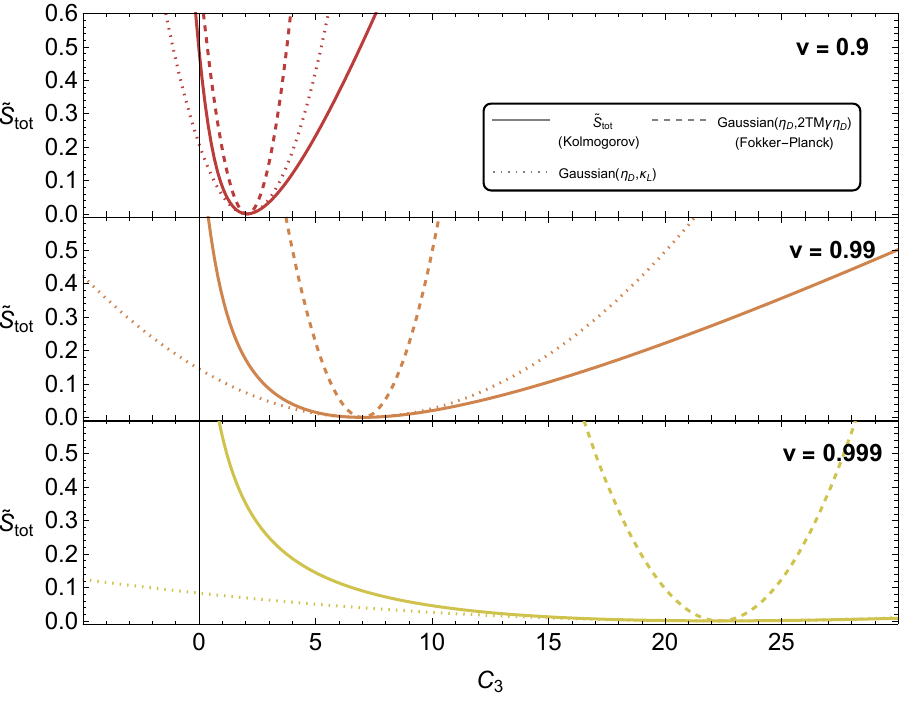}
    \caption{The magnitude of the logarithm of the momentum transfer probability distribution, $\tilde{S}_{\rm tot}$, and two possible truncations thereof, for three heavy quark velocities $v = 0.9, 0.99, 0.999$.  (The most probable momentum transfer $k_3=C_3\pi\sqrt{\lambda}T^2 t/2$ is
    that with smallest $\tilde{S}_{\rm tot}$.)
    The solid curve is the complete non-Gaussian result that governs the Kolmogorov dynamics. The dashed curve is the Gaussian distribution with a mean that matches the minimum of $\tilde{S}_{\rm tot}$ and variance chosen by hand so as to satisfy the Einstein relation, i.e., with $\kappa_L = 2 T M \gamma \eta_D$. This dashed curve is the ``Fokker-Planck truncation'' that we refer to as such throughout. As a further comparison, in the dotted curves we have also plotted a different ``direct'' Gaussian truncation of the non-Gaussian result \textit{without} modifying the variance of the momentum transfer distribution by hand.  Although this truncation keeps the correct value of the variance, since it violates the Einstein relation~\cite{Gubser:2006nz} it does \textit{not} lead to equilibration.}
    \label{fig:comparison-zero-change-probability}
\end{figure}

To illustrate our considerations in the preceding discussion, it is instructive to compare the size of these effects in the full Kolmogorov equation relative to Gaussian truncations of the dynamics. Fig.~\ref{fig:comparison-zero-change-probability} shows the complete result for $\tilde{S}_{\rm tot}$,
which specifies the momentum transfer probability distirbution, compared to its Fokker-Planck truncation and to a ``direct'' Gaussian truncation (discussed below) for three heavy quark velocities. 

In Fig.~\ref{fig:comparison-zero-change-probability}, the Gaussian that corresponds to the Fokker-Planck truncation (dashed curve) is much narrower than the complete non-Gaussian momentum transfer probability distribution (solid curve). That is, the non-Gaussian fluctuations in the Kolmogorov evolution make rare events where a heavy quark loses much less (or more) energy than is typical much more likely. Therefore, the effects of survivor bias get naturally enhanced in the Kolmogorov dynamics relative to its Fokker-Planck truncation, as may be easily verified by
observing from Fig.~\ref{fig:comparison-zero-change-probability} that in the former case the probability of losing no momentum at all is vastly larger (much less rare) than in the latter case. This makes it possible for those heavy quarks 
with some large momentum $u$ 
that lose little or no momentum to dominate $\mathscr{P}(u)$ at the next time step. We note that because the tails of the non-Gaussian momentum transfer probability distribution are, in fact, exponential~\cite{DuPlessis:2026pyr}, this survivor bias effect will be much stronger at larger velocities in the Kolmogorov evolution relative to in the Fokker-Planck truncation, as (per the dashed curves in Fig.~\ref{fig:comparison-zero-change-probability}) 
the latter unduly suppresses the no-energy-loss probability to a greater and greater extent at larger and larger velocities.

Mathematically, the survivor bias is encoded 
within the Kolmogorov dynamics via Eq.~\eqref{eqHJ16}, which maps the derivative of the logarithm of a physical heavy quark momentum space distribution to a momentum transfer rate via $\tfrac{\partial \tilde{S}_{\rm tot}}{\partial C_3} = \pi \partial_u f/2$. 
That is to say, the selected momentum transfer is determined via a specific value of the slope of $\tilde{S}_{\rm tot}$. The fact that the slope of $\tilde{S}_{\rm tot}$ as a function of $C_3$ is smaller for the full non-Gaussian dynamics than for the Fokker-Planck truncation at any given $C_3$ (as may be seen directly from Fig.~\ref{fig:comparison-zero-change-probability}) directly implies\footnote{Supplemented with the fact that $\tilde{S}_{\rm tot}$ is a convex function of the momentum transfer rate.} that the selected momentum transfer $\bar{C}_3$ will be further away from $\langle C_3 \rangle$ in Kolmogorov dynamics than in Fokker-Planck dynamics. In this way, rare events where heavy quarks survive for longer time with a large momentum are more common. This becomes the defining characteristic of the Kolmogorov dynamics of the large-momentum tails of a steeply falling heavy quark momentum space distribution $\mathscr{P}(u)$ --- as already evidenced in Fig.~\ref{fig:solutions-P-example}.  
Clearly, the values of $C_3$ determined in this way are closer to zero for the complete non-Gaussian result than for  the Fokker-Planck truncation. This is how the larger survivor bias of the complete non-Gaussian result gets encoded in the Kolmogorov equation.

Finally, 
the dotted curves in Fig.~\ref{fig:comparison-zero-change-probability} show the ``direct'' Gaussian truncation
of the momentum transfer probability distribution, where we eliminate all higher non-Gaussian moments but do not modify the variance by hand so as to satisfy the Einstein relation.
With the momentum transfer probability distribution depicted in the dotted curves,
the heavy quark momentum distribution does 
not equilibrate~\cite{Gubser:2006nz}.
The phenomenological consequences of this truncation were explored in Ref.~\cite{Horowitz:2015dta}, and it has also been used as a benchmark in Ref.~\cite{DuPlessis:2026pyr}.
The direct Gaussian truncation --- whose purpose in Fig.~\ref{fig:comparison-zero-change-probability} is to highlight what the variance of the momentum transfer probability distribution actually is --- overestimates rare events with small energy loss and underestimates rare events with large energy loss.   Heavy quark dynamics with this momentum transfer probability distribution (which does not equilibrate) would feature
even stronger survivor bias effects than 
in the Kolmogorov dynamics.

\subsection{The Heavy Quark Kolmogorov Equation in ${\mathcal N}=4$ SYM Theory} \label{sec:sym-validity}

In Section~\ref{sec:k-validity}, we concluded that the heavy quarks 
with momentum ${\bf p}$ 
at a time $\tau + \Delta \tau$ did not, on average, lose momentum $\langle C_3 \rangle \frac{\pi M}{2}\Delta \tau$ in the previous time step --- which is the average momentum that these heavy quarks will lose in the next time step. 
In fact, if the heavy quark momentum space distribution in the vicinity of the momentum ${\bf p}$ is near equilibrium, with $\partial f/\partial u \sim - v$, then the heavy quarks
with momentum ${\bf p}$ will typically have
{\it gained} momentum. And,  
for phenomenologically relevant distributions that have not yet equilibrated --- which satisfy $-v < \partial f / \partial u < 0 $ 
corresponding to distributions that fall with increasing momentum but falls less steeply than in equilibrium
--- heavy quarks with momentum ${\bf p}$ at time $\tau + \Delta \tau$ will on average have experienced a momentum transfer $\bar C_3 \frac{\pi M}{2} \Delta \tau < \langle C_3 \rangle \frac{\pi M}{2} \Delta \tau$ during the previous time step.
A crucial consequence is that in order to assess the validity of the calculation of the evolution of the heavy quark distribution at any momentum ${\bf p}$ where the distribution is a decreasing function thereof, 
we need not ask whether we can calculate the
average momentum transfer that these 
heavy quarks will experience in the next time step.  The important question
is whether
we trust our description of the momentum transfer dynamics for the value of the momentum transfer rate $\bar{C}_3$ that the heavy quarks that have arrived at the momentum bin ${\bf u}$ just experienced. We shall answer this question for 
strongly coupled $\mathcal{N}=4$ SYM theory in this subsection.

Note that this is very different than asking whether we trust our description of the momentum transfer around the mean of the full $P({\bf k};{\bf v})$. And, in fact, as long as there is only one maximum of the convolution between the momentum space distribution and the momentum transfer probability (which, as we established before, is at a value of the momentum transfer rate below the mean), for this purpose it doesn't matter what the mean of $P({\bf k};{\bf v})$ is. The heavy quarks in the  momentum bin ${\bf u}$ at time $\tau$ that
arrived in this bin having just experienced
the mean momentum transfer, meaning
having just lost $\langle C_3\rangle \frac{\pi}{2}\Delta\tau$,
will be outnumbered at their arrival momentum bin by heavy quarks that started out 
in a more highly populated bin
with a lower ${\bf u}$ and lost less momentum.
This means that we can predict the dynamics of
a phenomenologically relevant
distribution satisfying $-v < \partial f / \partial u < 0 $ 
without knowing
the mean of $P({\bf k};{\bf v})$, which is to say without knowing how much momentum an individual heavy quark will lose on average.
Furthermore, the relevant momentum transfer rate $\bar{C}_3$, is always smaller: $|\bar{C}_3| < \langle C_3 \rangle$. 
(We note that if the distribution were falling more steeply than a thermal distribution, or if the distribution were an increasing function of momentum, the opposite ($|\bar{C}_3| > \langle C_3 \rangle$) would be true.) In other words, the mean $\langle C_3 \rangle$, which is the crucial quantity 
if one asks what will happen next to an individual heavy quark, 
can play a completely secondary role in the 
evolution of the shape of the momentum distribution for an ensemble of heavy quarks.

As we will  illustrate at the end of this subsection, the difference between $\bar{C}_3$ and $\langle C_3 \rangle$ can be quite large for phenomenologically relevant heavy quark distributions. In fact, the dynamics of the heavy quark momentum distribution in the relativistic regime  $u\gg 1$
takes place close to $\bar{C}_3/\langle C_3 \rangle = 0$ for the phenomenologically-motivated initial conditions we consider in this work.
As we shall explain in what follows, this
makes our calculation of the evolution of the 
distribution robust even in a regime where $u$ exceeds the ``speed limit'' above which
the holographic description of the behavior of an individual heavy quark and the holographic calculation of $\langle C_3 \rangle$ breaks down.
This is so since the range of momentum transfer that is relevant to the evolution of
the distribution
is pushed closer to ${\bf k} = 0$, thus decreasing the size of the corrections to the HQET-based formalism on which our derivation relies as well as the size of corrections in the derivation of the Kolmogorov equation \eqref{eqU10}. In the present subsection, we shall illustrate these considerations quantitatively for the case of heavy quarks propagating through the strongly coupled plasma of $\mathcal{N} = 4$ SYM theory. We first discuss the range of validity of the momentum transfer probability $P({\bf k};{\bf v})$ as a function of ${\bf v}$, before drawing conclusions about the range of validity of the Kolmogorov equation and the 
heavy quark 
momentum distribution $\mathscr{P}$ obtained from it, including seeing how this range of validity depends on $\mathscr{P}$ itself.

\subsubsection{What is the largest $\gamma$ at which we can trust $P({\bf k};{\bf v})$ at a fixed ${\bf k}$? }

In {strongly coupled} $\mathcal{N}=4$ SYM theory, the momentum transfer probability distribution $P({\bf k};{\bf v})$ takes the explicit form given in Eq.~\eqref{eqU1}.  The moments of this distribution grow rapidly with the Lorentz boost factor; for 
 $\gamma \gg 1$, they scale like~\cite{Rajagopal:2025ukd}
\begin{equation}
	\langle k_\perp^{2m} \, k_3^n \rangle_c \propto \sqrt{\lambda} \T (\sqrt{\gamma} \,  T)^{n + 2m + 1} \times \gamma^{n - 1} \,  .
    \label{eq5.15}
\end{equation}
Using HQET to estimate the time $t$ after which the momentum transfer is too large for the $1/M$ expansion to be reliable, i.e., when the subleading terms in Eq.~\eqref{eq5.4} become important, while simultaneously requiring $t \gg 1/T$, one arrives at~\cite{Rajagopal:2025ukd}
\begin{equation}
	\sqrt{\gamma} \ll \frac{M}{\sqrt{\lambda} T} \, . 
    \label{eq5.16}
\end{equation}
This is exactly the ``speed limit'' that had been obtained before in (at least) three different ways in the description of heavy quark propagation through strongly coupled $\mathcal{N}=4$ plasma:
\begin{enumerate}
	\item If an external electric field is applied to maintain the heavy quark momentum~\cite{Casalderrey-Solana:2007ahi}, the critical electric field at which this set-up breaks down due to pair creation occurs when the heavy quark velocity reaches $\sqrt{\gamma} \sim \frac{M}{\sqrt{\lambda} T}$.
	\item Fluctuation-based calculations of the momentum diffusion coefficient break down if $\sqrt{\gamma} \sim \frac{M}{\sqrt{\lambda} T}$ because for larger values of $\gamma$ the position of the D7 brane that determines the mass of the heavy quarks in the holographic description~\cite{Liu:2006he,Gubser:2006nz}
    lies below the trailing string worldsheet horizon, rendering the meaning of signal propagation\footnote{Another argument that makes reference to the speed of propagation of the string was presented in~Ref.~\cite{Chernicoff:2008sa}.} (and thus fluctuation-based calculations) questionable.
    \item Decelerating particles generically emit \textit{Bremsstrahlung} radiation. 
    The velocity at which this radiative energy loss becomes comparable to energy loss via the drag force formula~\cite{Gubser:2006bz,Herzog:2006gh}, and the latter becomes thus unreliable, is equally set by 
    $\sqrt{\gamma} \sim \frac{M}{\sqrt{\lambda} T}$~\cite{BitaghsirFadafan:2008adl}.
\end{enumerate}

As we consider heavy quark transport in the absence of an external field, and as the momentum transfer probability in terms of $\langle W_{\bf v} \rangle_T({\bf L})$ explicitly projects pair creation events out of the final state, only the two latter points are relevant in the present context. Fittingly, these two ways in which the trailing string calculation has been shown to break down have precise counterparts in the HQET expansion in Eq.~\eqref{eq5.5}:
\begin{enumerate}[label=(\alph*)]
	\item $\mathcal{O}(t T^2/M)$ corrections are related to the position of the D7 brane (controlled by $T/M$), and are therefore mapped to the effects described in point 2 above.
	\item $\mathcal{O}(t k^2/M)$ corrections are related to the momentum change of the heavy quark (controlled by ${\bf k}$), and can be mapped to the effect described in point 3 above.
\end{enumerate}

To investigate corrections of the type (a) that are due to the position of the D7 brane, in close analogy to the result of Ref.~\cite{Rajagopal:2025ukd}  we write the logarithm of the momentum transfer distribution as a function of
the (dimensionless) position in the holographic coordinate, $z_b$, of the bottom of the D7 brane that characterizes a finite-mass heavy quark as follows:
\begin{align}
    \tilde{S}_{\rm tot}^M({\bf C}) &= \sqrt{1+{\bf C}^2} \frac{\pi}{32}  \frac{(w_+^4 - w_-^4 )^2 }{w_-^5 (1 - w_-^4) }  F_1 \! \left( \frac32 , \frac54 , 1 ; 3 ; -\frac{w_+^4 - w_-^4}{w_-^4} , \frac{w_+^4 - w_-^4}{1 - w_-^4} \right) + \frac{\pi}{2} |v C_3| \theta( - v C_3 )  
    \label{eq:Stot-finiteMass}
\end{align}
where 
\begin{equation}
    w_{\pm}^4 = {\rm max} \left\{ z_{\pm}^4 , z_b^4 \right\} \ , 
    \label{eq:turningPoints-finiteMass}
\end{equation}
{with $z_\pm$ as defined in Eq.~\eqref{eqU4}.
(With $M=\infty$ and hence $z_b=0$, Eq.~\eqref{eq:Stot-finiteMass} reduces
to Eq.~\eqref{eqU3}.)}
In order to write Eq.~\eqref{eq:Stot-finiteMass}, one needs to assume that the momentum transfer ${\bf k}$ of interest is small enough that it may be neglected in the calculation. That is to say, in writing this expression we have neglected corrections of type (b), although we have kept the mixed higher order corrections {(i.e., terms that are simultaneously suppressed by both $T/M$ and ${\bf k}$)} that come with the calculation of $\tilde{S}_{\rm tot}^M({\bf C})$ --- which is to say that in re-summing all corrections controlled by $T/M$ we have also introduced additional ${\bf k}$ dependence as compared to the $M = \infty$ result.

In the limit of large heavy quark mass $M$, we can set  $z_b = \frac{\sqrt{\lambda} T }{2M}$, as corrections to this relation start at $\mathcal{O}(z_b^7)$~\cite{Mateos:2007vn}. The motivation for introducing Eq.~\eqref{eq:Stot-finiteMass} becomes clear when we write (in the $\lambda \gg 1$ limit, omitting subleading terms in $1/\sqrt{\lambda}$)
\begin{align}
    P({\bf k};{\bf v}) &= e^{- \sqrt{\lambda} T t \left[ \tilde{S}_{\rm tot}({\bf C},{\bf v}) + \frac{T}{M} \tilde{F}_1({\bf k},{\bf v},T,M,t) + \frac{k^2 - (k \cdot v)^2}{M T} \tilde{F}_2({\bf k},{\bf v},T,M,t) \right]  } \nonumber \\
    &=  e^{- \sqrt{\lambda} T t \left[ \tilde{S}_{\rm tot}^M({\bf C},{\bf v};z_b) + \frac{ {\bf k}_\perp^2 }{M T} \tilde{G}_\perp({\bf k},{\bf v},T,M,t) + \frac{ k_L^2 }{M T} \tilde{G}_L({\bf k},{\bf v},T,M,t) \right]  } \, . \label{eq:P-finiteM-schematic}
\end{align}
Here, the first line on the right-hand side is simply the HQET expansion (featuring so far unknown corrections $\tilde{F}_{1,2}$ to the infinite mass limit), and the second line on the right-hand side is the rearrangement where we separate the piece $\tilde{S}_{\rm tot}^M({\bf C})$  calculable via holography from the rest. In each line, there is an arbitrary splitting between the functions $\tilde{F}_{1}$ and $\tilde{F}_{2}$ (respectively $\tilde{G}_{\perp}$ and $\tilde{G}_{L}$) due to the fact that mixed higher order corrections in $\tfrac{T}{M}$ and $\tfrac{k^2 - (k\cdot v)^2}{MT}$ (respectively $\tfrac{{\bf k}_\perp^2}{MT}$ and $\tfrac{k_L^2}{MT}$) could be included in either function. Note that a feature of these expressions is that $\tilde{F}_{1}$, $\tilde{F}_{2}$, $\tilde{G}_{\perp}$ and $\tilde{G}_{L}$ depend on all of the dimensionless combinations of their arguments, whereas $\tilde{S}_{\rm tot}$ and  $\tilde{S}_{\rm tot}^M$ only depend on the specific combinations ${\bf C}$ and $z_b$ (the latter only for $\tilde{S}_{\rm tot}^M$). The key reorganization that takes place when going from the first to the second line is that $\tilde{S}_{\rm tot}^M$ absorbs all terms from $\tilde{F}_1$ that are not suppressed by $\tfrac{{\bf k}_\perp^2}{MT}$ or $\tfrac{ k_L^2}{MT}$, plus some contributions coming from both $\tilde{F}_1$ and $\tilde{F}_2$ that only depend on ${\bf k}$ via the ratio ${\bf k}/t$ (if this dependence can be written as a function of ${\bf C}$ alone). 
What the second line of Eq.~\eqref{eq:P-finiteM-schematic} illustrates is that in order to obtain a reliable result
for $P({\bf k};{\bf v})$ in some regime of the momentum transfer ${\bf k}$ from the holographic calculation that 
we are able to perform explicitly,
it is sufficient to consider the resummation of those $T/M$ corrections of type (a) that are encoded in $\tilde{S}_{\rm tot}^M$ as long as ${\bf k}$ (both $k_\perp$ and $k_L$)  is small enough. However, after these corrections have been fully accounted for, one needs to check what is the value of ${\bf k}$ up to which they provide a consistent picture for the physics of the momentum transfer.

\begin{figure}
    \centering
    \includegraphics[width=\linewidth]{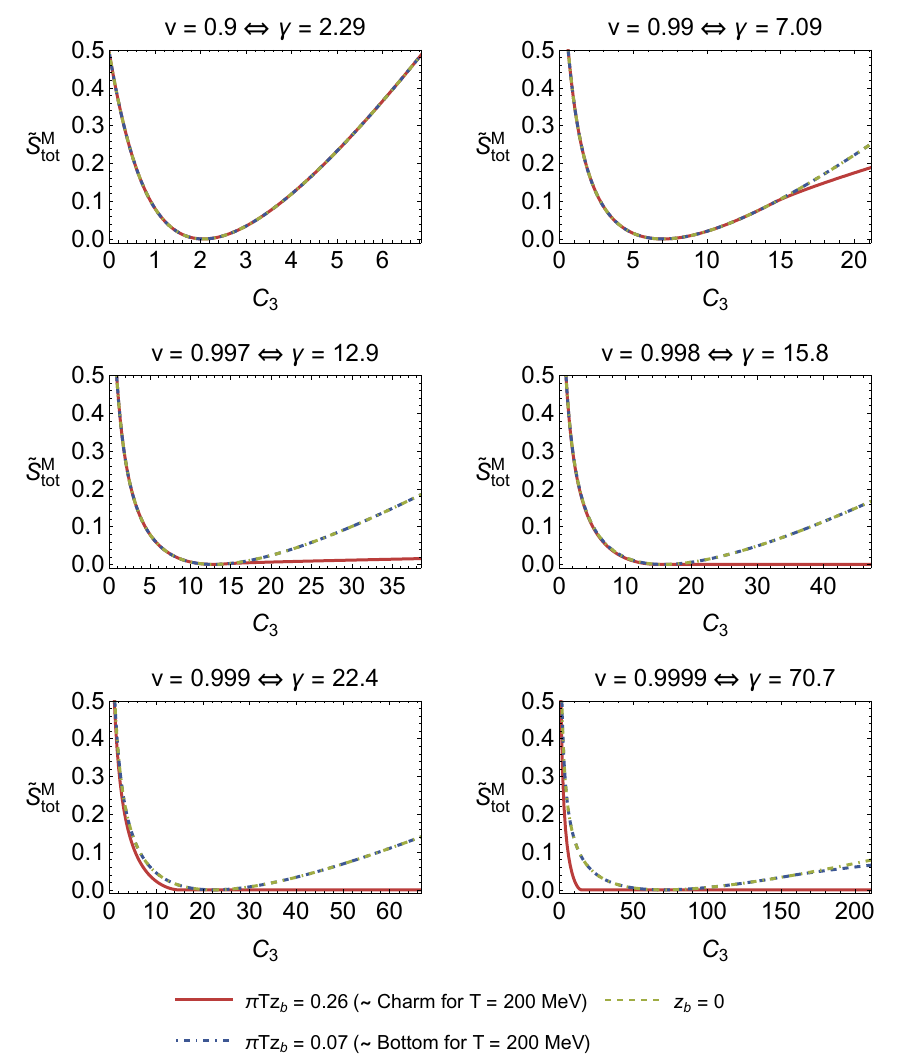}
    \caption{The logarithm of the momentum transfer function \eqref{eq:Stot-finiteMass} for finite mass heavy quarks ($z_b = \frac{\sqrt{\lambda} T }{2M}$) and in the infinite mass limit ($z_b =0$), as a function of the longitudinal momentum transfer rate $C_3$ with $C_\perp = 0$. Choosing values of the 't Hooft coupling $\lambda = 11$ and temperature $T=200$ GeV that would be reasonable for a QCD plasma, curves are plotted for the case of a charm-like ($M = 1.50$ GeV) and bottom-like  ($M = 4.75$ GeV) heavy quark, corresponding to values $z_b = 0.26$ and $z_b = 0.07$, respectively. From left to right and top to bottom the velocity increases;  the $C_3$ axis is rescaled in each plot so that the most likely momentum transfer in the $M \to \infty$ limit stays in the same relative position.
    }
    \label{fig:stot-as-function-of-mass-and-velocity}
\end{figure}

To gain more quantitative insights into correction (a), in Fig.~\ref{fig:stot-as-function-of-mass-and-velocity} we compare
the finite mass expression $\tilde{S}_{\rm tot}^M({\bf C},{\bf v}) $ to its infinite mass limit $\tilde{S}_{\rm tot}({\bf C},{\bf v})$ as a function of longitudinal momentum transfer $C_3$ and for different masses. The size of $T/M$ corrections can be read off as the difference from the infinite mass limit $z_b=0$. Deviations are clearly more pronounced for the lighter charm-like mass for bottom-like quarks, only the last panel displays a mild deviation at large $C_3$) and they increase with increasing velocity. For $\sqrt{\gamma} < \frac{2M}{\sqrt{\lambda} T}$, $\tilde{S}_{\rm tot}^M$ is identical to its infinite mass limit up to the value of $C_3$ at which $z_- = z_b$.  
For the case $\sqrt{\gamma} > \frac{2M}{\sqrt{\lambda} T}$ (that is realized for charm-like quarks in the last two plots of Fig.~\ref{fig:stot-as-function-of-mass-and-velocity}), the infinite mass limit and the finite mass expression disagree at all values of $C_3$ in the interval $0 < C_3 < v \gamma$ ($v\gamma$ is the mean defined by the $M \to \infty$ limit, $\tilde{S}_{\rm tot}$). 
Furthermore, for these large values of the heavy quark velocity there exists a value of $C_3$, smaller than the most likely value in the infinite mass limit, at which $\tilde{S}_{\rm tot}^M$ becomes $0$ for all larger values of the momentum transfer (to be precise, it is the value of $C_3$ at which $z_+ = z_b$). Therefore, while the aforementioned disagreement between $\tilde{S}_{\rm tot}$ and $\tilde{S}_{\rm tot}^M$ is comparatively mild at small $C_3$, it becomes stark as $C_3$ approaches this critical value from below. This also means that 
for $\sqrt{\gamma}>\frac{2M}{\sqrt{\lambda}T}$ the momentum transfer probability distribution
$P({\bf k};{\bf v})$ defined by $\tilde{S}_{\rm tot}^M$ is not normalizable.
This is yet another manifestation of the aforementioned speed limit for heavy quarks, as it confirms that we are unable to calculate $\langle C_3 \rangle$ for heavy quarks whose speed exceeds this limit, meaning that we are unable to calculate the mean momentum transfer experienced by an individual heavy quark whose speed exceeds
this limit.

For charm-quark-like parameters in Fig.~\ref{fig:stot-as-function-of-mass-and-velocity}, the velocity at which the momentum transfer probability distribution ceases to be normalizable (in the absence of corrections that we have not calculated) is around $\gamma \sim 15$. For bottom-quark-like parameters, it is at $\gamma > 100$. 
However, we emphasize that the fact that the $P({\bf k};{\bf v})$ defined solely by $\tilde{S}_{\rm tot}^M$ is not normalizable does not mean that the calculation of $\tilde{S}_{\rm tot}^M$ itself is not reliable at small momentum transfer --- it merely means that by the point $C_3$ reaches the value after which the distribution becomes flat, the corrections controlled by $\tilde{G}_T$ and $\tilde{G}_L$ in Eq.~\eqref{eq:P-finiteM-schematic} have taken over the result. Crucially, these corrections are small at small enough momentum transfer --- and hence $P({\bf k};{\bf v})$ is indeed well described by $\tilde{S}_{\rm tot}^M$ at small enough momentum transfer even for heavy quarks with $\sqrt{\gamma}>\frac{2M}{\sqrt{\lambda T}}$, above the speed limit.

In our calculations elsewhere in this work, we use $\tilde{S}_{\rm tot}$ in the infinite mass limit throughout. Therefore, starting from the corrections of type (a) to $\tilde{S}_{\rm tot}({\bf C},{\bf v})$, the observations in the preceding paragraphs illustrate that the velocity $v$ up to which our calculation is reliable depends on the value of the momentum transfer rate ${\bf C}$ at which $\tilde{S}_{\rm tot}({\bf C},{\bf v})$ is evaluated --- or, equivalently, that the momentum transfer rate up to which (starting from ${\bf C} = 0$) the calculation of $\tilde{S}_{\rm tot}({\bf C},{\bf v})$ is reliable depends on the velocity $v$ with which the heavy quark is propagating.\footnote{Following the logic of the HQET expansion, the regime of validity of $\tilde{S}_{\rm tot}$ should be assessed starting from ${\bf k} = 0$ and stopping at the value of the momentum transfer at which $\tilde{S}_{\rm tot}$ and $\tilde{S}_{\rm tot}^M$ begin to differ appreciably. Observing that $\tilde{S}_{\rm tot}$ and $\tilde{S}_{\rm tot}^M$ agree for a value of $C_3$ that is not in the continuously connected regime of validity around ${\bf C} = 0$ --- such as the minimum of the dashed lines in the lower right panel of Fig.~\ref{fig:stot-as-function-of-mass-and-velocity} --- should not be construed to mean that the calculation is trustworthy at that value of $C_3$.}

On the other hand, by following the logic of Ref.~\cite{BitaghsirFadafan:2008adl} we see that corrections of type (b) can be quantified directly by identifying the momentum change rate in the formula for vacuum Bremsstrahlung radiation~\cite{Mikhailov:2003er} with the momentum transfer rate as $dp/dt = - \pi \sqrt{\lambda} T^2 C_3/2$. This is a rather short calculation by comparison with what we just did for corrections of type (a). We give the result of the corresponding estimate for the domain of validity of the calculation of $\tilde{S}_{\rm tot}$ below.

There are thus two criteria to assess whether our calculation of $P({\bf k};{\bf v})$ for a given momentum transfer rate ${\bf k}$ is within the domain of validity of the calculation, 
depending on whether we start from the corrections (a) or (b): 
\begin{enumerate}[label=(\alph*)]
    \item For any given velocity, an upper limit to the regime of applicability of the calculation of $\tilde{S}_{\rm tot}({\bf C},{\bf v})$ (by comparing it to $\tilde{S}_{\rm tot}^M({\bf C},{\bf v})$ given by Eq.~\eqref{eq:Stot-finiteMass}) is given by the value of the momentum transfer at which either $z_-$ or $z_+$ is equal to $z_b$. From Eq.~\eqref{eqU4}, we see that for $C_\perp = 0$ this limit to the regime of applicability of our calculation of
    $P({\bf k};{\bf v})$ is given by
    \begin{equation}\label{eq:C3-criterion}
        \left( \frac{1}{1+C_3^2} \right)^{1/4} > z_b \iff |C_3| < \sqrt{\frac{16 M^4}{\lambda^2 T^4} - 1} \, ,
    \end{equation}
    independent of ${\bf v}$. This is a nontrivial condition on the momentum transfer that appears \textit{after} one completes the calculation of the $T/M$ corrections neglecting the effects of the ``bending'' of the HQ trajectory due to ${\bf k} \neq 0$.
    This condition should be contrasted with the $C_3$-independent criterion $\sqrt{\gamma} \ll M/(\sqrt{\lambda} T)$
    for the validity of the calculation of moments of the distribution, given by equation \eqref{eq5.16}. If $\gamma$ exceeds the speed limit set by Eq.~\eqref{eq5.16}, our calculation of $P({\bf k};{\bf v})$ remains reliable at low momentum transfer $k$, which is to say as long as $|C_3|$ satisfies Eq.~\eqref{eq:C3-criterion}. As long as $k$ is low enough that Eq.~\eqref{eq:C3-criterion} is satisfied, we can in fact trust our calculation of $P({\bf k};{\bf v})$ up to arbitrarily large values of $\gamma$ --- as long as corrections of type (b) do not change this conclusion.
    
    \item The size of the corrections of type (b) due to the Bremsstrahlung radiation resulting from the change ${\bf k}$ in the momentum of the heavy quark can be estimated by using the vacuum result for Bremsstrahlung radiation~\cite{Mikhailov:2003er}, and demanding that the resulting energy loss be smaller than the energy loss already implied by the medium-induced energy loss. This leads to
    \begin{equation}
        v \left| \frac{dp}{dt} \right| > \frac{\sqrt{\lambda}}{2\pi} \frac{1}{M^2} \left( \frac{dp}{dt} \right)^2 \implies |C_3| < v \frac{4M^2}{\lambda T^2} \, .\label{eq:Bremsstrahlung}
    \end{equation}
    In contrast to (a), even though the ratio $M/T$ appears explicitly in both cases, (b) is a condition derived without doing any calculation of how the finiteness of $M/T$ affects the dynamics of a heavy quark --- the temperature only enters Eq.~\eqref{eq:Bremsstrahlung} through the rewriting of the momentum loss rate $dp/dt$ in terms of $C_3$.
\end{enumerate}
Because in most practical situations we have $M > \sqrt{\lambda} T$, the distinction between these two criteria boils down to the factor of $v$ in the second one, meaning that
for large $\gamma$ the two criteria are equivalent in their consequences. This confirms that our calculation of $P({\bf k};{\bf v})$ is reliable at an arbitrarily large but finite $\gamma$ 
in the momentum transfer regime 
\begin{equation}\label{eq:C-limit}
\sqrt{|C_3|}=\sqrt{\frac{2|k_L|}{\pi M \Delta\tau}}<\frac{2 M}{\sqrt{\lambda} T}\ .
\end{equation}
Although it is the speed limit 
$\sqrt{\gamma}<\frac{2M}{\sqrt{\lambda}T}$ that 
determines where we can trust our calculation of 
$\langle C_3 \rangle$ and the mean momentum
transfer experienced by an individual heavy quark, we shall confirm in the next subsection that it is the momentum transfer limit \eqref{eq:C-limit} that determines where we can trust our calculation of the heavy quark momentum distribution $\mathscr{P}$ via the Kolmogorov equation.

We close this subsection with two comments.
First, although the analysis above reflects our ignorance about the size of the corrections, it is quantitatively conservative in the following ways:
\begin{itemize}
    \item Because $\tilde{S}_{\rm tot}^{M} \leq \tilde{S}_{\rm tot}$, and because $\tilde{S}_{\rm tot}^{M} = 0$ when both $\gamma$ and $C_3$ are large, the corrections to $\tilde{S}_{\rm tot}^{M}$ in $P({\bf k};{\bf v})$, that is to say, the terms proportional to $\tilde{G}_\perp$ and $\tilde{G}_L$, will tend to go in the direction of $\tilde{S}_{\rm tot}$ because $P({\bf k};{\bf v})$ %as defined in Eq.~\eqref{eq:P-formal} 
    is guaranteed to be a normalized probability distribution after all effects are taken into account because of quantum mechanical probability conservation (i.e., unitarity).
    \item The vacuum Bremsstrahlung radiation is not separable from the drag force in a thermal medium: energy-momentum conservation dictates that the energy lost by the heavy quark is the same as the sum of the energy that is radiated and deposited in the medium. Calculating them separately and adding them up will give an over-estimate, as the corresponding calculation for circular motion of a heavy quark shows~\cite{BitaghsirFadafan:2008adl}. 
\end{itemize}
And, second, we note that at large $M$ the bounds on $|C_3|$
that we have derived scale with a power of $M$ that would have been hard to guess only from formal HQET considerations, without doing a calculation.

\subsubsection{Validity of the Distribution $\mathscr{P}$ Obtained from the Kolmogorov Equation}

The validity of our calculation of the dynamics of the heavy quark distribution $\mathscr{P}$ should be assessed by asking whether the momentum transfer rate $C_3$ at which we evaluate $\tilde{S}_{\rm tot}$ in the course of solving the Kolmogorov equation 
to determine the evolution of $\mathscr{P}$ is within the bounds we just found or not. And, this momentum transfer rate is of order $\bar C_3$, where $\bar C_3\frac{\pi M}{2}\Delta \tau$ is the most likely momentum transfer that heavy quarks with some momentum ${\bf u}$ experienced in the previous $\Delta \tau$. (Recall that $\bar C_3$ depends on $u$ because it depends on the slope of the distribution $\mathscr{P}$ 
through $\partial f/\partial u$.)
Note that because the exponential time dependence $P({\bf k};{\bf v}) \propto \exp(- \sqrt{\lambda} T t \tilde{S}_{\rm tot}({\bf C},{\bf v}) )$ holds in a given range of ${\bf C}$ around zero momentum transfer, determining $\bar{\bf C}$ from a given (log of the) distribution function $f$ is \textit{not} sensitive to the overall normalization of the distribution, but rather only to $\partial \tilde{S}_{\rm tot}({\bf C},{\bf v})/\partial {\bf C}$ --- provided that $\bar{\bf C}$ is within the aforementioned range. In other words, the determination of $\bar{\bf C}$ only depends on the appropriate convolution between the momentum change probability $P({\bf k};{\bf v})$ and the momentum space distribution $\mathscr{P}$ having a maximum, and on that maximum being located at a low enough value of $|{\bf k}|$. Crucially, this can be assessed locally in momentum space, by comparing the derivative of $\tilde{S}_{\rm tot}$ with the derivative of $f$ starting from $C_3 = 0$ and smoothly increasing the magnitude of $C_3$ until the maximum is found, without needing to know 
$P({\bf k};{\bf v})$ for larger momentum transfer and hence without needing to know the mean momentum transfer. The only assumption in this argument is that such a maximum --- the saddle point in the derivation in Eq.~\eqref{eqU10} --- is unique (which is a mild assumption for falling momentum space distributions $\mathscr{P}$).

In summary, while the moments of the leading-order momentum transfer probability distribution $P({\bf k};{\bf v})$ (as determined by $\tilde{S}_{\rm tot}$) become unreliable for heavy quarks that exceed the speed limit $\sqrt{\gamma} > \tfrac{2M}{\sqrt{\lambda} T}$, the reliability of 
our calculation of the heavy quark momentum space distribution $\mathscr{P}({\bf u})$ via the Kolmogorov equation depends only on our knowledge of the shape of $P$ at low enough ${\bf k}$ and not on its moments --- where what ${\bf k}$ is low enough depends on the logarithmic derivative of $\mathscr{P}({\bf u})$ at ${\bf u}$.
Based on the arguments above, we can set a bound on the validity of our calculation of the evolution of the heavy quark distribution using the Kolmogorov equation~\eqref{eqU12} based on the validity of the expression for $\tilde{S}_{\rm tot}({\bf C}, {\bf v})$ we use in its derivation Eq.~\eqref{eqU10} --- which is that of the $M\to \infty$ limit --- in terms of a ratio of the dynamically selected momentum transfer rate $\bar{C}_3$ (for heavy quarks with mass $M$) to the mean momentum transfer rate $\langle C_3 \rangle$ of the $M\rightarrow \infty$ probability distribution defined by $\tilde{S}_{\rm tot}({\bf C}, {\bf v})$ by recasting Eq.~\eqref{eq:C-limit} as
\begin{equation}
    \left| \frac{\bar{C}_3 }{ \langle C_3 \rangle } \right| =  \left| \frac{\bar{C}_3 }{v \gamma } \right| < \frac{1}{\gamma} \frac{4M^2}{\lambda T^2} \, . \label{eq:limit-C3}
\end{equation}
Although this addresses a different question than that answered by the previous speed limit, we can nevertheless note that if we interpret Eq.~\eqref{eq:limit-C3} as a limit on $\gamma$, namely 
\begin{equation}
\gamma< \left| \frac{\langle C_3 \rangle}{\bar{C}_3}\right| \frac{4M^2}{\lambda T^2}  \ ,
\end{equation}
it serves to increase the speed limit by a factor of $|\langle C_3 \rangle/\bar C_3|$, as we stated in Eq.~\eqref{eq:new-speed-limit}.

\subsubsection{Momentum Transfer in the Kolmogorov Equation for ${\cal N}=4$ SYM Theory}

The most likely rate of momentum transfer $\bar{C}_3$ that a heavy quark underwent before reaching its present state in the $\mathcal{N}=4$ Kolmogorov equation has a precise mathematical characterization. It is the value of $C_3$ that solves Eq.~\eqref{eqHJ16}, which we repeat here in terms of $f'$:
\begin{equation}
    f'(u) = \frac{2}{\pi} \left. \frac{\partial \tilde{S}_{\rm tot} }{\partial C_3} \right|_{C_3 = \bar{C}_3(x=f'(u),u) } \, . \label{eq:sec4-fp-Stot-map}
\end{equation}
The value of $C_3$ determined in this way, which manifestly depends on the state of the heavy quark momentum space distribution, has to be examined in multiple ways. First, regarding the assumptions made in order to derive the Kolmogorov equation, and second, regarding the limiting condition~\eqref{eq:limit-C3} on the momentum transfer stemming from the calculation of $P({\bf k};{\bf v})$.

\begin{figure}   
    \includegraphics[width=0.98\linewidth,center]{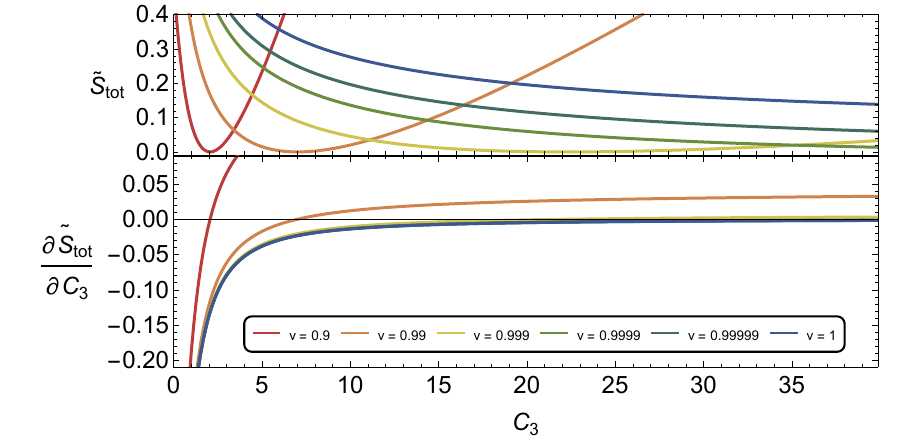}
    \caption{Top panel: $\tilde{S}_{\rm tot}$ for different values of the heavy quark velocity $v$. The location of the minimum of this curve, namely $\langle C_3\rangle$ which encodes the most likely momentum transfer that a heavy quark with velocity $v$ will undergo at the next time step, grows linearly with $\gamma$. Bottom panel: first derivative of the top panel with respect to $C_3$. This quantity enters Eq.~\eqref{eq:sec4-fp-Stot-map} to determine the momentum transfer $\bar{C}_3$ that specifies the evolution of the heavy quark momentum distribution $\mathscr{P}$. The figure illustrates that the latter quantity is much less sensitive to $\gamma$ in the ultra-relativistic limit than is the former.}
    \label{fig:Stot-as-f-of-v}
\end{figure}

For the first, we have carried out a numerical examination of the approximations involved in deriving the Kolmogorov equation, whose details we present in Appendix~\ref{app:Kolmogorov-checks}. The conclusion from this analysis is that the approximations made in Eqs.~\eqref{eq5.8} and~\eqref{eq5.9} have a negligible impact in the large momentum regime $u \gg 1$, but induce a comparatively larger error in the small momentum regime $u < 1$. Since our main purpose is to understand the equilibration dynamics starting far from equilibrium, this does not affect our main conclusions --- all it does is to alter the details of the dynamics in the regime that is already close to thermal (kinetic) equilibrium, which is bound to lead to a thermal steady state.

Another way to understand this is to note that the selected value of $C_3$ at which the RHS of Eq.~\eqref{eq:sec4-fp-Stot-map} is evaluated, namely $\bar C_3$, is not very sensitive to the value of $\gamma$ once $\gamma$ is large. That is to say, past a certain value of $u$ (equivalently, $\gamma$), the selected momentum transfer will be mostly sensitive to the value of $f'$, rather than how relativistic the heavy quarks are. This is apparent from Fig.~\ref{fig:Stot-as-f-of-v}. This Figure illustrates that, even if the value of $C_3$ at the maximum of the probability distribution (minimum of $\tilde{S}_{\rm tot}$) is \textit{not} slowly varying with $\gamma$ --- it in fact grows linearly  --- the dynamics in the ultra-relativistic limit is largely independent of the boost factor $\gamma$ for falling distributions.

\begin{figure}
    \centering
    \includegraphics[width=0.49\linewidth]{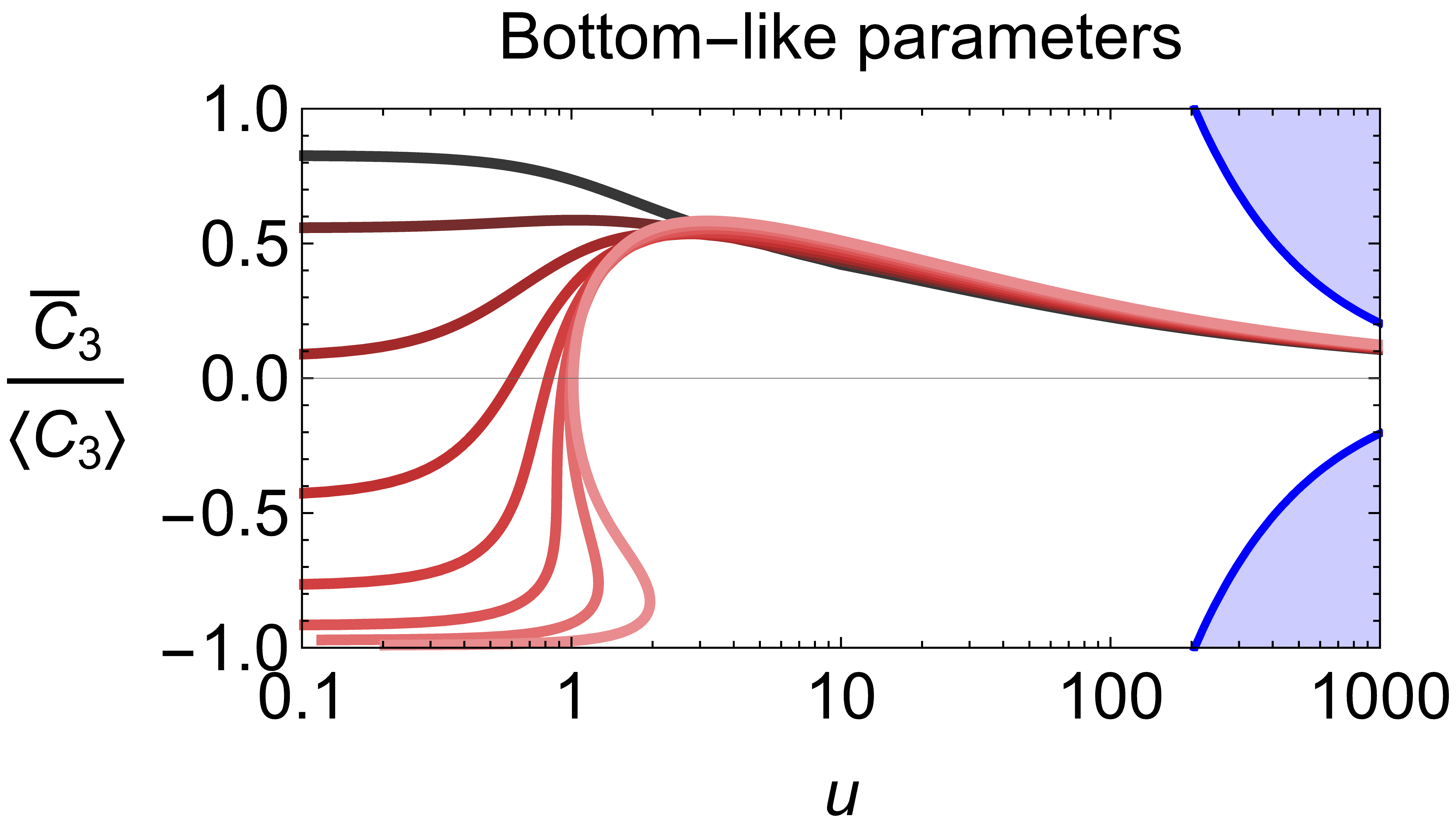}
    \includegraphics[width=0.49\linewidth]{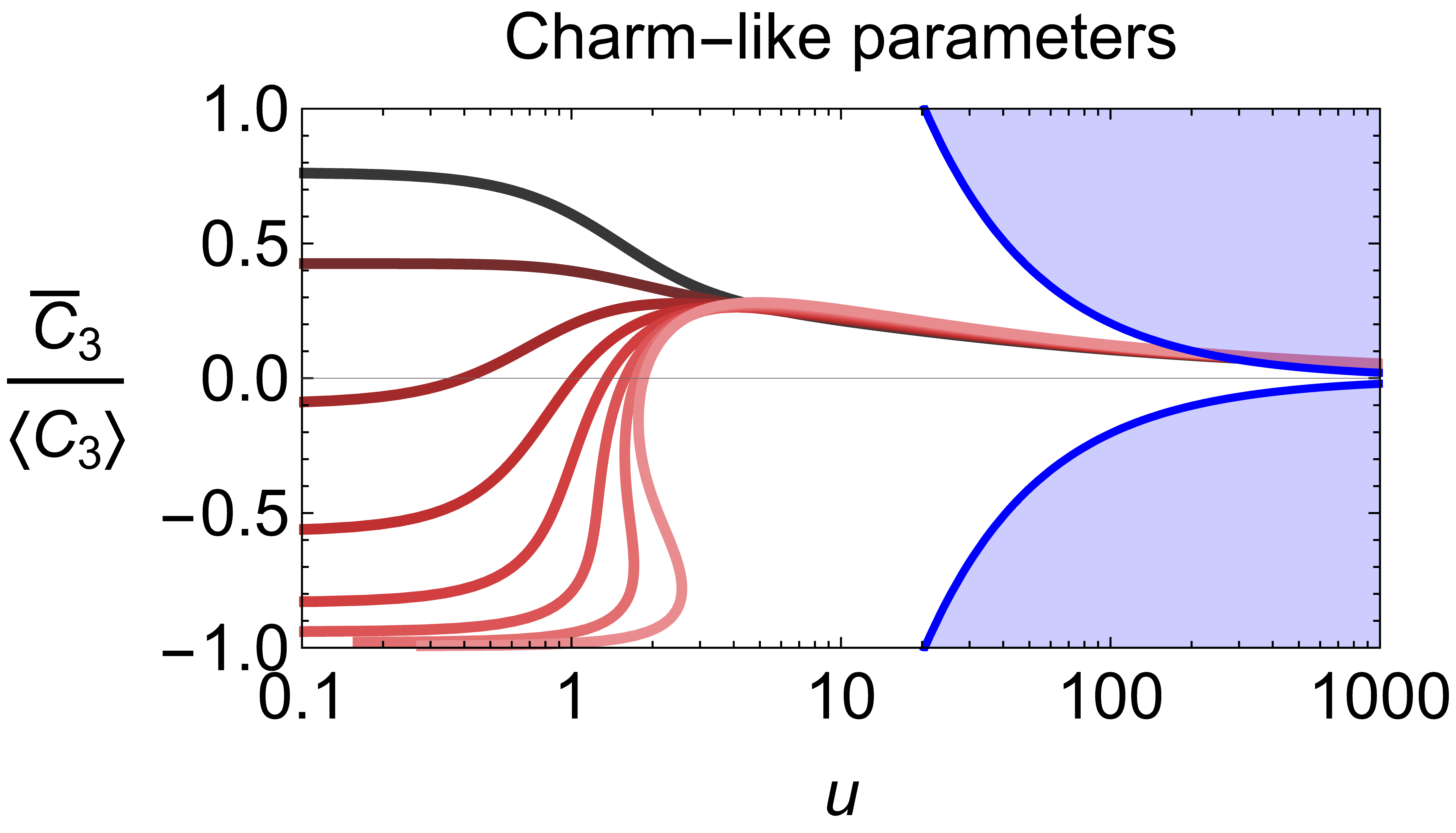}
    \caption{The ratio $\bar C_3/\langle C_3 \rangle$ as a function of $u$ at different times, using the same color code for time as in Figs.~\ref{fig:K-FP-solution-sample} and \ref{fig:solutions-P-example}. Constraints on the validity of the calculation of $P({\bf k};{\bf v})$ at the dynamically relevant momentum transfer rate $\bar{C}_3$ are shown as blue-shaded regions and discussed in the text. Left: bottom-like parameters $M_b/T = 4.75/0.2$. Right: charm-like parameters $M_c/T = 1.50/0.2$. We have taken $\lambda = 11$ in both panels. }
    \label{fig:C3-limit}
\end{figure}

The second way in which we need to examine the selected momentum transfer $\bar{C}_3$, as discussed before, pertains to the regime of validity of the calculation of $P({\bf k};{\bf v})$ itself, which we quantify via the limiting condition~\eqref{eq:limit-C3}. This examines whether we expect the closed, self-consistent equation we have, to be an accurate description of the physics for quarks of different masses. To quantify this, we plot in Fig.~\ref{fig:C3-limit} the left hand side of~\eqref{eq:limit-C3} as calculated from the solutions we found for bottom-like parameters in the left panel of Fig.~\ref{fig:solutions-P-example} and from the solutions we will find for charm-like parameters in the next Section
for the same time slices as in the plots we just referred to. In shaded blue we display the excluded region by virtue of the inequality in Eq.~\eqref{eq:limit-C3}.

We see that in both panels of Fig.~\ref{fig:C3-limit} (left for bottom-like parameters, right for charm-like parameters) the shape of the region not excluded by  the inequality in Eq.~\eqref{eq:limit-C3} is funnel-like, always allowing for some interval, even if narrow, of $\bar{C}_3$ where the calculation is valid. More generally, for a given tolerance, one should consider the calculation to be valid provided there is enough distance between the $\bar{C}_3/\langle C_3 \rangle$ curves and the shaded regions. Regardless of the chosen tolerance, it is clear that the allowed region is larger than what it would be if all of the heavy quarks lost momentum according to the mean of $P({\bf k};{\bf v})$. For bottom(-like) quarks, this means that, for all practical purposes, the calculation is far from entering the excluded region even up to momentum of $p/M \sim 100$. A similar statement can be made about charm(-like) quarks up to $p/M \sim 10$; however, the interval $10 < u < 100$ is also of interest.
Here, in $\mathcal{N} = 4$ SYM theory, one can expect that the trend slightly above $u = 10$ will still be well-described by our results, but eventually (and certainly around $u \sim 100$) the corrections due to effects that are not accounted for in our calculation will become important.
Recalling from Eq.~\eqref{eq:Cbar-equals-minus-meanC3} that $\bar C_3/
\langle C_3 \rangle =-1$ in equilibrium, we see that over the range of time depicted in Fig.~\ref{fig:C3-limit} the heavy quark momentum distribution has approached 
its equilibrium shape for $u\lesssim 2$, with our Kolmogorov equation analysis valid out to much larger values of $u$ than that.

\section{Heavy Quark Dynamics and Solutions} \label{sec:dynamics-solutions}
\label{sec5}

We are now in a position to discuss the solutions of the Kolmogorov equation~\eqref{eqU12} directly in terms of the heavy quark momentum distribution $\mathscr{P}(u,\tau)$, and discuss their qualitative and quantitative features.

\subsection{Initial conditions} \label{sec:init-conds}

In a collider experiment, single-inclusive heavy-quark distributions depend (at least) on two variables: transverse momentum $p_T$ and 
longitudinal momentum rapidity $y$.
In sufficiently high-energy collisions, momentum distributions are 
approximately boost invariant 
over a few units of rapidity around mid-rapidity,
and much of the interesting physics  --- ranging from heavy-quark
production in elementary collisions through hydrodynamic evolution of 
the droplets of quark-gluon plasma produced in heavy ion collisions --- are studied primarily via measuring transverse 
momentum distributions.
The initial condition for the distribution of heavy quarks in heavy-ion collisions
depends separately on $y$ and $p_T$ and
is set by the heavy quark production cross-section. 
It is determined by the convolution of perturbatively calculable cross-sections for parton-parton collisions dominated by large momentum transfers
and parton distribution functions of the nuclei (functions of the longitudinal momenta of the incident partons involved in the hard collision).
Such calculations were first performed in Refs.~\cite{Cacciari:1998it,Cacciari:2001td} at ``FONLL'' (Fixed Order with a Next to Leading large transverse momentum Logarithm resummation) accuracy.

Throughout this paper, the 
heavy quark momentum distribution $\mathscr{P}$
whose evolution we analyze 
is spherically symmetric in momentum 
space, meaning that it depends on $u=p/M$ with $p$ the magnitude of the heavy quark three-momentum. We leave the extension of the formalism for solving the Kolmogorov equation that we have developed to momentum distributions that depend separately on $p_T$ and $y$ to future work.
This means that we cannot literally use FONLL 
heavy-quark momentum distributions to set the initial conditions
for our calculations.
What we can do, however, is to choose
the initial heavy quark momentum distribution $\mathscr{P}_0(u)$ in our calculation such that $d\sigma/dp_T^2$ in any plane in momentum-space that includes the origin is 
similar to $d\sigma/dp_T^2$ at $y=0$ obtained from FONLL calculations~\cite{Cacciari:2012ny,Cacciari:2015fta}.  
We shall refer to such initial conditions as ``FONLL-motivated''.

Since the initial conditions for our spherically symmetric calculations can at best be FONLL-motivated, in this Section and throughout this paper we choose a simple parametrized form
\begin{equation}
    \mathscr{P}_0(u) = \frac{A}{\left[1+(u/s_0)^2\right]^3}  \label{eq:init-cond-form}
\end{equation}
for the initial heavy-quark momentum 
distribution that has the same 
power-law fall-off at large $u$ that characterizes 
the FONLL $d\sigma/dp_T^2$ at large $p_T$ and that has a similar shape. 
Since the overall normalization $A$ doesn't enter the equations that we solve, the only 
parameter in Eq.~\eqref{eq:init-cond-form} that is meaningful for our purposes is $s_0$. Since $u/s_0=p/(Ms_0)$, the parameter $s_0$ defines a momentum scale $Ms_0$.  
The simple form~\eqref{eq:init-cond-form} 
does a reasonable job of describing the 
$p_T$-distributions for single-inclusive  
bottom and charm quarks
produced in 
proton-proton collisions with $\sqrt{s}=5.03$~TeV~\cite{Cacciari:2012ny,Cacciari:2015fta}, obtained from FONLL calculations~\cite{Cacciari:2012ny,Cacciari:2015fta} with 
NNPDF3.0 Next-to-Leading-Order
parton distribution functions~\cite{NNPDF:2014otw}
(in turn obtained from
www.lpthe.jussieu.fr/$\sim$cacciari/fonll/fonllform.html)
if we use
\begin{align}
    s_0^b = 1.7 
    \label{eq:FONLL-m-b}
\end{align}
for bottom quarks with $M=4.75$~GeV~\cite{Cacciari:2012ny,Cacciari:2015fta}
and 
\begin{align}
    s_0^c = 2.6  \label{eq:FONLL-m-c}
\end{align}
for charm quarks with $M=1.5$~GeV~\cite{Cacciari:2012ny,Cacciari:2015fta}.
This means that $\mathscr{P}_0(u)$ has a transition
from a broad maximum at $u=0$ 
to a power-law fall-off at large 
$u$ that takes place at $u\sim s_0$, meaning 
at $p \sim ({\rm few}) M$.

Throughout this work, we study how 
the distribution $\mathscr{P}$ evolves in time
as the heavy quarks in the ensemble propagate through the strongly coupled plasma of 
${\cal N}$=4~SYM theory at a constant temperature $T$. In making plots, we shall take $T=0.2$~GeV, meaning that $M/T=4.75/0.2$ for bottom quarks and $1.5/0.2$ for charm quarks.
We defer the extension of our analysis to the evolution of heavy quarks in a droplet of strongly coupled plasma with finite extent that expands and cools hydrodynamically to future work. For this purpose, reformulating the Kolmogorov evolution that we have analyzed in this paper in terms of a non-Gaussian, Langevin-like, stochastic description of an individual heavy quark, using the momentum change probability distribution in Eq.~\eqref{eqU1} to evolve the heavy quark momentum in discrete time could be advantageous.
As we already noted 
in Section~\ref{sec:comparing-K-FP-conceptual},
we leave this as an avenue to be pursued in the future. 

Before turning to solutions to the Kolmogorov equation with initial conditions
of the form \eqref{eq:init-cond-form},
we note that 
later in this Section we will also 
tweak the form of Eq.~\eqref{eq:init-cond-form}
to make it fall off with
different power laws at large momentum,
so as to investigate the consequences of
varying the steepness of the initial 
distribution.

\subsection{Solutions}

We solve the Kolmogorov equation as described in Section~\ref{sec:K-as-HJ}. For our FONLL-motivated initial conditions with bottom-quark-like parameters, the solutions were already presented in that Section as part of the description of the solution method, in particular in Fig.~\ref{fig:solutions-P-example}. In what follows, we further explore the properties of these solutions, and discuss with the same level of detail the solutions obtained by using charm-quark-like parameters.

\subsubsection{Bottom-like parameters}

The evolution of the heavy quark momentum distribution $\mathscr{P}(u,\tau)$ obtained by
solving the Kolmogorov equation 
starting from our FONLL-motivated initial conditions was presented in Fig.~\ref{fig:solutions-P-example}, where the key features of the dynamics of the Kolmogorov equation are apparent and can be 
compared to 
the evolution obtained via the Fokker-Planck truncation of the Kolmogorov equation 
with the Einstein relation imposed artificially. 
Most importantly:
\begin{enumerate}
    \item Thermalization of the low-momentum regime is not very sensitive to whether the dynamics includes all of the non-Gaussian fluctuations (Kolmogorov) or is simply the truncated version with only the mean value matched to the microscopic dynamics (Fokker-Planck).
    \item Thermalization of the high-momentum regime is qualitatively different in the two cases: the full non-Gaussian dynamics takes much longer to fully equilibrate the distribution at large momentum, preserving memory of the initial condition for a much longer time. The clearest manifestation of this difference is that for the Fokker-Planck case the high-momentum tail in Fig.\ref{fig:solutions-P-example} gets uniformly shifted towards lower momentum (effectively losing energy in a way that is well-described by the mean of the momentum change distribution), whereas in the Kolmogorov dynamics fluctuations sustain the occupation of the higher momentum regime in such a way that a description only in terms of the mean becomes inadequate. In particular, the 
    non-Gaussianity of the fluctuations in the momentum change described by Eq.~\eqref{eqU1} mean that the probability that the momentum change during some time step is near zero is much larger
    in the Kolmogorov dynamics than
    it is in the Fokker-Planck case where the momentum change fluctuations are Gaussian. Consequently, the survivor bias effects that are naturally encoded in our framework as described at length in Section~\ref{sec:interplay} mean that at any $u$ where the heavy quark momentum distribution is falling steeply, the distribution evolves much more slowly than does the momentum of a typical heavy quark with that $u$ in the ensemble.
\end{enumerate}

\begin{figure}
    \centering
    \includegraphics[width=0.49\linewidth]{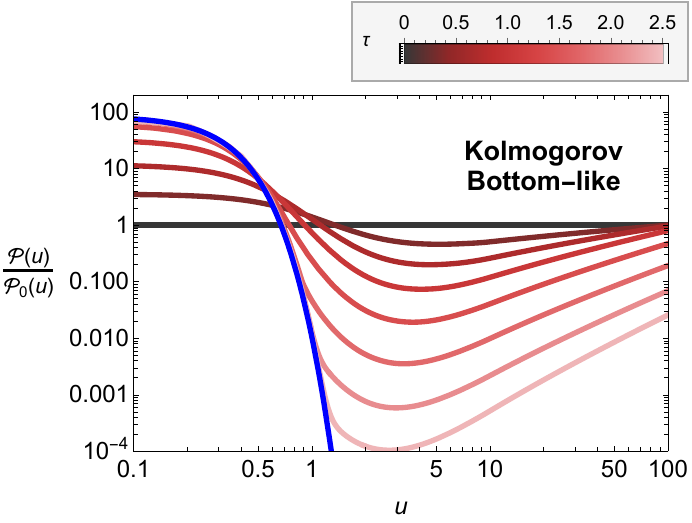}
    \includegraphics[width=0.49\linewidth]{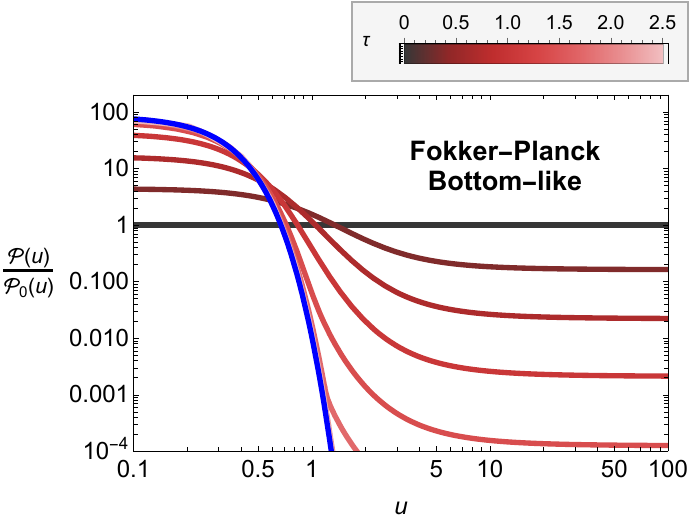}
    \caption{Ratio of the heavy-quark momentum distribution functions $\mathscr{P}(u,\tau)$ to their initial conditions $\mathscr{P}_0(u) = \mathscr{P}(u,\tau=0)$ resulting from time evolution starting from our FONLL-motivated initial conditions with bottom-quark-like parameters, including $M/T=4.75/0.2$, via the $\mathcal{N}=4$ SYM Kolmogorov equation (left) and its Fokker-Planck truncation (right). The late-time ratio, determined by the equilibrium (thermal) distribution, is displayed in blue. Snapshots of the time evolution of the distribution function are shown from darker to lighter tones of red (same color scheme as for the dashed lines in Figs.~\ref{fig:K-FP-solution-sample} and \ref{fig:solutions-P-example}, starting from the same FONLL-motivated initial condition at $\tau = 0$, and spaced in time by $\Delta \tau = 0.35$.)}
    \label{fig:ratios-P-example}
\end{figure}

In Fig.~\ref{fig:ratios-P-example} we present another way to examine these features, this time in terms of the ratio between the heavy quark momentum space distribution at a time $\tau$ and its initial condition at $\tau = 0$. This is a way to quantify how different the final state of the heavy quark distribution is as a consequence of the heavy quarks
in the ensemble having propagated through a thermal medium with temperature $T$ for a time $\tau$.

The blue curves (same in both panels of Fig.~\ref{fig:ratios-P-example}) highlight how different the heavy quark momentum distribution in thermal equilibrium (for our calculation in a uniform, motionless, constant-temperature, thermal medium) is from our FONLL-motivated initial distribution: in equilibrium, there are many more very soft heavy quarks, more by almost a factor of 100 than in our starting distribution, and at large momenta the equilibrium distribution is Boltzmann-suppressed. The evolution from our FONLL-motivated initial distribution to equilibrium in the low momentum $u\lesssim 1$ regime is quite similar in the full Kolmogorov evolution (left panel of Fig.~\ref{fig:ratios-P-example}) and the truncated Fokker-Planck evolution (right panel). 

As we have already seen in our discussion
of Fig.~\ref{fig:solutions-P-example}, the difference between how the Kolmogorov and Fokker-Planck evolution of the heavy quark momentum distribution is most apparent at high momentum, say $u\gtrsim (5-10)$.  In the right-panel of Fig.~\ref{fig:solutions-P-example}, Fokker-Planck evolution causes the high-momentum tail of the distribution to step uniformly leftward with each time step, governed by the typical momentum lost by individual heavy quarks.  In the right-panel of 
Fig.~\ref{fig:ratios-P-example}, this corresponds to the fact that with each time step the ratio plotted drops uniformly downward, heading towards the blue curve far below.  The distribution behaves differently at high momentum when evolved according to the Kolmogorov equation as in the left panels of Figs.~\ref{fig:solutions-P-example} and \ref{fig:ratios-P-example}. 
The difference at high momentum can be understood in terms of the size of the momentum fluctuations, which are much larger for the full Kolmogorov equation than for its Fokker-Planck truncation, since $\kappa_L$ grows as $\gamma^{5/2}$ in the former and as $\gamma$ in the latter. Larger fluctuations make events where a heavy quark loses significantly less energy than the mean of $P({\bf k};{\bf v})$ much less rare. In combination with  the survivor bias 
effect that we have described at length in Section~\ref{sec:interplay} (which means that the distribution at any given large value of $u$ is populated by those heavy quarks that lost the least energy in the previous time step), this allows the high momentum tail 
of the heavy quark distribution to remain occupied for 
much longer.  We see this manifest in the 
left panel of Fig.~\ref{fig:ratios-P-example} 
for $u\gtrsim (5-10)$, where the ratio plotted drops more slowly at higher $u$.  Indeed, at the highest values of $u$ plotted, the ratio in fact lingers close to unity for the first few time steps before it begins to drop.

It is tempting to think of the curves plotted in the left panel of Fig.~\ref{fig:ratios-P-example}, describing the ratio between the solution to the Kolmogorov equation at a given time and its initial condition, as the analogue in our calculation to the nuclear modification 
factor $R_{\rm AA}$ --- namely the ratio between the heavy-quark momentum distribution after evolution through the QGP produced in a heavy ion collision to that in a proton-proton collision in which no QGP is produced.
As we shall explain, this analogy may be instructive at high momentum, but it is misleading at lower momenta.  
The most important differences between the $R_{\rm AA}$ ratio in a heavy ion collision and the ratio we have plotted in Fig.~\ref{fig:ratios-P-example} arise because 
in a heavy ion collision the strongly coupled medium is not spherically symmetric and is expanding and flowing --- with boost invariant longitudinal expansion plus radial expansion in the transverse plane --- whereas in our calculation the strongly coupled medium is spherically symmetric, homogeneous, and at rest.  
In an expanding fluid, at late time heavy quarks come into local equilibrium in the moving fluid, meaning that they end up flowing along with the medium while diffusing in it, with lab-frame momenta that they inherit from the flow of the fluid. This 
means that if heavy quarks were to equilibrate in the (longitudinally and transversely) expanding droplet of strongly coupled QGP produced in a heavy ion collision, the ratio of their final momentum distribution to their initial FONLL momentum distribution would look nothing at all like the blue curve in Fig.~\ref{fig:solutions-P-example}.  The momentum carried by the heavy quarks 
riding along with the flowing medium would greatly reduce the blue curve at small $u$, and would push the tail of the blue curve outwards in $u$.  This would result in qualitative changes to  the shape of the red-brown curves in both panels of Fig.~\ref{fig:ratios-P-example} in the low and intermediate momentum regimes, where the Kolmogorov and Fokker-Planck evolutions are similar, that describe the approach to equilibrium.

At high momentum, the qualitative behavior of the time-evolution of the ratios plotted in both panels of 
Fig.~\ref{fig:ratios-P-example} does not depend 
on any detail of the shape of the blue curve; all that matters is that the blue curve is many orders of magnitude below unity. This is also the case for the analogue ratio in a heavy ion collision.  Focusing in particular on the Kolmogorov evolution at high momentum in the left panel of
Fig.~\ref{fig:ratios-P-example}, we furthermore note that the shape of the momentum transfer probability 
distribution $P({\bf k};{\bf v})$, with non-Gaussian moments enhanced at large $\gamma$, is strikingly similar when calculated perturbatively in QCD~\cite{DuPlessis:2026pyr} to its shape in strongly coupled ${\cal N}=4$ SYM theory~\cite{Rajagopal:2025ukd} that we have employed in our calculation. 
Given the importance of the non-Gaussian higher moments of $P({\bf k};{\bf v})$ in any quantum field theory~\cite{Rajagopal:2025rxr} and given
the similarity of the shape of $P({\bf k};{\bf v})$ in  the weakly coupled QCD plasma and
the strongly coupled ${\cal N}=4$~SYM~\cite{DuPlessis:2026pyr}, it is reasonable to expect it to be similar in strongly coupled QGP.
The non-Gaussian shape of $P({\bf k};{\bf v})$ together with the survivor bias phenomenon that we have described are the key ingredients that result in the qualitative behavior of the Kolmogorov evolution
of the heavy quark distribution in the high-momentum regime  in the left panel of
Fig.~\ref{fig:ratios-P-example}.
And, the survivor bias effect is also operative in a heavy ion collision setting at any momentum high enough that the heavy quark spectrum is steeply falling.  
The evolution of the ratios plotted in Fig.~\ref{fig:ratios-P-example} 
may thus yield qualitative insights 
into the evolution of the analogous ratio 
in a heavy ion collision --- at high momentum.
For many additional reasons, such insights can only be qualitative. For example, there is no analogue of either hadronization or jet showers in our calculation, and the $R_{\rm AA}$ that is measured in QCD is that for $B$ (or $D$) mesons or for $b$-jets, not for $b$-quarks.  Also,  $R_{\rm AA}$ describes ratios between distributions at freezeout in a heavy ion collision and distributions in a proton-proton collision, and the heavy quarks at the freezeout surface in any individual heavy ion collision have travelled for different distances/durations through QGP with different temperatures and fluid velocities, whereas each curve in the left panel of Fig.~\ref{fig:ratios-P-example} describes an ensemble of heavy quarks that have all traveled for the same distance/duration through a medium 
at rest with the same temperature.

In light of the discussion above, it is interesting to compare the results in the 
right panel of Fig.~\ref{fig:ratios-P-example}
to the earliest Fokker-Planck calculations
of $R_{\rm AA}$ for heavy quarks in an expanding
hydrodynamic fluid modeling the droplet of QGP formed
in a heavy ion collision from Refs.~\cite{Moore:2004tg,vanHees:2004gq}.  
In these calculations, at high $p_T$ $R_{\rm AA}$ is roughly flat as a function of $p_T$ and drops with time~\cite{Moore:2004tg} (time as measured in units of the diffusion constant), just as in the right 
panel of Fig.~\ref{fig:ratios-P-example}.  
(However, at low $p_T$ $R_{\rm AA}$ is $\sim 1$ in the calculations of Ref.~\cite{Moore:2004tg}; we have explained above why it is nowhere near the factor of $\sim 100$ seen at low $u$ in 
Fig.~\ref{fig:ratios-P-example}.)
This comparison suggests that the high-momentum region of the right panel of Fig.~\ref{fig:ratios-P-example} would yield qualitative insights 
into the evolution of the analogous ratio 
in a heavy ion collision --- if that evolution were governed by Fokker-Planck dynamics.

The comparison above emboldens us to 
note that the $D$-meson $R_{\rm AA}$ ratio measured by CMS~\cite{CMS:2017qjw} and ALICE~\cite{ALICE:2021rxa,ALICE:2021mgk} in heavy ion collisions at the LHC
rises with increasing $p_T$ at large $p_T$, 
qualitatively consistent with what we see at large $u$ in the left panel of Fig.~\ref{fig:ratios-P-example} --- where we have evolved the heavy quark momentum momentum distribution according to the Kolmogorov equation.  
As we have already discussed,
the $D$-meson $R_{\rm AA}$ in LHC heavy ion collisions behaves quite differently at low momentum
than what we find there.
Although it is near unity, more than an order of magnitude smaller than what we find at low momentum, being near unity there means that
the $R_{\rm AA}$ ratio in heavy ion collisions 
has a minimum at intermediate momenta --- 
as does the ratio that we have plotted
in the left panel of Fig.~\ref{fig:ratios-P-example}. 

Finally, it is striking that the experimental measurements of the $D$-meson $R_{\rm AA}$ ratio~\cite{CMS:2017qjw,ALICE:2021rxa,ALICE:2021mgk},
including in particular the way that this ratio rises with increasing $p_T$ at high $p_T$,
can also be described qualitatively (and even semi-quantitatively)~\cite{Beraudo:2025nvq} by assuming that
sufficiently relativistic charm quarks (which are found in jets) behave like light quarks in jets in strongly coupled plasma~\cite{Chesler:2014jva,Chesler:2015nqz}.
(These calculations describe an $R_{\rm AA}$ for both jets and light hadrons that increases with increasing $p_T$ at large $p_T$~\cite{Casalderrey-Solana:2014bpa,Casalderrey-Solana:2018wrw}.)
We leave further investigation of the similarity of the Kolmogorov evolution of the 
heavy quark momentum distribution at high-$p_T$ 
seen in the left panels of Figs.~\ref{fig:solutions-P-example} and \ref{fig:ratios-P-example} 
to the dynamics of light quarks in strongly coupled plasma to future work.

\subsubsection{Charm-like parameters}

\begin{figure}
    \centering
    \includegraphics[width=0.49\linewidth]{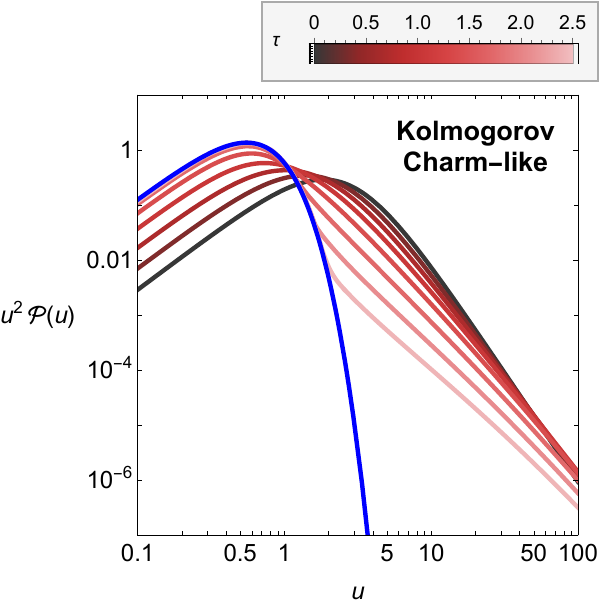}
    \includegraphics[width=0.49\linewidth]{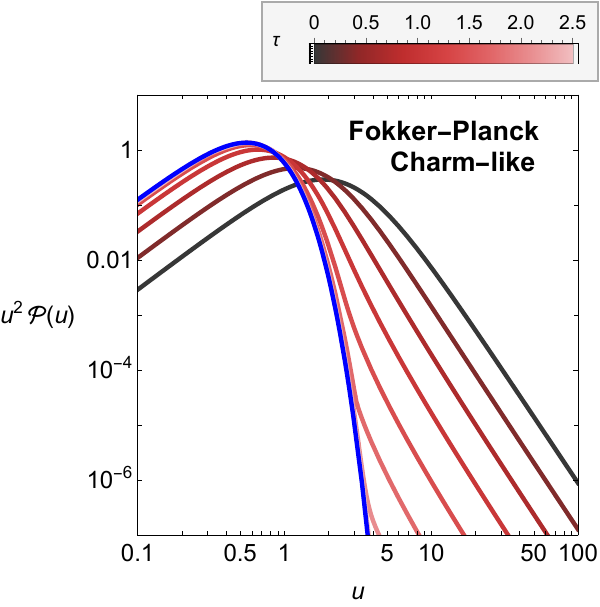}
    \caption{Physical momentum space distribution functions ${\mathscr P}(u)$ resulting from the time evolution 
    of our FONLL-motivated initial distribution (dark grey curve) using the Kolmogorov equation 
    for $\mathcal{N}=4$ SYM theory (left panel) and its Fokker-Planck truncation (right panel), using $M/T = 1.5/0.2$. The equilibrium configuration is displayed in blue. Snapshots of the time evolution of the distribution function are shown from darker to lighter tones of red (same color scheme as for the dashed lines in Fig.~\ref{fig:K-FP-solution-sample}), starting from the same initial condition at $\tau = 0$, and spaced in time by $\Delta \tau = 0.35$.
     These results for charm-like heavy quarks can be compared to the results for bottom-like heavy quarks in Fig.~\ref{fig:solutions-P-example}.}
    \label{fig:solutions-P-charm}
\end{figure}

\begin{figure}
    \centering
    \includegraphics[width=0.49\linewidth]{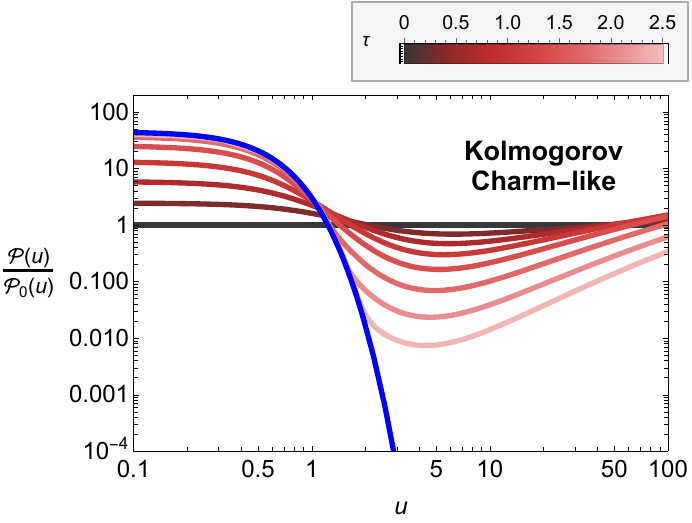}
    \includegraphics[width=0.49\linewidth]{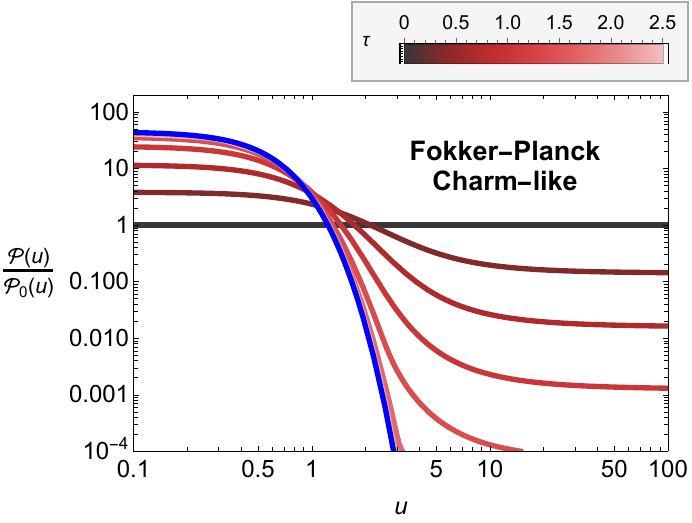}
    \caption{Same as Fig.~\ref{fig:ratios-P-example}, but for the charm-like case, using $M/T = 1.50/0.2$.}
    \label{fig:ratios-P-charm}
\end{figure}

We now explore what, if anything, changes
in our results when we use a ratio of heavy quark mass to temperature motivated by the charm quark in heavy ion collisions. A lower ratio between these quantities reduces the extent to which several of the approximations we made are under control, but still allows for these approximations to be useful organizing principles and to provide physically sensible dynamics.

Figs.~\ref{fig:solutions-P-charm} and~\ref{fig:ratios-P-charm} show the charm-like dynamics counterparts to the bottom-like dynamics in Figs.~\ref{fig:solutions-P-example} and~\ref{fig:ratios-P-example}, respectively. Compared to the  heavier mass case, charm-like dynamics exhibits an even sharper qualitative contrast between the Kolmogorov and Fokker-Planck solutions. Equilibration of the high momentum sector is much slower for the full Kolmogorov equation than for its Fokker-Planck truncation. There are also quantitative differences in the low momentum sector, where the distribution approaches the Boltzmann distribution at a later time when evolved according to the Kolmogorov equation. As for the bottom-like case, the blue curves
in Fig.~\ref{fig:ratios-P-charm} and the red-brown curves that approach them bear no resemblance to what one would see in a heavy ion collision setting.  Also as for the bottom-like case, at high momenta we see the heavy quark momentum distribution ratio rising with $p_T$  at large $p_T$ when evolved according to the Kolmogorov equation as in the left panel of Fig.~\ref{fig:ratios-P-charm}, whereas it is roughly flat with $p_T$ and dropping much more rapidly with time when the Fokker-Planck truncation is used in the right panel.

Apart from the relaxation/equilibration time scale, the Fokker-Planck dynamics show no qualitative difference between the two quark masses. In contrast, the two cases of Kolmogorov dynamics exhibit larger differences, most clearly seen by comparing the left panels of Figs.~\ref{fig:ratios-P-charm} and Fig.~\ref{fig:ratios-P-example}. The suppression at intermediate momentum is smaller for charm than for bottom quarks at the time when the low momentum part of the distribution is nearly fully thermalized: the minimum of the ratio in the bottom-like case is $\sim 0.01$ when the soft sector has almost thermalized, whereas the corresponding minimum in the charm-like case is $\sim 0.1$. Furthermore, the large non-Gaussian fluctuations in the momentum transfer between heavy quarks with large momenta (large $\gamma$) and the medium have a greater effect in the charm-like case, as the ratio between the heavy quark distribution and its initial condition even exceeds unity for momenta of $p \sim 50 M$ and above. 
At such large momenta, however, in the charm-like case the heavy quark momentum distribution is in a regime where our calculation of the momentum transfer probability is not under control and
may receive large corrections from effects that we have neglected in this work (see the right panel of Fig.~\ref{fig:C3-limit}).

\subsection{Sensitivity to the steepness of the initial condition}

\begin{figure}
    \centering
    \includegraphics[width=0.8\linewidth]{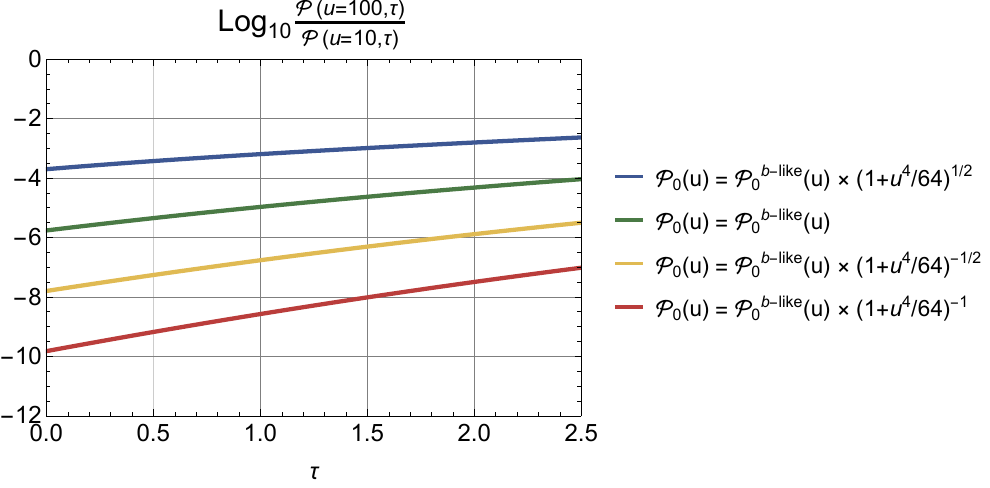}
    \caption{Plot of the ratio between the momentum distribution $\mathscr{P}(u,\tau)$ at $u = 100$ and at $u = 10$, for bottom-like parameters and four different initial conditions, as a function of time. The green curve describes Kolmogorov evolution at large $u$ starting from $\mathscr{P}_0^{\rm b-like}$, the FONLL-motivated initial condition in Eqs.~\eqref{eq:init-cond-form} and \eqref{eq:FONLL-m-b}. It starts near $-6$ at $\tau=0$, corresponding to the $u^{-6}$-dependence of the initial condition \eqref{eq:init-cond-form} and its time dependence shows how Kolmogorov evolution makes the distribution become less steep with time, see the left panel of Fig.~\ref{fig:solutions-P-example}. 
    The blue, yellow and red curves are analogous, but for Kolmogorov evolution starting from an initial distribution that has been multiplied by $(1+u^4/64)^a$, with $a=1/2$, $-1/2$, $-1$ respectively, as in Eq.~\eqref{eq:modified-init-cond-form}.  We see similar behavior in all cases: the high-momentum power-law characterizing $\mathscr{P}(u,\tau)$ at large $u$ becomes less steep with time, as $\mathscr{P}(u,\tau)$ is pushed downward more rapidly at $u=10$ than at $u=100$.}
    \label{fig:slope-b-like}
\end{figure}

\begin{figure}
    \centering
    \includegraphics[width=0.8\linewidth]{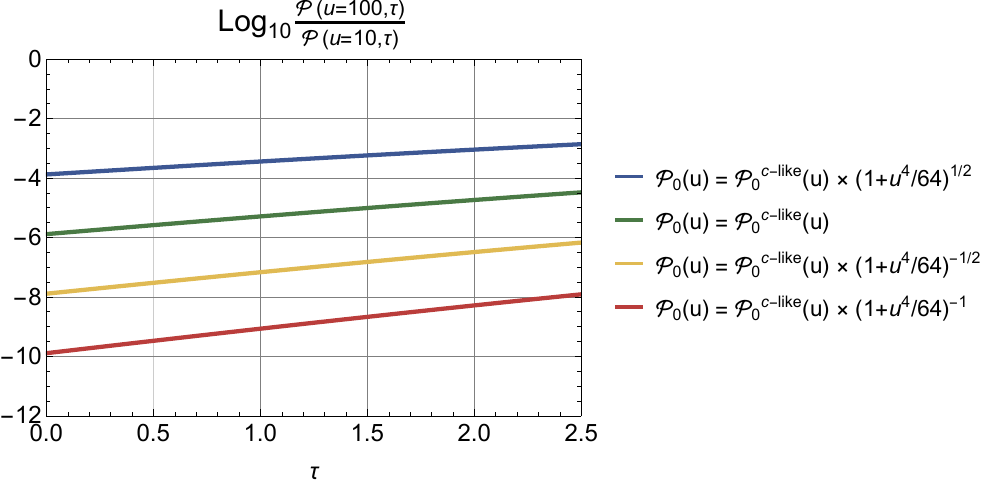}
    \caption{Same as Fig.~\ref{fig:slope-b-like}, but here the green curve describes Kolmogorov evolution at large $u$ starting from $\mathscr{P}_0^{\rm c-like}$, the FONLL-motivated initial condition in Eq.~\eqref{eq:init-cond-form} with charm-like parameters~\eqref{eq:FONLL-m-c}. The blue, yellow and red curves describe Kolmogorov evolution starting from initial distributions that have been modified in the same fashion as in Fig.~\ref{fig:slope-b-like}. We see similar behavior in all cases: the high-momentum power-law characterizing $\mathscr{P}(u,\tau)$ at large $u$ becomes less steep with time, as $\mathscr{P}(u,\tau)$ is pushed downward more rapidly at $u=10$ than at $u=100$.}
    \label{fig:slope-c-like}
\end{figure}

It is also worth exploring how 
the behavior of the momentum space distribution function $\mathscr{P}(u,\tau)$ at large $u$ that we have obtained by solving either the Kolmogorov equation or its Fokker-Planck truncation 
would differ if the steepness of the initial
distribution differs from that in our FONLL-motivated 
initial distribution.
To this end, we have checked how the time-evolution of  $\mathscr{P}(u,\tau)$ is altered if,
instead of taking the initial condition to be $\mathscr{P}_0(u)$ 
from
Eq.~\eqref{eq:init-cond-form}, we initialize it as
\begin{equation}\label{eq:modified-init-cond-form}
    \mathscr{P}(u,\tau = 0) = \mathscr{P}_0(u) \times \left(1 + \frac{u^4}{64}\right)^{a} \, ,
\end{equation}
with $a = 1/2$, $0$, $-1/2$, $-1$. The purpose of using this form is to change the steepness of the power law at high momentum without modifying the low momentum behavior of the initial condition.
In Figs.~\ref{fig:slope-b-like} and~\ref{fig:slope-c-like}, we focus on the time-evolution of the power-law that characterizes 
$\mathscr{P}(u,\tau)$ at large $u$, namely high momentum.  In all cases (bottom-like in Fig.~\ref{fig:slope-b-like}; charm-like in Fig.~\ref{fig:slope-c-like}; all four initial conditions with differing initial steepness), we see the same qualitative behavior.
The power law characterizing the high-momentum tail of the distribution $\mathscr{P}(u,\tau)$ becomes less steep (less negative) with time; $\mathscr{P}(u,\tau)$ is pushed downward more rapidly at $u=10$ than at $u=100$.  This demonstrates the robustness of our conclusion that when we evolve the heavy quark momentum distribution with the Kolmogorov equation, the ratio $\mathscr{P}(u,\tau)/\mathscr{P}(u,0)$ is generically an increasing function of $u$ at large momentum $u$.

\section{Conclusions and Outlook}
\label{sec:ConclusionsOutlook}

In this work we have studied the dynamics of a heavy quark momentum space distribution $\mathscr{P}(u,\tau)$ in a strongly coupled gauge theory plasma. To our knowledge, this is the first study of its kind. It stands apart from previous studies in that:
\begin{enumerate}
    \item it includes all of the non-Gaussian statistics of the momentum transfer generated by the quantum and thermal fluctuations of a strongly coupled quantum field theory, and
    \item the time evolution operator $K$ that drives this dynamics is directly determined from a calculation in said quantum field theory in terms of a Wilson loop.
\end{enumerate} 
That is to say, we have completed a first-principles study of heavy quark transport at strong coupling.

The dynamics we presented describes the same variables as a conventional transport description --- either Langevin or Boltzmann --- and is subject to the same Markovian assumptions. It describes the evolution of the momentum distribution of
heavy quarks in a continuous, coarse-grained-in-time, manner 
which is self-consistent provided the propagation time to attain a substantial momentum change is sufficiently long compared to $1/T$, the microscopic time-scale of the strongly coupled medium. 
Only in this regime a Markovian process emerges. 

Unlike Langevin or Boltzmann dynamics, the approach we have presented here is systematically improvable in powers of $1/M$
regardless of the coupling strength of the medium. 
While technically challenging, calculating $(1/M)$-corrections to our results is conceptually well-defined. As we discussed in Section~\ref{sec:general-validity}, the HQET Lagrangian formulation of the dynamics (which is a top-down effective field theory of QCD) precisely determines the field-theoretic expectation values that need to be calculated if precision at subleading $1/M$ powers is desired. Corrections to the derivation of the Kolmogorov equation as presented in Section~\ref{sec:novel-derivation}, which we examined in detail throughout Section~\ref{sec:interplay} by dissecting the various approximations that come into play, can also in principle be incorporated, as all of them are directly calculable from the momentum transfer probability~\eqref{eqU1}. 

In this work, we have studied heavy quark dynamics in a strongly coupled medium that is infinite in extent,  spatially homogeneous, and unchanging in time. And, we have only considered heavy quark distributions that are isotropic in three-dimensional momentum space, initially and throughout the evolution.
In Figs.~\ref{fig:solutions-P-example}
and \ref{fig:ratios-P-example} 
(Figs.~\ref{fig:solutions-P-charm} and
\ref{fig:ratios-P-charm}) 
we have shown how bottom-like (charm-like) heavy quarks with $M/T=4.75/0.2$ ($M/T=1.5/0.2)$ with an initial $|\vec{p}|$-distribution motivated by 
QCD calculations for the $p_T$-distribution of bottom (charm) quarks in an LHC collision equilibrate in this simplified setup. Despite these simplifications, our setup allows us to study the central aspects of the equilibration process ---
including the effects of non-Gaussian fluctuations, the evolution of the high-momentum tail of the distribution, and the approach to equilibrium ---
directly from first principles in the strongly coupled quantum field theory. 
The shape of the equilibrium heavy quark momentum distribution is quite different in this simplified setup than what the shape of the (non-stationary) late-time distribution would be in a more realistic hydrodynamic droplet of fluid that is expanding and cooling. However, at high momentum, 
say $|\vec{p}|/T\equiv u\gtrsim (5-10)$, all that matters is that the equilibrium distribution is many orders of magnitude below the initial distribution --- which is the case in our setup just as in more realistic contexts.
This indicates that the most striking features of our 
results that we recall below, features that are immediately apparent 
for $u\gtrsim (5-10)$,
will be just as relevant and important
in any more complete, more phenomenological, future treatment of non-Gaussian heavy quark transport as they are in the simplified setup that we have employed here.

Upon solving the Kolmogorov equation that describes the complete non-Gaussian dynamics of heavy quark 
momentum fluctuations, we see in our results in Figs.~\ref{fig:solutions-P-example}, \ref{fig:ratios-P-example},~\ref{fig:solutions-P-charm} and
\ref{fig:ratios-P-charm} that 
for $u\gtrsim (5-10)$ the evolution of the initial heavy quark momentum
distribution $\mathscr{P}(u,\tau)$ downwards (toward the equilibrium distribution far below) is glacially slow relative to when the distribution is evolved via  truncated, Gaussian, Fokker-Planck dynamics.
That is, the ``survivor bias'' effect that is responsible for having a less severe suppression of the high momentum tail of the heavy quark momentum space distribution 
is much stronger 
in the complete non-Gaussian dynamics than in the Fokker-Planck truncation thereof. Rare events --- whose probability is much smaller than those that make up the bulk of the probability --- 
in which the heavy quark loses almost no momentum in a time step
give the dominant contribution to the dynamics of the tail of the distribution, precisely because measuring the properties of the tail is equivalent to conditioning the partons in it to have lost much less energy than the mean.
And, these rare events are much less rare in the complete non-Gaussian dynamics described by the Kolmogorov equation than they would be if the fluctuations about the mean momentum loss were Gaussian.
As a consequence of the strong survivor bias intrinsic to our calculation that is a direct consequence of the non-Gaussian dynamics, as illustrated in Figs.~\ref{fig:ratios-P-example} 
and \ref{fig:ratios-P-charm} 
we find in our calculation that the ratio of $\mathscr{P}(u,\tau)$ after 
time evolution according 
to the Kolmogorov evolution 
to the initial momentum distribution $\mathscr{P}_0(u)$
is an increasing function of $u$ 
for $u\gtrsim (5-10)$.
This is reminiscent of  experimental measurements of the $D$-meson $R_{\rm AA}$ ratio~\cite{CMS:2017qjw,ALICE:2021rxa,ALICE:2021mgk},
which rises with increasing $p_T$ at large $p_T$. It is striking that this feature 
can also be described semi-quantitatively~\cite{Beraudo:2025nvq} by assuming that
sufficiently relativistic charm quarks (which are found in jets) behave like light quarks in jets in strongly coupled plasma whose rate of energy loss~\cite{Chesler:2014jva,Chesler:2015nqz} describes an $R_{\rm AA}$ for both jets and light hadrons that also
increases with increasing $p_T$ at large $p_T$~\cite{Casalderrey-Solana:2014bpa,Casalderrey-Solana:2018wrw}.
We leave investigating the similarity of the Kolmogorov evolution of the 
momentum distribution 
for relativistic heavy quarks
seen in the left panels of Figs.~\ref{fig:solutions-P-example} and \ref{fig:ratios-P-example} 
to the dynamics of light quarks in strongly coupled plasma to future work.

We have emphasized that the medium in which we have done our calculation, with a temperature $T$ that is constant throughout space and time, is unrealistic. 
Furthermore, we have focused on initial conditions with spherical symmetry in three-dimensional momentum space. This symmetry is preserved by the dynamics in an infinite homogeneous medium, and thus in our setup the problem becomes effectively one-dimensional, providing a direct view into the dynamics of the fluctuations in, and loss of, energy (and momentum in the direction of the heavy quark)
without the need to keep track of 
momentum broadening.
In all these respects, our analysis is also far from realistic, as the production of heavy quarks in any collider process will only have rotational symmetry around the collision axis, rather than full spherical symmetry. That being said,
in a realistic, hydrodynamic, medium that expands and cools, 
and with a realistic momentum distribution,
the non-Gaussianity in the momentum transfer fluctuations experienced by a heavy quark will be just as important as in our calculation, as will the resulting survivor bias effect at large $u$.  
The logic behind the survivor bias effect is the same in any Monte Carlo description of jet quenching or heavy quark transport in a QCD medium, flowing hydrodynamically or not. 
This means that 
we expect that our qualitative conclusions regarding heavy quark dynamics at large momentum, namely the enhanced survivor bias that results from non-Gaussian momentum
transfer fluctuations making 
loss-of-little-momentum less improbable,
will remain valid.
As such, our results open up the possibility to connect the dynamics of non-relativistic and ultra-relativistic heavy quarks
treated from first principles in quantum field theory
in a single framework, from diffusive-like dynamics at low momentum to jet quenching physics at high momentum, as has recently been done at strong coupling via a more phenomenological 
approach~\cite{Beraudo:2025nvq}.

The considerations of the preceding paragraph regarding the simplified medium and initial heavy quark momentum distribution in the present setup, together with hadronization and QGP initial state effects, will all need to be addressed before meeting the state of the art of heavy ion collision phenomenology.
Generalizing our calculation to
an initial momentum distribution that is not spherically symmetric and
a medium whose  temperature varies as a function of both position and time 
will require generalizing the 
heavy quark momentum distribution
$\mathscr{P}(u,\tau)$ whose evolution we
have described here to a phase-space distribution --- as the momentum transfer probability $P({\bf k};{\bf v})$ will be different at different positions and times. In any more realistic setting, the heavy quark phase-space distribution would be anisotropic
in momentum space and inhomogeneous in position space, both in the initial conditions and at all subsequent times.  We noted in Section~\ref{sec:comparing-K-FP-conceptual} that 
an interesting avenue to pursue in future work would be to recast our calculation by using the Kolmogorov kernel $K$ as input to a \textit{non-Gaussian}, Langevin-like, stochastic description of an individual heavy quark, using the momentum change probability distribution $P({\bf k};{\bf v})$ in Eq.~\eqref{eqU1} to evolve the 
momenta of individual heavy quarks in an ensemble in discrete time steps.  
Although there is no apparent advantage to such an approach in our simplified setup, it could be advantageous when the medium is finite in extent, expanding, and cooling, and when the heavy quark phase-space distribution is anisotropic in momentum space and inhomogeneous in position space.

In summary, by means of our present calculation, we have outlined a path to obtaining a characterization of heavy quark dynamics in heavy-ion collisions directly rooted in first-principles calculations, where the complete momentum change probability $P({\bf k};{\bf v})$ 
would be used to define a Kolmogorov equation that retains all of the non-Gaussian information specified by QCD, simultaneously describing relativistic and non-relativistic heavy quarks. The biggest challenge to be surmounted is that, to our knowledge, the only calculations of $P({\bf k};{\bf v})$ that are available for gauge theories at nonzero temperature are i) the strongly coupled $\mathcal{N}=4$ SYM theory result from Ref.~\cite{Rajagopal:2025ukd} that we have used here and ii) the recent weakly coupled calculation from Ref.~\cite{DuPlessis:2026pyr} --- and for good reason: even at leading order in HQET, $P({\bf k};{\bf v})$ is a highly nontrivial object. 
While a direct QCD calculation of $P({\bf k};{\bf v})$ at realistic values of the coupling is out of reach with current methods, leveraging existing knowledge from perturbative QCD, lattice QCD determinations of the heavy quark diffusion coefficient, and strongly coupled gauge theory calculations via holographic methods, may already be enough to construct a field-theory-informed parametrization of the momentum change probability of intrinsic phenomenological interest, to be used either in the Kolmogorov dynamics form we have presented here, or in a Monte Carlo description where each heavy quark is evolved individually. Doing so will provide phenomenological studies with a direct connection to fundamental QCD quantities that may then be extracted from experimental data without the limitations of existing approaches.

\acknowledgments

We gratefully acknowledge helpful conversations with Joerg Aichelin, Andrea Beraudo, Pol Bernard Gossiaux, Vincenzo Greco, Dani Pablos, Jean Francois Du Plessis and Xiaojun Yao. KR is grateful to the Kavli Institute for Theoretical Physics (KITP) for hospitality and support as this work was completed; this research was supported in part by grant NSF PHY-2309135 to the KITP. BSH is additionally supported by grant 994312 from the Simons Foundation.
Research supported in part by by the U.S.~Department of Energy, Office of Science, Office of Nuclear Physics under grant Contract Number DE-SC0011090.

\appendix

\section{Validation of the Approximations Employed in Deriving the Kolmogorov equation} \label{app:Kolmogorov-checks}

In this Appendix, we complete our analysis of the validity of the key approximations
we made in the derivation of the Kolmogorov equation in Eq.~\eqref{eqU10}.  We focus specifically on the approximations that allowed us to neglect the $\mathcal{O}((\Delta \tau)^2)$ terms in Eq.~\eqref{eqU10}, which we have restated and discussed in more detail in Eqs.~\eqref{eq5.8} and~\eqref{eq5.9}. To be concrete, we shall assess the magnitude of the error in these approximations by plotting the ratio
\begin{align}
    r &\equiv \frac{f \! \left( {\bf u} , \tau \right) + \Delta \tau \frac{\pi}{2} {\bf C} \cdot \frac{\partial f}{\partial {\bf u} } \left( {\bf u} , \tau \right) - \Delta \tau \tilde{S}_{\rm tot} \! \left( {\bf C} ; {\bf v} \left({\bf u}  \right)  \right)}{f \!  \left( {\bf u} + \Delta \tau \frac{\pi}{2} {\bf C} , \tau \right) - \Delta \tau \tilde{S}_{\rm tot} \! \left( {\bf C} ; {\bf v} \left({\bf u} + \Delta \tau \frac{\pi}{2} {\bf C}  \right)  \right)} \, , \label{eq:error-estimate}
\end{align}
evaluated on the value of the momentum transfer rate ${\bf C} = \bar{\bf C}({\bf u})$ that determines the saddle point in Eq.~\eqref{eqU10} that serves to specify the subsequent dynamics of the Kolmogorov equation. Due to the spherical symmetry of the solutions we study, this value of ${\bf C}$ is fully specified by its radial component $C_3$. 
If the approximation~\eqref{eq5.8} holds, 
the ratio of the terms in the numerator and denominator
of Eq.~\eqref{eq:error-estimate} that are not proportional to $\Delta\tau$ is close to unity;
if the approximation~\eqref{eq5.9} holds, 
the ratio of the terms in the numerator and denominator
of Eq.~\eqref{eq:error-estimate} that are proportional to $\Delta\tau$ is close to unity.
This means that the ratio $r$ quantifies the validity of the Taylor expansion at small $\Delta \tau$ of the exponent in Eq.~\eqref{eqU10} as a whole --- and, as such, quantifies the overall error incurred in that derivation.

\begin{figure}
    \centering
    \includegraphics[width=0.7\linewidth]{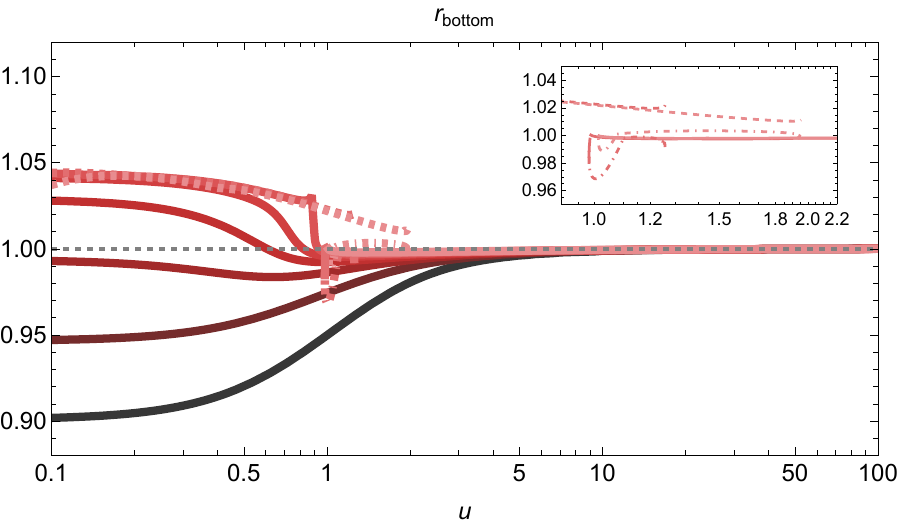}
    \caption{Plot of the ratio $r$ defined in Eq.~\eqref{eq:error-estimate}  for the same solution to the Kolmogorov equation shown in Fig.~\ref{fig:solutions-P-example}, starting from our FONLL-motivated bottom-like initial conditions. 
    The deviation of $r$ away from unity is a quantitative measure of the error in the approximations made in deriving the Kolmogorov equation.  The curves show $r$ as a function of $u$ at different times, with the colors corresponding to times as in Fig.~\ref{fig:solutions-P-example}. For times before the solution for $f(u)$ becomes multivalued, we display the ratio $r$ with a solid line. For later times, we use a solid line for the ratio in the large $u$ region, a dot-dashed line for $r$ from the $f$ component that only exists in the multivalued interval, and a dashed line for $r$ in the small $u$ region. See Eq.~\eqref{eq:P-solution-explicit-reco} for the definitions of these regions. The grey dashed line shows $r=1$ for reference.  The inset shows the interval where the solution is multivalued in more detail (only the later times are shown). Because the denominator in Eq.~\eqref{eq:error-estimate} depends on $f({\bf u} + \Delta \tau \tfrac{\pi}{2} \bar{\bf C}({\bf u}),\tau )$, the curves in this figure can be discontinuous even if $\bar{\bf C}({\bf u})$ is continuous as a function of ${\bf u}$ (as it is, see Fig.~\ref{fig:K-FP-solution-sample}). This happens because $f$ gets evaluated at values of its (rescaled) momentum argument on different branches --- they would only be guaranteed to agree if $\bar{\bf C}({\bf u})$ were to vanish at the corresponding edge of the multivalued interval.
    }
    \label{fig:K-approximations}
\end{figure}

\begin{figure}
    \centering
    \includegraphics[width=0.7\linewidth]{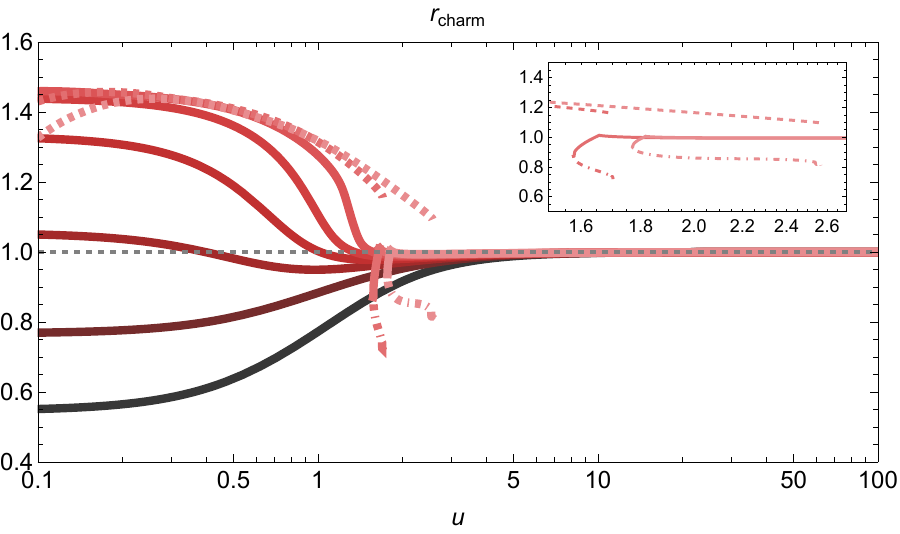}
    \caption{Plot of the ratio $r$ defined in Eq.~\eqref{eq:error-estimate}  for the  solution to the Kolmogorov equation shown in Fig.~\ref{fig:solutions-P-charm}, starting from our FONLL-motivated charm-like initial conditions. 
    The curves show $r$ at different times, with the colors corresponding to times as in Fig.~\ref{fig:solutions-P-charm}. For times before the solution for $f(u)$ becomes multivalued, we display the ratio $r$ with a solid line. For later times, we use a solid line for the ratio in the large $u$ region, a dot-dashed line for $r$ from the $f$ component that only exists in the multivalued interval, and a dashed line for $r$ in the small $u$ region. See Eq.~\eqref{eq:P-solution-explicit-reco} for the definitions of these regions. The grey dashed line shows $r=1$ for reference. The inset shows the interval where the solution is multivalued in more detail (only the later times are shown). Because the denominator in Eq.~\eqref{eq:error-estimate} depends on $f({\bf u} + \Delta \tau \tfrac{\pi}{2} \bar{\bf C}({\bf u}),\tau )$, the curves in this figure can be discontinuous even if $\bar{\bf C}({\bf u})$ is continuous as a function of ${\bf u}$ (as it is, see Fig.~\ref{fig:K-FP-solution-sample}). This happens because $f$ gets evaluated at values of its (rescaled) momentum argument on different branches.}
    \label{fig:K-approximations-charm}
\end{figure}

In Figs.~\ref{fig:K-approximations} and~\ref{fig:K-approximations-charm}
we have plotted the ratio $r$ as a function of $u$ at various times.  
In making these plots, we have
chosen the smallest value of $\Delta \tau$ that is consistent with the result of the calculation of $P({\bf k};{\bf v})$ that we use~\cite{Rajagopal:2025ukd}. Concretely, in evaluating $\Delta \tau = \Delta t \sqrt{\lambda} T^2/M$ we have set $\Delta t = 1/T$ and $\lambda = 11$ and have employed
the $M/T$ ratios we used throughout the main text ($M/T = 4.75/0.2$ for bottom-like parameters 
in Fig.~\ref{fig:K-approximations}
and $M/T = 1.5/0.2$ for charm-like parameters in
Fig.~\ref{fig:K-approximations-charm}).
In Fig.~\ref{fig:K-approximations} (Fig.~\ref{fig:K-approximations-charm}) we use our bottom-like (charm-like) FONLL-motivated initial conditions, see Eqs.~\eqref{eq:init-cond-form},
\eqref{eq:FONLL-m-b}, \eqref{eq:FONLL-m-c}.
As we can see, both ratios $r_{\rm bottom}$ and $r_{\rm charm}$ are very close to unity at large momentum. This is so because the function $f$ varies more and more slowly as a function of $u$ at large momentum, as does $\tilde{S}_{\rm tot}$. Conversely, at small momentum the ratios are somewhat away from unity in both cases: up to $\sim 10 \%$ for bottom and up to $\sim 50\%$ for charm. Note that the fact that the approximations are worse for charm-like parameters than for bottom-like parameters is consistent with the fact that the minimal applicable value of $\Delta \tau$ is bigger for charm-like parameters.

The approximation is worse at small momentum (small $u$) because the relative momentum change of an individual heavy quark is such that its dynamics at the next time step will be very different than that in the previous time step. Even if we are only looking at those in a fixed, low, momentum bin, momentum fluctuations will be constantly removing quarks from that bin and replacing them with new quarks that come from a different momentum bin (that may be far away in momentum, even one where their momentum has opposite sign), meaning that the $P({\bf k};{\bf v})$ that they experienced in the previous time step will be very different than what they will experience in the next. 
The error introduced at small $u$ by the approximations that we have made has no consequence at very late time since, as we have shown in Section~\ref{sec:K-as-HJ}, the Kolmogorov dynamics is guaranteed to reach kinetic equilibrium.
It may merit further investigation if a 
precise characterization of the path to equilibration at low momentum in full detail is desired.
We see from Figs.~~\ref{fig:K-approximations} and~\ref{fig:K-approximations-charm} that the approximations we have made are {\it very} good at large $u$, say $u\gtrsim 5$. This can
be understood by noting that
the heavy quarks in any specified large momentum bin  --- in particular, those 
that stay in the large-$u$ tail of the 
heavy quark momentum 
distribution $\mathscr{P}(u,\tau)$
for a long period of time --- get momentum updates sampled from very similar distributions because their velocity doesn't change that much. This means that they came from bins that have very similar momentum change probabilities $P({\bf k};{\bf v})$.

\bibliography{main.bib}

\providecommand{\href}[2]{#2}\begingroup\raggedright\begin{thebibliography}{100}

\bibitem{Rapp:2009my}
R.~Rapp and H.~van Hees, {\it {Heavy Quarks in the Quark-Gluon Plasma}},  pp.~111--206, 2010.
\newblock \href{http://arxiv.org/abs/0903.1096}{{\tt arXiv:0903.1096}}.

\bibitem{Andronic:2015wma}
A.~Andronic et~al., {\it {Heavy-flavour and quarkonium production in the LHC era: from proton{\textendash}proton to heavy-ion collisions}},  {\em Eur. Phys. J. C} {\bf 76} (2016), no.~3 107, [\href{http://arxiv.org/abs/1506.03981}{{\tt arXiv:1506.03981}}].

\bibitem{Aarts:2016hap}
G.~Aarts et~al., {\it {Heavy-flavor production and medium properties in high-energy nuclear collisions - What next?}},  {\em Eur. Phys. J. A} {\bf 53} (2017), no.~5 93, [\href{http://arxiv.org/abs/1612.08032}{{\tt arXiv:1612.08032}}].

\bibitem{Prino:2016cni}
F.~Prino and R.~Rapp, {\it {Open Heavy Flavor in QCD Matter and in Nuclear Collisions}},  {\em J. Phys. G} {\bf 43} (2016), no.~9 093002, [\href{http://arxiv.org/abs/1603.00529}{{\tt arXiv:1603.00529}}].

\bibitem{Rapp:2018qla}
A.~Beraudo et~al., {\it {Extraction of Heavy-Flavor Transport Coefficients in QCD Matter}},  {\em Nucl. Phys. A} {\bf 979} (2018) 21--86, [\href{http://arxiv.org/abs/1803.03824}{{\tt arXiv:1803.03824}}].

\bibitem{Cao:2018ews}
S.~Cao et~al., {\it {Toward the determination of heavy-quark transport coefficients in quark-gluon plasma}},  {\em Phys. Rev. C} {\bf 99} (2019), no.~5 054907, [\href{http://arxiv.org/abs/1809.07894}{{\tt arXiv:1809.07894}}].

\bibitem{Dong:2019byy}
X.~Dong, Y.-J. Lee, and R.~Rapp, {\it {Open Heavy-Flavor Production in Heavy-Ion Collisions}},  {\em Ann. Rev. Nucl. Part. Sci.} {\bf 69} (2019) 417--445, [\href{http://arxiv.org/abs/1903.07709}{{\tt arXiv:1903.07709}}].

\bibitem{Apolinario:2022vzg}
L.~Apolin{\'a}rio, Y.-J. Lee, and M.~Winn, {\it {Heavy quarks and jets as probes of the QGP}},  {\em Prog. Part. Nucl. Phys.} {\bf 127} (2022) 103990, [\href{http://arxiv.org/abs/2203.16352}{{\tt arXiv:2203.16352}}].

\bibitem{PHENIX:2006iih}
{\bf PHENIX} Collaboration, A.~Adare et~al., {\it {Energy Loss and Flow of Heavy Quarks in Au+Au Collisions at s(NN)**(1/2) = 200-GeV}},  {\em Phys. Rev. Lett.} {\bf 98} (2007) 172301, [\href{http://arxiv.org/abs/nucl-ex/0611018}{{\tt nucl-ex/0611018}}].

\bibitem{PHENIX:2010xji}
{\bf PHENIX} Collaboration, A.~Adare et~al., {\it {Heavy Quark Production in $p+p$ and Energy Loss and Flow of Heavy Quarks in Au+Au Collisions at $\sqrt{s_{NN}}=200$ GeV}},  {\em Phys. Rev. C} {\bf 84} (2011) 044905, [\href{http://arxiv.org/abs/1005.1627}{{\tt arXiv:1005.1627}}].

\bibitem{STAR:2014wif}
{\bf STAR} Collaboration, L.~Adamczyk et~al., {\it {Observation of $D^0$ Meson Nuclear Modifications in Au+Au Collisions at $\sqrt{s_{NN}}=200$ GeV}},  {\em Phys. Rev. Lett.} {\bf 113} (2014), no.~14 142301, [\href{http://arxiv.org/abs/1404.6185}{{\tt arXiv:1404.6185}}]. [Erratum: Phys.Rev.Lett. 121, 229901 (2018)].

\bibitem{STAR:2017kkh}
{\bf STAR} Collaboration, L.~Adamczyk et~al., {\it {Measurement of $D^0$ Azimuthal Anisotropy at Midrapidity in Au+Au Collisions at $\sqrt{s_{NN}}$=200 GeV}},  {\em Phys. Rev. Lett.} {\bf 118} (2017), no.~21 212301, [\href{http://arxiv.org/abs/1701.06060}{{\tt arXiv:1701.06060}}].

\bibitem{STAR:2018zdy}
{\bf STAR} Collaboration, J.~Adam et~al., {\it {Centrality and transverse momentum dependence of $D^0$-meson production at mid-rapidity in Au+Au collisions at ${\sqrt{s_{\rm NN}} = \rm{200\,GeV}}$}},  {\em Phys. Rev. C} {\bf 99} (2019), no.~3 034908, [\href{http://arxiv.org/abs/1812.10224}{{\tt arXiv:1812.10224}}].

\bibitem{Shi:2024eyk}
{\bf sPHENIX} Collaboration, Z.~Shi, {\it {Heavy Flavor Physics at the sPHENIX Experiment}},  {\em Universe} {\bf 10} (2024), no.~3 126, [\href{http://arxiv.org/abs/2401.11036}{{\tt arXiv:2401.11036}}].

\bibitem{ALICE:2013olq}
{\bf ALICE} Collaboration, B.~Abelev et~al., {\it {D meson elliptic flow in non-central Pb-Pb collisions at $\sqrt{s_{\rm NN}}$ = 2.76TeV}},  {\em Phys. Rev. Lett.} {\bf 111} (2013) 102301, [\href{http://arxiv.org/abs/1305.2707}{{\tt arXiv:1305.2707}}].

\bibitem{ALICE:2014qvj}
{\bf ALICE} Collaboration, B.~B. Abelev et~al., {\it {Azimuthal anisotropy of D meson production in Pb-Pb collisions at $\sqrt{s_{\rm NN}} = 2.76$ TeV}},  {\em Phys. Rev. C} {\bf 90} (2014), no.~3 034904, [\href{http://arxiv.org/abs/1405.2001}{{\tt arXiv:1405.2001}}].

\bibitem{ALICE:2017pbx}
{\bf ALICE} Collaboration, S.~Acharya et~al., {\it {$D$-meson azimuthal anisotropy in midcentral Pb-Pb collisions at $\mathbf{\sqrt{s_{\rm NN}}=5.02}$ TeV}},  {\em Phys. Rev. Lett.} {\bf 120} (2018), no.~10 102301, [\href{http://arxiv.org/abs/1707.01005}{{\tt arXiv:1707.01005}}].

\bibitem{CMS:2017qjw}
{\bf CMS} Collaboration, A.~M. Sirunyan et~al., {\it {Nuclear modification factor of D$^0$ mesons in PbPb collisions at $\sqrt{s_\mathrm{NN}} = 5.02$ TeV}},  {\em Phys. Lett. B} {\bf 782} (2018) 474--496, [\href{http://arxiv.org/abs/1708.04962}{{\tt arXiv:1708.04962}}].

\bibitem{CMS:2017uuv}
{\bf CMS} Collaboration, A.~M. Sirunyan et~al., {\it {Measurement of prompt and nonprompt charmonium suppression in $\text {PbPb}$ collisions at 5.02 $\,\text {Te}\text {V}$}},  {\em Eur. Phys. J. C} {\bf 78} (2018), no.~6 509, [\href{http://arxiv.org/abs/1712.08959}{{\tt arXiv:1712.08959}}]. [Erratum: Eur.Phys.J.C 83, 145 (2023)].

\bibitem{ATLAS:2018hqe}
{\bf ATLAS} Collaboration, M.~Aaboud et~al., {\it {Prompt and non-prompt $J/\psi $ and $\psi (2\mathrm {S})$ suppression at high transverse momentum in $5.02~\mathrm {TeV}$ Pb+Pb collisions with the ATLAS experiment}},  {\em Eur. Phys. J. C} {\bf 78} (2018), no.~9 762, [\href{http://arxiv.org/abs/1805.04077}{{\tt arXiv:1805.04077}}].

\bibitem{ALICE:2018gif}
{\bf ALICE} Collaboration, S.~Acharya et~al., {\it {Event-shape engineering for the D-meson elliptic flow in mid-central Pb-Pb collisions at $\sqrt{s_{\rm NN}} =5.02$ TeV}},  {\em JHEP} {\bf 02} (2019) 150, [\href{http://arxiv.org/abs/1809.09371}{{\tt arXiv:1809.09371}}].

\bibitem{ATLAS:2020yxw}
{\bf ATLAS} Collaboration, G.~Aad et~al., {\it {Measurement of azimuthal anisotropy of muons from charm and bottom hadrons in Pb+Pb collisions at sNN=5.02 TeV with the ATLAS detector}},  {\em Phys. Lett. B} {\bf 807} (2020) 135595, [\href{http://arxiv.org/abs/2003.03565}{{\tt arXiv:2003.03565}}].

\bibitem{ALICE:2020iug}
{\bf ALICE} Collaboration, S.~Acharya et~al., {\it {Transverse-momentum and event-shape dependence of D-meson flow harmonics in Pb{\textendash}Pb collisions at $\sqrt {s_{NN}}$ = 5.02 TeV}},  {\em Phys. Lett. B} {\bf 813} (2021) 136054, [\href{http://arxiv.org/abs/2005.11131}{{\tt arXiv:2005.11131}}].

\bibitem{CMS:2020bnz}
{\bf CMS} Collaboration, A.~M. Sirunyan et~al., {\it {Measurement of prompt ${\mathrm{D^0}}$ and ${\mathrm{\overline{D}}{}^0}$ meson azimuthal anisotropy and search for strong electric fields in PbPb collisions at $\sqrt{s_\mathrm{NN}} =$ 5.02 TeV}},  {\em Phys. Lett. B} {\bf 816} (2021) 136253, [\href{http://arxiv.org/abs/2009.12628}{{\tt arXiv:2009.12628}}].

\bibitem{ATLAS:2021xtw}
{\bf ATLAS} Collaboration, G.~Aad et~al., {\it {Measurement of the nuclear modification factor for muons from charm and bottom hadrons in Pb+Pb collisions at 5.02 TeV with the ATLAS detector}},  {\em Phys. Lett. B} {\bf 829} (2022) 137077, [\href{http://arxiv.org/abs/2109.00411}{{\tt arXiv:2109.00411}}].

\bibitem{ALICE:2021rxa}
{\bf ALICE} Collaboration, S.~Acharya et~al., {\it {Prompt D$^{0}$, D$^{+}$, and D$^{*+}$ production in Pb{\textendash}Pb collisions at $ \sqrt{s_{\mathrm{NN}}} $ = 5.02 TeV}},  {\em JHEP} {\bf 01} (2022) 174, [\href{http://arxiv.org/abs/2110.09420}{{\tt arXiv:2110.09420}}].

\bibitem{ALICE:2021kfc}
{\bf ALICE} Collaboration, S.~Acharya et~al., {\it {Measurement of prompt $D_s^+$-meson production and azimuthal anisotropy in Pb{\textendash}Pb collisions at $\sqrt {s_{NN}}$=5.02TeV}},  {\em Phys. Lett. B} {\bf 827} (2022) 136986, [\href{http://arxiv.org/abs/2110.10006}{{\tt arXiv:2110.10006}}].

\bibitem{CMS:2022sxl}
{\bf CMS} Collaboration, A.~Tumasyan et~al., {\it {Observation of the $B_c^+$ Meson in Pb-Pb and pp Collisions at $\sqrt{s_{NN}}$=5.02{\,}{\,}TeV and Measurement of its Nuclear Modification Factor}},  {\em Phys. Rev. Lett.} {\bf 128} (2022), no.~25 252301, [\href{http://arxiv.org/abs/2201.02659}{{\tt arXiv:2201.02659}}].

\bibitem{ALICE:2022tji}
{\bf ALICE} Collaboration, S.~Acharya et~al., {\it {Measurement of beauty production via non-prompt D$^{0}$ mesons in Pb-Pb collisions at $ \sqrt{{\textrm{s}}_{\textrm{NN}}} $= 5.02 TeV}},  {\em JHEP} {\bf 12} (2022) 126, [\href{http://arxiv.org/abs/2202.00815}{{\tt arXiv:2202.00815}}].

\bibitem{CMS:2022vfn}
{\bf CMS} Collaboration, A.~Tumasyan et~al., {\it {Measurements of azimuthal anisotropy of nonprompt D0 mesons in PbPb collisions at sNN=5.02TeV}},  {\em Phys. Lett. B} {\bf 850} (2024) 138389, [\href{http://arxiv.org/abs/2212.01636}{{\tt arXiv:2212.01636}}].

\bibitem{ALICE:2023gjj}
{\bf ALICE} Collaboration, S.~Acharya et~al., {\it {Measurement of non-prompt ${{\textrm{D}}^{0}}$-meson elliptic flow in Pb{\textendash}Pb collisions at $\sqrt{s_{\textrm{NN}}} = 5.02$~TeV}},  {\em Eur. Phys. J. C} {\bf 83} (2023), no.~12 1123, [\href{http://arxiv.org/abs/2307.14084}{{\tt arXiv:2307.14084}}].

\bibitem{ALICE:2023hou}
{\bf ALICE} Collaboration, S.~Acharya et~al., {\it {Prompt and non-prompt~J$/\psi$ production at midrapidity in Pb{\textendash}Pb collisions at $ \sqrt{s_{\textrm{NN}}} $ = 5.02 TeV}},  {\em JHEP} {\bf 02} (2024) 066, [\href{http://arxiv.org/abs/2308.16125}{{\tt arXiv:2308.16125}}].

\bibitem{CMS:2024vip}
{\bf CMS} Collaboration, A.~Hayrapetyan et~al., {\it {Bottom quark energy loss and hadronization with B$^{+}$ and $ {\textrm{B}}_{\textrm{s}}^0 $ nuclear modification factors using pp and PbPb collisions at $ \sqrt{s_{\textrm{NN}}} $ = 5.02 TeV}},  {\em JHEP} {\bf 02} (2025) 195, [\href{http://arxiv.org/abs/2409.07258}{{\tt arXiv:2409.07258}}].

\bibitem{ALICE:2022wwr}
{\bf ALICE} Collaboration, {\it {Letter of intent for ALICE 3: A next-generation heavy-ion experiment at the LHC}},  \href{http://arxiv.org/abs/2211.02491}{{\tt arXiv:2211.02491}}.

\bibitem{ALICE3Scoping2025}
{\bf ALICE Collaboration} Collaboration, {\it Scoping document for alice 3: Alice phase iib upgrade for the lhc long shutdown 4},  tech. rep., 2025.

\bibitem{Svetitsky:1987gq}
B.~Svetitsky, {\it {Diffusion of charmed quarks in the quark-gluon plasma}},  {\em Phys. Rev. D} {\bf 37} (1988) 2484--2491.

\bibitem{Moore:2004tg}
G.~D. Moore and D.~Teaney, {\it {How much do heavy quarks thermalize in a heavy ion collision?}},  {\em Phys. Rev. C} {\bf 71} (2005) 064904, [\href{http://arxiv.org/abs/hep-ph/0412346}{{\tt hep-ph/0412346}}].

\bibitem{vanHees:2004gq}
H.~van Hees and R.~Rapp, {\it {Thermalization of heavy quarks in the quark-gluon plasma}},  {\em Phys. Rev. C} {\bf 71} (2005) 034907, [\href{http://arxiv.org/abs/nucl-th/0412015}{{\tt nucl-th/0412015}}].

\bibitem{vanHees:2005wb}
H.~van Hees, V.~Greco, and R.~Rapp, {\it {Heavy-quark probes of the quark-gluon plasma at RHIC}},  {\em Phys. Rev. C} {\bf 73} (2006) 034913, [\href{http://arxiv.org/abs/nucl-th/0508055}{{\tt nucl-th/0508055}}].

\bibitem{vanHees:2007me}
H.~van Hees, M.~Mannarelli, V.~Greco, and R.~Rapp, {\it {Nonperturbative heavy-quark diffusion in the quark-gluon plasma}},  {\em Phys. Rev. Lett.} {\bf 100} (2008) 192301, [\href{http://arxiv.org/abs/0709.2884}{{\tt arXiv:0709.2884}}].

\bibitem{Xu:2017obm}
Y.~Xu, J.~E. Bernhard, S.~A. Bass, M.~Nahrgang, and S.~Cao, {\it {Data-driven analysis for the temperature and momentum dependence of the heavy-quark diffusion coefficient in relativistic heavy-ion collisions}},  {\em Phys. Rev. C} {\bf 97} (2018), no.~1 014907, [\href{http://arxiv.org/abs/1710.00807}{{\tt arXiv:1710.00807}}].

\bibitem{He:2011qa}
M.~He, R.~J. Fries, and R.~Rapp, {\it {Heavy-Quark Diffusion and Hadronization in Quark-Gluon Plasma}},  {\em Phys. Rev. C} {\bf 86} (2012) 014903, [\href{http://arxiv.org/abs/1106.6006}{{\tt arXiv:1106.6006}}].

\bibitem{Cao:2013ita}
S.~Cao, G.-Y. Qin, and S.~A. Bass, {\it {Heavy-quark dynamics and hadronization in ultrarelativistic heavy-ion collisions: Collisional versus radiative energy loss}},  {\em Phys. Rev. C} {\bf 88} (2013) 044907, [\href{http://arxiv.org/abs/1308.0617}{{\tt arXiv:1308.0617}}].

\bibitem{Das:2013kea}
S.~K. Das, F.~Scardina, S.~Plumari, and V.~Greco, {\it {Heavy-flavor in-medium momentum evolution: Langevin versus Boltzmann approach}},  {\em Phys. Rev. C} {\bf 90} (2014) 044901, [\href{http://arxiv.org/abs/1312.6857}{{\tt arXiv:1312.6857}}].

\bibitem{Berrehrah:2014tva}
H.~Berrehrah, P.~B. Gossiaux, J.~Aichelin, W.~Cassing, J.~M. Torres-Rincon, and E.~Bratkovskaya, {\it {Transport coefficients of heavy quarks around $T_c$ at finite quark chemical potential}},  {\em Phys. Rev. C} {\bf 90} (2014) 051901, [\href{http://arxiv.org/abs/1406.5322}{{\tt arXiv:1406.5322}}].

\bibitem{Xu:2018gux}
Y.~Xu et~al., {\it {Resolving discrepancies in the estimation of heavy quark transport coefficients in relativistic heavy-ion collisions}},  {\em Phys. Rev. C} {\bf 99} (2019), no.~1 014902, [\href{http://arxiv.org/abs/1809.10734}{{\tt arXiv:1809.10734}}].

\bibitem{Gossiaux:2019mjc}
P.~B. Gossiaux, {\it {Open Heavy Flavors in Nuclear Collisions: Theory Overview}},  {\em Nucl. Phys. A} {\bf 982} (2019) 113--119, [\href{http://arxiv.org/abs/1901.01606}{{\tt arXiv:1901.01606}}].

\bibitem{Andronic:2024oxz}
A.~Andronic et~al., {\it {Comparative study of quarkonium transport in hot QCD matter}},  {\em Eur. Phys. J. A} {\bf 60} (2024), no.~4 88, [\href{http://arxiv.org/abs/2402.04366}{{\tt arXiv:2402.04366}}].

\bibitem{Krishna:2025bll}
T.~Krishna, R.~Rapp, Y.~Fu, S.~A. Bass, and W.~Ke, {\it {Nonperturbative heavy-flavor transport approach for hot QCD matter}},  {\em Phys. Lett. B} {\bf 871} (2025) 139999, [\href{http://arxiv.org/abs/2509.13881}{{\tt arXiv:2509.13881}}].

\bibitem{Beraudo:2025nvq}
A.~Beraudo, J.~F. Du~Plessis, D.~Pablos, and K.~Rajagopal, {\it {Heavy Quark Energy Loss in the Hybrid Model}},  \href{http://arxiv.org/abs/2510.24847}{{\tt arXiv:2510.24847}}.

\bibitem{Burnier:2010rp}
Y.~Burnier, M.~Laine, J.~Langelage, and L.~Mether, {\it {Colour-electric spectral function at next-to-leading order}},  {\em JHEP} {\bf 08} (2010) 094, [\href{http://arxiv.org/abs/1006.0867}{{\tt arXiv:1006.0867}}].

\bibitem{Eller:2019spw}
A.~M. Eller, J.~Ghiglieri, and G.~D. Moore, {\it {Thermal Heavy Quark Self-Energy from Euclidean Correlators}},  {\em Phys. Rev. D} {\bf 99} (2019), no.~9 094042, [\href{http://arxiv.org/abs/1903.08064}{{\tt arXiv:1903.08064}}]. [Erratum: Phys.Rev.D 102, 039901 (2020)].

\bibitem{Scheihing-Hitschfeld:2022xqx}
B.~Scheihing-Hitschfeld and X.~Yao, {\it {Gauge Invariance of Non-Abelian Field Strength Correlators: The Axial Gauge Puzzle}},  {\em Phys. Rev. Lett.} {\bf 130} (2023), no.~5 052302, [\href{http://arxiv.org/abs/2205.04477}{{\tt arXiv:2205.04477}}].

\bibitem{Scheihing-Hitschfeld:2023tuz}
B.~Scheihing-Hitschfeld and X.~Yao, {\it {Real time quarkonium transport coefficients in open quantum systems from Euclidean QCD}},  {\em Phys. Rev. D} {\bf 108} (2023), no.~5 054024, [\href{http://arxiv.org/abs/2306.13127}{{\tt arXiv:2306.13127}}]. [Erratum: Phys.Rev.D 109, 099902 (2024)].

\bibitem{delaCruz:2024cix}
D.~de~la Cruz, A.~M. Eller, and G.~D. Moore, {\it {QCD field-strength correlators on a Polyakov loop with gradient flow at next-to-leading order}},  {\em Phys. Rev. D} {\bf 110} (2024), no.~9 094057, [\href{http://arxiv.org/abs/2410.01578}{{\tt arXiv:2410.01578}}].

\bibitem{Caron-Huot:2007rwy}
S.~Caron-Huot and G.~D. Moore, {\it {Heavy quark diffusion in perturbative QCD at next-to-leading order}},  {\em Phys. Rev. Lett.} {\bf 100} (2008) 052301, [\href{http://arxiv.org/abs/0708.4232}{{\tt arXiv:0708.4232}}].

\bibitem{Caron-Huot:2008dyw}
S.~Caron-Huot and G.~D. Moore, {\it {Heavy quark diffusion in QCD and N=4 SYM at next-to-leading order}},  {\em JHEP} {\bf 02} (2008) 081, [\href{http://arxiv.org/abs/0801.2173}{{\tt arXiv:0801.2173}}].

\bibitem{Petreczky:2005nh}
P.~Petreczky and D.~Teaney, {\it {Heavy quark diffusion from the lattice}},  {\em Phys. Rev. D} {\bf 73} (2006) 014508, [\href{http://arxiv.org/abs/hep-ph/0507318}{{\tt hep-ph/0507318}}].

\bibitem{Caron-Huot:2009ncn}
S.~Caron-Huot, M.~Laine, and G.~D. Moore, {\it {A Way to estimate the heavy quark thermalization rate from the lattice}},  {\em JHEP} {\bf 04} (2009) 053, [\href{http://arxiv.org/abs/0901.1195}{{\tt arXiv:0901.1195}}].

\bibitem{Meyer:2011gj}
H.~B. Meyer, {\it {Transport Properties of the Quark-Gluon Plasma: A Lattice QCD Perspective}},  {\em Eur. Phys. J. A} {\bf 47} (2011) 86, [\href{http://arxiv.org/abs/1104.3708}{{\tt arXiv:1104.3708}}].

\bibitem{Banerjee:2011ra}
D.~Banerjee, S.~Datta, R.~Gavai, and P.~Majumdar, {\it {Heavy Quark Momentum Diffusion Coefficient from Lattice QCD}},  {\em Phys. Rev. D} {\bf 85} (2012) 014510, [\href{http://arxiv.org/abs/1109.5738}{{\tt arXiv:1109.5738}}].

\bibitem{Francis:2015daa}
A.~Francis, O.~Kaczmarek, M.~Laine, T.~Neuhaus, and H.~Ohno, {\it {Nonperturbative estimate of the heavy quark momentum diffusion coefficient}},  {\em Phys. Rev. D} {\bf 92} (2015), no.~11 116003, [\href{http://arxiv.org/abs/1508.04543}{{\tt arXiv:1508.04543}}].

\bibitem{Brambilla:2020siz}
N.~Brambilla, V.~Leino, P.~Petreczky, and A.~Vairo, {\it {Lattice QCD constraints on the heavy quark diffusion coefficient}},  {\em Phys. Rev. D} {\bf 102} (2020), no.~7 074503, [\href{http://arxiv.org/abs/2007.10078}{{\tt arXiv:2007.10078}}].

\bibitem{Brambilla:2022xbd}
{\bf TUMQCD} Collaboration, N.~Brambilla, V.~Leino, J.~Mayer-Steudte, and P.~Petreczky, {\it {Heavy quark diffusion coefficient with gradient flow}},  {\em Phys. Rev. D} {\bf 107} (2023), no.~5 054508, [\href{http://arxiv.org/abs/2206.02861}{{\tt arXiv:2206.02861}}].

\bibitem{Altenkort:2023eav}
{\bf HotQCD} Collaboration, L.~Altenkort, D.~de~la Cruz, O.~Kaczmarek, R.~Larsen, G.~D. Moore, S.~Mukherjee, P.~Petreczky, H.-T. Shu, and S.~Stendebach, {\it {Quark Mass Dependence of Heavy Quark Diffusion Coefficient from Lattice QCD}},  {\em Phys. Rev. Lett.} {\bf 132} (2024), no.~5 051902, [\href{http://arxiv.org/abs/2311.01525}{{\tt arXiv:2311.01525}}].

\bibitem{HotQCD:2025fbd}
{\bf HotQCD} Collaboration, D.~Bollweg, J.~L. Dasilva~Gol{\'a}n, O.~Kaczmarek, R.~N. Larsen, G.~D. Moore, S.~Mukherjee, P.~Petreczky, H.-T. Shu, S.~Stendebach, and J.~H. Weber, {\it {Temperature dependence of heavy quark diffusion from (2+1)-flavor lattice QCD}},  {\em JHEP} {\bf 09} (2025) 180, [\href{http://arxiv.org/abs/2506.11958}{{\tt arXiv:2506.11958}}].

\bibitem{Casalderrey-Solana:2006fio}
J.~Casalderrey-Solana and D.~Teaney, {\it {Heavy quark diffusion in strongly coupled N=4 Yang-Mills}},  {\em Phys. Rev. D} {\bf 74} (2006) 085012, [\href{http://arxiv.org/abs/hep-ph/0605199}{{\tt hep-ph/0605199}}].

\bibitem{Gubser:2006nz}
S.~S. Gubser, {\it {Momentum fluctuations of heavy quarks in the gauge-string duality}},  {\em Nucl. Phys. B} {\bf 790} (2008) 175--199, [\href{http://arxiv.org/abs/hep-th/0612143}{{\tt hep-th/0612143}}].

\bibitem{Casalderrey-Solana:2007ahi}
J.~Casalderrey-Solana and D.~Teaney, {\it {Transverse Momentum Broadening of a Fast Quark in a N=4 Yang Mills Plasma}},  {\em JHEP} {\bf 04} (2007) 039, [\href{http://arxiv.org/abs/hep-th/0701123}{{\tt hep-th/0701123}}].

\bibitem{Casalderrey-Solana:2011dxg}
J.~Casalderrey-Solana, H.~Liu, D.~Mateos, K.~Rajagopal, and U.~Achim~Wiedemann, {\em {Gauge/String Duality, Hot QCD and Heavy Ion Collisions}}.
\newblock Cambridge University Press, 2014.

\bibitem{Rajagopal:2025ukd}
K.~Rajagopal, B.~Scheihing-Hitschfeld, and U.~A. Wiedemann, {\it {Dynamics of heavy quarks in strongly coupled $ \mathcal{N} $ = 4 SYM plasma}},  {\em JHEP} {\bf 07} (2025) 013, [\href{http://arxiv.org/abs/2501.06289}{{\tt arXiv:2501.06289}}].

\bibitem{Rajagopal:2025rxr}
K.~Rajagopal, B.~Scheihing-Hitschfeld, and U.~A. Wiedemann, {\it {Universal Equilibration Condition for Heavy Quarks}},  {\em Phys. Rev. Lett.} {\bf 135} (2025), no.~24 242301, [\href{http://arxiv.org/abs/2504.21139}{{\tt arXiv:2504.21139}}].

\bibitem{Akamatsu:2008ge}
Y.~Akamatsu, T.~Hatsuda, and T.~Hirano, {\it {Heavy Quark Diffusion with Relativistic Langevin Dynamics in the Quark-Gluon Fluid}},  {\em Phys. Rev. C} {\bf 79} (2009) 054907, [\href{http://arxiv.org/abs/0809.1499}{{\tt arXiv:0809.1499}}].

\bibitem{Beraudo:2009pe}
A.~Beraudo, A.~De~Pace, W.~M. Alberico, and A.~Molinari, {\it {Transport properties and Langevin dynamics of heavy quarks and quarkonia in the Quark Gluon Plasma}},  {\em Nucl. Phys. A} {\bf 831} (2009) 59--90, [\href{http://arxiv.org/abs/0902.0741}{{\tt arXiv:0902.0741}}].

\bibitem{He:2013zua}
M.~He, H.~van Hees, P.~B. Gossiaux, R.~J. Fries, and R.~Rapp, {\it {Relativistic Langevin Dynamics in Expanding Media}},  {\em Phys. Rev. E} {\bf 88} (2013) 032138, [\href{http://arxiv.org/abs/1305.1425}{{\tt arXiv:1305.1425}}].

\bibitem{DuPlessis:2026pyr}
J.~F. Du~Plessis and B.~Scheihing-Hitschfeld, {\it {Heavy Quark Transport is Non-Gaussian Beyond Leading Log}},  \href{http://arxiv.org/abs/2604.21895}{{\tt arXiv:2604.21895}}.

\bibitem{Gossiaux:2008jv}
P.~B. Gossiaux and J.~Aichelin, {\it {Towards an understanding of the RHIC single electron data}},  {\em Phys. Rev. C} {\bf 78} (2008) 014904, [\href{http://arxiv.org/abs/0802.2525}{{\tt arXiv:0802.2525}}].

\bibitem{Uphoff:2014hza}
J.~Uphoff, O.~Fochler, Z.~Xu, and C.~Greiner, {\it {Elastic and radiative heavy quark interactions in ultra-relativistic heavy-ion collisions}},  {\em J. Phys. G} {\bf 42} (2015), no.~11 115106, [\href{http://arxiv.org/abs/1408.2964}{{\tt arXiv:1408.2964}}].

\bibitem{Cao:2016gvr}
S.~Cao, T.~Luo, G.-Y. Qin, and X.-N. Wang, {\it {Linearized Boltzmann transport model for jet propagation in the quark-gluon plasma: Heavy quark evolution}},  {\em Phys. Rev. C} {\bf 94} (2016), no.~1 014909, [\href{http://arxiv.org/abs/1605.06447}{{\tt arXiv:1605.06447}}].

\bibitem{Li:2019wri}
S.~Li, C.~Wang, R.~Wan, and J.~Liao, {\it {Probing the transport properties of Quark-Gluon Plasma via heavy-flavor Boltzmann and Langevin dynamics}},  {\em Phys. Rev. C} {\bf 99} (2019), no.~5 054909, [\href{http://arxiv.org/abs/1901.04600}{{\tt arXiv:1901.04600}}].

\bibitem{Liu:2021dpm}
F.-L. Liu, W.-J. Xing, X.-Y. Wu, G.-Y. Qin, S.~Cao, and X.-N. Wang, {\it {QLBT: a linear Boltzmann transport model for heavy quarks in a quark-gluon plasma of quasi-particles}},  {\em Eur. Phys. J. C} {\bf 82} (2022), no.~4 350, [\href{http://arxiv.org/abs/2107.11713}{{\tt arXiv:2107.11713}}].

\bibitem{Liu:2006he}
H.~Liu, K.~Rajagopal, and U.~A. Wiedemann, {\it {Wilson loops in heavy ion collisions and their calculation in AdS/CFT}},  {\em JHEP} {\bf 03} (2007) 066, [\href{http://arxiv.org/abs/hep-ph/0612168}{{\tt hep-ph/0612168}}].

\bibitem{BitaghsirFadafan:2008adl}
K.~Bitaghsir~Fadafan, H.~Liu, K.~Rajagopal, and U.~A. Wiedemann, {\it {Stirring Strongly Coupled Plasma}},  {\em Eur. Phys. J. C} {\bf 61} (2009) 553--567, [\href{http://arxiv.org/abs/0809.2869}{{\tt arXiv:0809.2869}}].

\bibitem{Chernicoff:2008sa}
M.~Chernicoff and A.~Guijosa, {\it {Acceleration, Energy Loss and Screening in Strongly-Coupled Gauge Theories}},  {\em JHEP} {\bf 06} (2008) 005, [\href{http://arxiv.org/abs/0803.3070}{{\tt arXiv:0803.3070}}].

\bibitem{Herzog:2006gh}
C.~P. Herzog, A.~Karch, P.~Kovtun, C.~Kozcaz, and L.~G. Yaffe, {\it {Energy loss of a heavy quark moving through N=4 supersymmetric Yang-Mills plasma}},  {\em JHEP} {\bf 07} (2006) 013, [\href{http://arxiv.org/abs/hep-th/0605158}{{\tt hep-th/0605158}}].

\bibitem{Gubser:2006bz}
S.~S. Gubser, {\it {Drag force in AdS/CFT}},  {\em Phys. Rev. D} {\bf 74} (2006) 126005, [\href{http://arxiv.org/abs/hep-th/0605182}{{\tt hep-th/0605182}}].

\bibitem{Georgi:1990um}
H.~Georgi, {\it {An Effective Field Theory for Heavy Quarks at Low-energies}},  {\em Phys. Lett. B} {\bf 240} (1990) 447--450.

\bibitem{Chesler:2014jva}
P.~M. Chesler and K.~Rajagopal, {\it {Jet quenching in strongly coupled plasma}},  {\em Phys. Rev. D} {\bf 90} (2014), no.~2 025033, [\href{http://arxiv.org/abs/1402.6756}{{\tt arXiv:1402.6756}}].

\bibitem{Chesler:2015nqz}
P.~M. Chesler and K.~Rajagopal, {\it {On the Evolution of Jet Energy and Opening Angle in Strongly Coupled Plasma}},  {\em JHEP} {\bf 05} (2016) 098, [\href{http://arxiv.org/abs/1511.07567}{{\tt arXiv:1511.07567}}].

\bibitem{Baier:2001yt}
R.~Baier, Y.~L. Dokshitzer, A.~H. Mueller, and D.~Schiff, {\it {Quenching of hadron spectra in media}},  {\em JHEP} {\bf 09} (2001) 033, [\href{http://arxiv.org/abs/hep-ph/0106347}{{\tt hep-ph/0106347}}].

\bibitem{Horowitz:2015dta}
W.~A. Horowitz, {\it {Fluctuating heavy quark energy loss in a strongly coupled quark-gluon plasma}},  {\em Phys. Rev. D} {\bf 91} (2015), no.~8 085019, [\href{http://arxiv.org/abs/1501.04693}{{\tt arXiv:1501.04693}}].

\bibitem{Mateos:2007vn}
D.~Mateos, R.~C. Myers, and R.~M. Thomson, {\it {Thermodynamics of the brane}},  {\em JHEP} {\bf 05} (2007) 067, [\href{http://arxiv.org/abs/hep-th/0701132}{{\tt hep-th/0701132}}].

\bibitem{Mikhailov:2003er}
A.~Mikhailov, {\it {Nonlinear waves in AdS / CFT correspondence}},  \href{http://arxiv.org/abs/hep-th/0305196}{{\tt hep-th/0305196}}.

\bibitem{Cacciari:1998it}
M.~Cacciari, M.~Greco, and P.~Nason, {\it {The $p_T$ spectrum in heavy-flavour hadroproduction.}},  {\em JHEP} {\bf 05} (1998) 007, [\href{http://arxiv.org/abs/hep-ph/9803400}{{\tt hep-ph/9803400}}].

\bibitem{Cacciari:2001td}
M.~Cacciari, S.~Frixione, and P.~Nason, {\it {The p(T) spectrum in heavy flavor photoproduction}},  {\em JHEP} {\bf 03} (2001) 006, [\href{http://arxiv.org/abs/hep-ph/0102134}{{\tt hep-ph/0102134}}].

\bibitem{Cacciari:2012ny}
M.~Cacciari, S.~Frixione, N.~Houdeau, M.~L. Mangano, P.~Nason, and G.~Ridolfi, {\it {Theoretical predictions for charm and bottom production at the LHC}},  {\em JHEP} {\bf 10} (2012) 137, [\href{http://arxiv.org/abs/1205.6344}{{\tt arXiv:1205.6344}}].

\bibitem{Cacciari:2015fta}
M.~Cacciari, M.~L. Mangano, and P.~Nason, {\it {Gluon PDF constraints from the ratio of forward heavy-quark production at the LHC at $\sqrt{S}=7$ and 13 TeV}},  {\em Eur. Phys. J. C} {\bf 75} (2015), no.~12 610, [\href{http://arxiv.org/abs/1507.06197}{{\tt arXiv:1507.06197}}].

\bibitem{NNPDF:2014otw}
{\bf NNPDF} Collaboration, R.~D. Ball et~al., {\it {Parton distributions for the LHC Run II}},  {\em JHEP} {\bf 04} (2015) 040, [\href{http://arxiv.org/abs/1410.8849}{{\tt arXiv:1410.8849}}].

\bibitem{ALICE:2021mgk}
{\bf ALICE} Collaboration, S.~Acharya et~al., {\it {Measurement of beauty and charm production in pp collisions at $ \sqrt{s} $ = 5.02 TeV via non-prompt and prompt D mesons}},  {\em JHEP} {\bf 05} (2021) 220, [\href{http://arxiv.org/abs/2102.13601}{{\tt arXiv:2102.13601}}].

\bibitem{Casalderrey-Solana:2014bpa}
J.~Casalderrey-Solana, D.~C. Gulhan, J.~G. Milhano, D.~Pablos, and K.~Rajagopal, {\it {A Hybrid Strong/Weak Coupling Approach to Jet Quenching}},  {\em JHEP} {\bf 10} (2014) 019, [\href{http://arxiv.org/abs/1405.3864}{{\tt arXiv:1405.3864}}]. [Erratum: JHEP 09, 175 (2015)].

\bibitem{Casalderrey-Solana:2018wrw}
J.~Casalderrey-Solana, Z.~Hulcher, G.~Milhano, D.~Pablos, and K.~Rajagopal, {\it {Simultaneous description of hadron and jet suppression in heavy-ion collisions}},  {\em Phys. Rev. C} {\bf 99} (2019), no.~5 051901, [\href{http://arxiv.org/abs/1808.07386}{{\tt arXiv:1808.07386}}].

\end{thebibliography}\endgroup

\end{document}